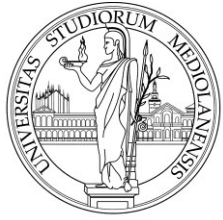

# Università degli Studi di Milano

**SCUOLA DI DOTTORATO IN INFORMATICA**

Tesi di Dottorato

XXIV Ciclo

# A Distributed Approach to Privacy on the Cloud

## Francesco Pagano

Relatore: Prof. Ernesto Damiani

Correlatore: Prof. Stelvio Cimato

Direttore della Scuola di Dottorato in Informatica:

                                                    Prof. Ernesto Damiani

Anno Accademico 2010/2011

# Abstract

The increasing adoption of Cloud-based data processing and storage poses a number of privacy issues. Users wish to preserve full control over their sensitive data and cannot accept it to be fully accessible to an external storage provider. Previous research in this area was mostly addressed at techniques to protect data stored on untrusted database servers; however, I argue that the Cloud architecture presents a number of specific problems and issues. This dissertation contains a detailed analysis of open issues. To handle them, I present a novel approach where confidential data is stored in a highly distributed partitioned database, partly located on the Cloud and partly on the clients.

In my approach, data can be either private or shared; the latter is shared in a secure manner by means of simple grant-and-revoke permissions. I have developed a proof-of-concept implementation using an in-memory RDBMS with row-level data encryption in order to achieve fine-grained data access control. This type of approach is rarely adopted in conventional outsourced RDBMSs because it requires several complex steps. Benchmarks of my proof-of-concept implementation show that my approach overcomes most of the problems.

# Acknowledgements

I want to thank many people who contributed to the final result of this work:
- My family: Antonella, Davide, and Riccardo, who has endured my long absences for the PhD;
- My uncle Augusto Pagano, who has corrected lots of English errors;
- My tutors, the professors Ernesto Damiani, Alessandro Provetti, and Stelvio Cimato, for their continuous support;
- The professors Giacomo Fiumara and Luigia Puccio, who helped me in numerical analysis;
- The professors Elisa Bertino, Sabrina De Capitani di Vimercati, and Pierangela Samarati, who gave me their seminal papers on privacy in outsourced servers and in the Cloud.

# Table of contents

# List of Figures

# List of Tables

# 1 INTRODUCTION

Cloud computing, today, provides users with readily available, pay-as-you-go computing and storage power, allowing them to dynamically adapt their Information Technology (IT) costs to their needs [23]. With Cloud computing, users need neither huge investments in the start-up phase nor costly competence in IT system management.

While the Cloud computing concept is drawing much interest, several obstacles remain to its widespread adoption [23], including:

- Current limits of ICT infrastructure: network availability, reliability, and quality of service;
- Differences in the development process of Cloud applications with respect to ordinary ones; and, importantly,
- Privacy risks for confidential information residing in the Cloud.

Thanks to increasing network availability, the first obstacle is becoming lesser and lesser of an issue over time; the second will tend to disappear with the training of new developers and by introducing new software development methods; the third obstacle, however, is far from being solved and may seriously impair the real prospects of Cloud computing.

This is particularly the case for Public Clouds [3], i.e., those available, upon subscription, to the general public (as opposed to Private Clouds operated by large enterprises and organizations for their employees).

Nowadays, the use of Public Clouds is becoming more and more frequent, thanks to productivity and collaboration tools (e.g. Google Apps[1], Zimbra[2], etc.), online social networks (such as Facebook[3] and LinkedIn[4]), etc. Those applications/services store a lot of information on the Cloud, many of which is confidential. Hence, a strong data access control is needed to prevent unauthorized uses. Typically, external access to shared data held in the Cloud goes through the usual authentication, authorization, and communication phases.

---

[1] www.google.com/a

The access control problem is well acknowledged in the database literature and the available approaches guarantee a high degree of assurance. For the traditional datacenters' scenario, where storage is internal to the Enterprise and considered trusted, the attacks are expected from the outside. When data is outsourced, though, even the data storage and its administrators are external to the Enterprise. Therefore, the requirement that the maintainer of the datastore may not access or alter the outsourced data is not easily met. This is especially the case for commercial Public Clouds like Google App Engine for Business, Microsoft Azure Platform or Amazon EC2 platform.

The privacy issue of Cloud environment is not only related to the datastore protection in untrusted servers. In a Cloud-based Data Center, a number of serious questions arise about the available services and the stored data [58]:

- Who controls them?
- Where are the servers located?
- Where is the data stored?
- Which legislation applies to them?
- Who backs up the servers?
- Who has access to the servers?
- How resilient is the service?
- How do auditors observe?
- How does users' security-team engage?

Hence, new categories of risks come up:

---

[2] www.zimbra.com

[3] www.facebook.com

[4] www.linkedin.com

- Control: many organizations are uncomfortable with having their data managed by systems and staff they neither know nor control; therefore the providers must offer a high degree of reputation, security, and transparency to win users' confidence.
- Privacy: migrating data to a shared computing infrastructure increases the potential for unauthorized exposure, privacy breaches, and data loss. Authentication and access control techniques, and data replication are crucial.
- Compliance: often the keeping of sensitive data (e.g., customers' identities) is regulated by strict national/international laws, such as the Health Insurance Portability and Accountability Act (HIPAA)[5] and the Sarbanes–Oxley Act (SOX)[6]. These and other regulations may severely limit the use of Clouds in practice.
- Security Management: even the simplest task may end up behind layers of abstraction or performed by someone else. Providers must supply easy-to-use controls to manage security settings for application and run-time environments.

In a well-attended keynote address at ICWS 2010 [58], Elisa Bertino has summarized the most important security threats emerged in some of the last conferences about Cloud computing:

- Abuse and Nefarious Use of Cloud Computing
- Insecure Application Programming Interfaces
- Malicious Insiders
- Shared Technology Vulnerabilities
- Data Loss/Leakage
- Account, Service & Traffic Hijacking
- Unknown Risk Profile
- Loss of governance
- Lock-in
- Isolation failure

---

[5] http://www.gpo.gov/fdsys/pkg/PLAW-104publ191/content-detail.html

[6] http://www.gpo.gov/fdsys/pkg/PLAW-107publ204/content-detail.html

- Compliance risks
- Management interface compromise
- Data protection
- Insecure or incomplete data deletion
- Malicious Insider
- Privileged user access
- Regulatory compliance
- Data location
- Data segregation
- Recovery
- Investigative support
- Long-term viability.

To complicate the scenario, the availability of many alternatives for Cloud sharing (Private, Public, Federated, and Hybrid) and delivery models (IaaS, PaaS, and SaaS) makes it difficult, if not impossible, to find universal solutions to privacy issues.

Although there are a number of techniques for preserving privacy of outsourced data on untrusted database servers (see chapter 2-Background on data protection), I will argue that they cannot be applied directly to the Cloud, particularly in the case of Public Cloud. First, these techniques were designed for RDBMS, while newer and less structured data models are increasingly used on the Cloud (see Section 6.3 - Semantic datastore). Second, besides managing data storage, the Cloud hosts all application logic up to the presentation layer. Hence, if a Cloud supplier even becomes untrustworthy, it can intercept communications between data storage and application logic, can monitor the user application memory and can modify (e.g., using Aspect Programming) presentation software components used to display application output to the final users.

Essentially, encryption-based techniques for safely outsourcing data to untrusted DBMS cannot guarantee the confidentiality of data in the Cloud. Even if the data layer is secured through encryption, at some points, on its path toward the user, the data will be in plaintext form, i.e., unprotected and vulnerable.

***To maintain data privacy, this dissertation proposes to move the entire presentation layer of Cloud applications from Server to Client side.***

However, separating the data layer (which would stay in the Cloud) from the presentation logics (which would stay in the client) may lead to an inefficient cooperation between the two parts. For this reason, I think it is better to move also the data layer into the client.

To ensure seamless cooperation between the data layer and the presentation logic, I propose a highly distributed architecture that is composed of a set of local nodes that run the applications and store all the data. Data is replicated when needed on all the nodes that access it. The local copies shall be synchronized using a central service on the Cloud using encryption, in the protocol which I designed, to guarantee data protection during the synchronization phase.

Moreover, the architecture I propose allows fine-grained data access control and has the capability to revoke the access rights to a local node. To achieve this result, a row-level encrypted database is used. Since this type of encryption is rarely adopted in conventional RDBMSs, as it requires several complex steps, I propose an extension to add this functionality to In-Memory DataBases.

A test implementation of the architecture and of an extended IMDB was realized and used to benchmark the system.

Finally, I studied and utilized a standard email server on the Cloud as Synchronizer, evaluating the impact of this choice in performance and scalability.

## 1.1 Structure of this dissertation

This thesis mainly consists of three parts. In the first one (Chapters 2-6) I want to give a survey of the state of the art, in particular of:

- Data protection in outsourced database (protection techniques on untrusted servers, data access control, etc.), in Chap. 2;
- Cryptography in databases (the various level of encryption and the granularity in database level encryption), in Chap. 3;

- In-memory databases, in Chap. 4;
- Broadcast in communication networks and Online Social Networks, in Chap. 5;
- An analysis of Cloud Computing and the peculiarity of Privacy within the Cloud, with particular attention to the differences between the Cloud environment and the untrusted outsourced data servers, in Chap. 6.

In the second part (Chapters 7-10), I present the innovative work:

- The UML complete layout of a distributed secure architecture for data sharing, in Chap. 7;
- The Scenarios where the proposed system fits, in Chap. 8;
- The structure, and the custom algorithms for secure data sharing, in Chap. 9;
- A self-assessment of what has been achieved, and a comparison with the original goals, in Chap. 10.

The third part (Chapters 11-16) is dedicated to the pilot implementation and its testing, as a way to validate the framework:

- Implementation and benchmarking, in Chap. 11;
- Further enhancement to improve scalability of the system, in Chap. 12;
- Comparison with other approaches, in particular with Broadcast Encryption, in Chap. 13;
- Vulnerability assessment, a key to prevent possible attacks, in Chap. 14;
- The scenarios for a successful deployment of my system, in Chap. 15, and
- Conclusion and future works, in Chap. 16.

# Part I - Background and state of the art

# 2 BACKGROUND ON DATA PROTECTION

*This chapter presents the most important privacy issues in databases outsourcing on untrusted servers. I illustrate the main techniques for providing data protection and for securing the confidentiality of data stored at external honest-but-curious servers. Then, I discuss the state of the art of access control to outsourced data, describing how data can be selectively accessed from users and how access rights may change over time.*

## 2.1   Outsourced storage on untrusted servers

Data storage involves high costs because it requires physical resources (disks, servers, etc.), frequent management procedures (backup, tuning, etc.), and skilled administrative staff.
A cheaper solution is data outsourcing to a specialized provider that takes care of the storage within its own specialized structure to offer high availability and disaster protection [108]. Delegating data management to outsourced untrusted entity implies a risk for confidentiality and privacy, and a potential improper use of database information (harvesting, targeting, reselling, etc.) by the provider itself [15], since traditional access control techniques prevent data access by external users, but not by internal administrators (DBA).
The main issues for guaranteeing proper protection and access to outsourced data are [6]:
- Data protection: the storage server is responsible of data management but should not be authorized to know the actual data content. To achieve this goal, almost all the proposed approaches in literature employ encryption to secure customers' data [95][106][109], or on splitting (fragments of the original data) across several or tables [110][111][112].

- Query execution: if the server stores encrypted information, it is not able to execute the users' queries, at most it could send the encrypted tables involved in the query to the requester, which, then, needs to decrypt and query by herself [15]. Also in case of fragmentation, the client needs some logic to recombine different subqueries [6]. To decrease the client overhead, we try to process the request as far as possible on the server, leaving the client with a task of finishing. For this purpose, a set of indexes may be added to the encrypted information [114][115][109]. The response of server is a trade-off between precision (the fraction of the information retrieved that is relevant to the user's information need[113]) and privacy, because a precision too close to 1 allows statistical data mining, while a precision too far from 1 overloads the client.

- Private access: the server may not inference information storing and analyzing the queries. This lead to the concept of oblivious searches on public key encrypted data [116].

- Data integrity and correctness: in addiction to honest-but-curious servers, which are untrusted w.r.t. data access but trusted w.r.t. properly enforcing data storage and management, we have also to consider totally untrusted server, which can alter stored data or queries' results. In this case, a mechanism to check integrity and correctness of data is needed. It may be based on signatures attached to tuples in the database [117][118], or on chain structures as skip lists [119] that allow the client to the integrity of the returned tuples.

- Access control enforcement: usually, the existing database access control mechanisms assume that the server is in charge of defining and enforcing access control. In the case of outsourced data, instead, it is unfeasible since the access control policy itself might be sensitive (and so it needs to be hidden to the server), access control restrictions might depend on the data content (that the server may not read), and an untrusted server may alter the access control. At the same time, it is not possible to give in charge the access control to data owner, to filter out from the query result the tuples that a client cannot access, because it is too expensive and it brings to bottlenecks. A feasible way is the use of different encryption keys for different data as proposed, for example, for XML documents [121]. To access such encrypted data, users have to decrypt them by using the appropriate key. If different users know different keys, they have different access rights [120].

- Support for selective write privileges: the previous problem considers only the "read" access control, but the same happens for "write" access.
- Data publication and utility: given a database instance containing sensitive information, it needs to be "anonymized" to obtain a view such that attackers cannot learn any sensitive information from the view, and legitimate users can use it to compute useful statistics [122]. Data protection and utility can be seen as conflicting goals: the more data are encrypted or obfuscated, the less the ability to withdraw knowledge and inferences from them.
- Private collaborative computation: sometimes, the data that comes to a client is the result of interaction between many servers that collaborate to accomplish a service. In this case, the goal is to perform the computation on each server without revealing the data used to the extern [123][124].

## 2.2 Data protection techniques

To prevent a server from accessing data stored on its own machines, the literature reports three major families of data protection techniques on untrusted servers [6]:

- Data encryption [15];
- Data fragmentation and encryption [16], which in turn can be classified into two major techniques
    - non-communicating servers [17][18];
    - unlinkable fragments [19], and
- Data fragmentation with owner involvement [20].

The goal of each of these techniques is to store data on the untrusted server in an inaccessible format, using encryption and/or fragmentation, to prevent provider's access to stored information. In all the scenarios, we can see four actors [15], as represented in Figure 1:

1. An outsourced server that operates on behalf of one or more data owners that outsource their data,
2. A data owner that store information into the server,
3. A human user that queries the database, and
4. A client software that elaborates locally the queries.

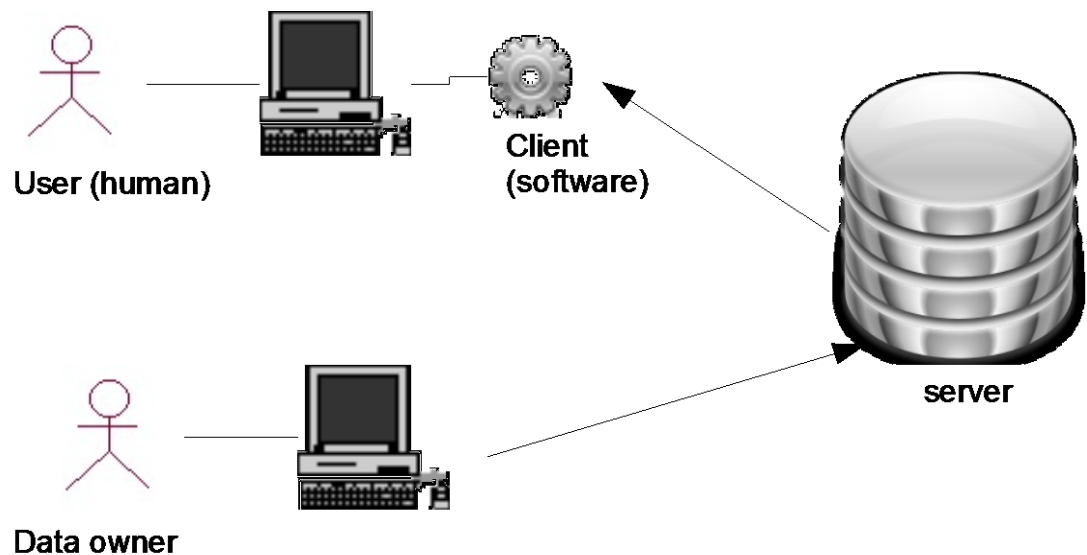

Figure 1　　The actors in protected outsourced storage

The process is:

1. The data owner stores information into the server accordingly with the used technique (e.g. after having encrypted the information);
2. The server stores this data;
3. When a user wants to access data, she creates a Db query $Q_1$. The query is processed by the client software that manipulates it to obtain an equivalent query $Q_2$ that is processable by the server (that cannot directly access data content). The latter elaborates $Q_2$ and sends the result set (a superset of the result set of $Q_1$), to the client, which post-processes it to delete the spurious results and to obtain the right result set of $Q_1$.

### 2.2.1 Data encryption

If data is outsourced to an untrusted RDBMS (Relational Data Base Management Systems), to prevent unauthorized access by the datastore manager (DM) of the outsourced, it may be stored in encrypted form. The encryption happens on client side, before sending data for storage to the external server.

#### *2.2.1.1 Data Model*

While data can be encrypted at various granularity levels (see Section 3), for balancing client workload and query execution efficiency, most proposals assume that the database is encrypted at tuple level [6]. For the same reason, symmetric encryption is usually preferred over asymmetric encryption since it is cheaper [125].

Obviously, the DM does not know the encryption keys, which are stored apart from the data; the RDBMS receives an encrypted database and works on bit-streams that only the clients, who hold the decryption keys, can interpret correctly.

In current systems, decryption keys are generated and distributed to trusted clients by the data owner or by a trusted delegate [110].

It is important to remark that, since data is encrypted, the DBMS cannot index it based on plaintext and therefore it cannot resolve all queries.

Some proposals [24][95][109] solve this problem by providing, for each encrypted field to be indexed, an additional indexable field, obtained by applying a non-injective transformation $f$ to plaintext values (e.g., a hash of the field's content).

Formally [15], given a plaintext database $B$, each table $r_i$ over schema $R_i(A_{i1},..., A_{in})$, where $R_i$ is the i-th relation and $A_j$ the j-th attribute, in $B$ is mapped onto a table $r_i^k$ over schema $R_i^k$ (ID, Etuple, $I_1,..., I_n$) in the corresponding encrypted database $B^k$, where:

- ID is a numerical attribute added as primary key of the encrypted table;
- Etuple is the attribute containing the encrypted tuple whose value is obtained applying an encryption function $E_k$ to the plaintext tuple, where $k$ is the secret key; and

- $I_j$ is the index associated with the j-th attribute in $R_i$.

Encrypted tuples and indexes can be stored in the same or in a separate table [13]. Conventionally, index values are represented with Greek letters.

Figure 2 shows the transformation of a plain text tuple into an encrypted one.

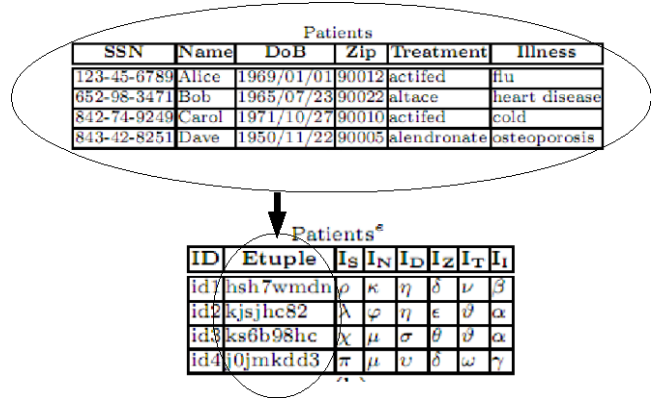

Figure 2    Data encryption, source: [6]

### 2.2.1.2  Query execution

The presence of indexes allows doing most of the work to answer a query at server side. Referring to Figure 1, when the user *U*, which needs not be aware that data have been outsourced to a third party, issues a query *Q*, the latter is passed to the client *C*, which splits it in $Q_s$, which will be executed on the server, and $Q_c$, which will be executed on the client. $Q_s$ works on the encrypted tables through indexes, and produces a result set $RS_e$ (a set of the encrypted tuples) that is a superset of *RS*, the result set for *Q*. However, due to index collisions the transformed query may return spurious tuples, that is, tuples that do not belong to *RS*; for this reason, *C* receives $RS_e$ and, using $Q_c$, filters the spurious tuples and performs the needed projections, obtaining *RS*, which is returned to *U* [15].

### 2.2.1.3  Indexing techniques

Since a client may have a limited storage and computation capacity, one of the primary goals of the query execution process is to minimize the workload at the client side, while maximizing the operations that can be computed at the server side [120][109]. Different index approaches allow the execution of different types of queries at server-side. In the following, I discuss the main type of indexes.

###### 1.1.1.1.1 *Index by encryption*

Each search key value is encrypted using an invertible encryption function $E_k()$. A query on a plaintext relation has to be transformed into a query on the corresponding encrypted relation by simply applying the encryption function on each value specified in the original query. This technique is simple and has the advantage of preserving the distinction between values, but it is often possible to guess the correspondence between plaintext and encrypted values based on frequency analysis, that is, by comparing the distributions of the plaintext values in the plain-text relations with the corresponding distributions of the encrypted values [120].

###### 1.1.1.1.2 *Bucket-Based Index*

This is a variant of the previous technique that reduces, however without solving, the risk of inference analysis increasing the number of collisions among search values. The domain $D$ of search values is mapped into a set of non overlapping partitions $P=\{p_1,..,p_k\}$ whose union covers all $D$. The index does not contain the single value of a tuple, but its equivalence class (partition) [109].

Using Bucket-Based Index, equality queries can be performed easily, although the result set needs to have a precision index < 1 to prevent statistical data mining. The trusted client, after receiving the encrypted result set, can decrypt it and exclude spurious tuples. Also, range queries are difficult to compute, since the transformation $f$ in general shall not (and should not) preserve the order relation of the original plaintext data. Specifically, it will be impossible for the outsourced RDBMS to answer range queries that cannot be reduced to multiple equality conditions (e.g., 1<=x<=3 can be translated into x=1 or x=2 or x=3) unless specific techniques are applied [109]. An efficient way for partitioning the domain of attributes, to minimize the number of spurious tuples in the result of a range/equality query, is shown in [126].

###### 1.1.1.1.3 *Index by hashing*

Bucketing preserves the relation between two adjacent values. Instead, using a secure one-way hash function *f* [7], which takes as input the clear values of an attribute and returns the corresponding index values, we obtain the same result, but without the proximity relationship [13]. *f* has two interesting properties:

- Since y=*f*(x) has a smaller bit-length than x, we still have collisions; and
- A secure hash function uniformly covers its range (i.e., the output probabilities from the hash function are uniform).

The resulting distribution of hash values is more dispersive, making frequency-based attacks ineffective.

### 1.1.1.1.1. *Auxiliary B+-tree*

To handle range queries, a solution, among others, is to use an encrypted version of a B+-tree to store plaintext values and to maintain the ordering [106].

Definition [127]: A B+-tree of order $m$ is a tree where each internal node contains up to m branches (children nodes) and thus store up to $m$ -1 search key values. $m$ is also known as the branching factor or the fanout of the tree.

- The B+-tree stores records (or pointers to actual records) only at the leaf nodes, which are all found at the same level in the tree, so the tree is always height balanced.
- All internal nodes, except the root, have between $\lceil m/2 \rceil$ and $m$ children
- The root is either a leaf or has at least two children.
- Internal nodes store search key values, and are used only as placeholders to guide the search. The number of search key values in each internal node is one less than the number of its non-empty children, and these keys partition the keys in the children in the fashion of a search tree. The keys are stored in non-decreasing order (i.e. sorted in lexicographical order).

---

[7] *f* must be deterministic and non-injective

- Depending on the size of a record as compared to the size of a key, a leaf node in a B+-tree of order m may store more or less than m records. Typically this is based on the size of a disk block, the size of a record pointer, etcetera. The leaf pages must store enough records to remain at least half full.
- The leaf nodes of a B+-tree are linked together to form a linked list. This is done so that the records can be retrieved sequentially without accessing the B+-tree index. This also supports fast processing of range-search queries as will be described later.

Figure 3 represents an example of a B+-tree.

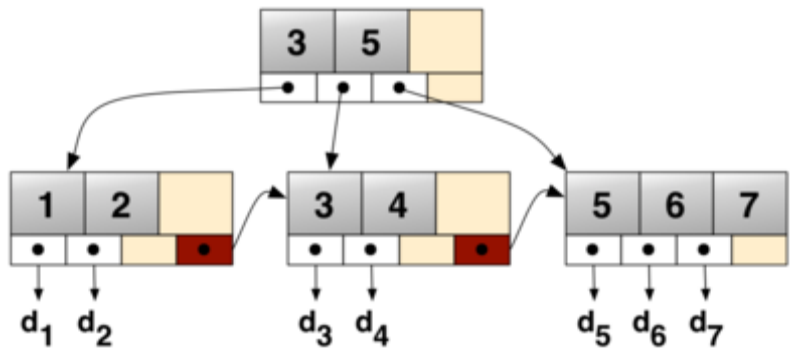

Figure 3    A B+-tree sample

To adapt to encrypted databases, the B+-tree is built as plaintext structure, and then each node is encrypted (to protect the sensitive data) as in Figure 4:

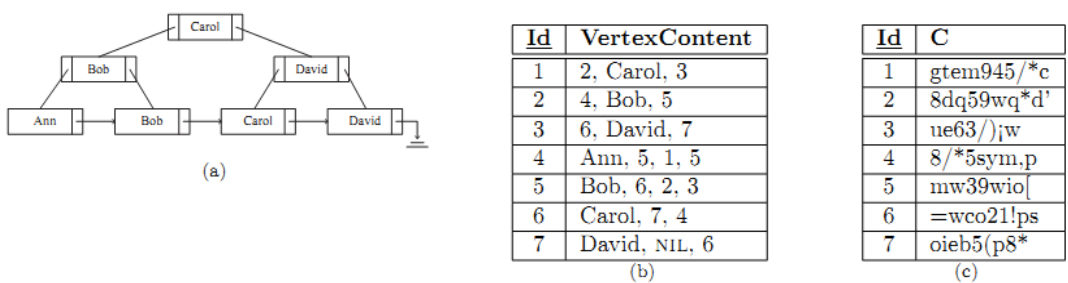

Figure 4    An encrypted B+-tree, source: [15]

Since data values are encrypted, the tree is managed at the Client side and it is read-only in the Server side.

This structure allows the client to make range queries. The client, starting from the root node, traverses the index. At each step, the client receives an encrypted node of the index, decrypts it, evaluates its content and then traverses the tree as usual for BST, asking the server for the next node, until reaching a leaf [14].

###### 1.1.1.1.1 *Other Approaches*

The previous indexing methods are not the only proposals. Table 1 summarizes other approaches that try to better support SQL clauses or to reduce the amount of spurious tuples in the result produced by the remote server, and the supported queries:

Table 1     Indexing methods supporting queries [15]

| Index | Query[8] | | |
|---|---|---|---|
| | Equality | Range | Aggregation |
| Bucket-based | + | o | - |
| Hash-based | + | - | o |
| B+ Tree | + | + | + |
| Character oriented | + | o | - |
| Privacy homomorphism | + | - | + |
| Partition Plaintext and Ciphertext (PPC) | + | + | + |
| Order Preserving Encryption Schema (OPES) | + | + | o |
| Secure index data structures | + | o | - |

## 2.2.2 Data fragmentation

Often, not all the outsourced data, but only some columns and/or some relations are confidential, e.g. the relation between patient and illness, in medical field. In this case, it is possible to split the outsourced information in two parts, one confidential and the other public. The aim of this solution is to minimize the computational load of encryption/decryption.

A confidentiality constraint *c* over relational schema $R(A_1,...,A_n)$ is a subset of attributes of *R*, i.e. $c \subseteq R$. Confidentiality constraints can contain a single attribute that is sensitive (singleton constraints) or a group of attributes that need to be never stored together since their joint visibility is sensitive (association constraints).

Figure 5 shows an example of medical data and a set of well-defined constraints over it.

---

[8] +=fully supported, o=partially supported, -=not supported.

| MEDICALDATA | | | | | | |
|---|---|---|---|---|---|---|
| SSN | Name | DoB | ZIP | Illness | Treatment | |
| 123456789 | Alice | 84/07/31 | 26015 | Pharyngitis | Antibiotic | $c_0$={SSN} |
| 231546586 | Bob | 82/05/20 | 26010 | Flu | Aspirin | $c_1$={Name, DoB} |
| 378565241 | Carol | 20/01/30 | 50010 | Gastritis | Antacid | $c_2$={Name, ZIP} |
| 489754278 | David | 80/07/02 | 20015 | Broken Leg | Physiotherapy | $c_3$={Name, Illness} |
| 589076542 | Emma | 75/02/07 | 26018 | Gastritis | None | $c_4$={Name, Treatment} |
| 675445372 | Fred | 75/02/17 | 26013 | Asthma | Bronchodilator | $c_5$={DoB, ZIP, Illness} |
| 719283746 | Gregory | 70/05/04 | 26020 | Diabetes | Insulin | $c_6$={DoB, ZIP, Treatment} |
| 812345098 | Henrik | 65/12/08 | 20010 | Cancer | Chemoterapy | |

(a)  (b)

Figure 5   A fragmentation sample (a) and a set of well-defined constraints over it (b), source: [128]

### 2.2.2.1 Non-communicating servers

In this technique, two split databases are stored each in a different untrusted server (called, say, $S_1$ and $S_2$). The two untrusted servers need to be independent and non-communicating to prevent their alliance and reconstruction of the complete information. In this scenario, the information may be stored encoded or as plaintext at each server [110]. Basically, sensitive attributes (singleton constraints) need to be encrypted, while sensitive associations can be protected by splitting (fragmenting) the involved attributes among the two servers.

Given a relational schema $R$, a fragmentation of $R$ is then a triple ($F_1$, $F_2$, $E$), where fragments $F_1$ and $F_2$ contain a set of attributes in the clear (including a tuple identifier to ensure lossless decomposition) and a set E of attributes encrypted (i.e., E ⊆ $F_1$ and E ⊆ $F_2$).

Figure 6 shows a correct fragmentation for the sample in Figure 5.

| tid | SSN | Name | ZIP | tid | DoB | Illness | Treatment |
|---|---|---|---|---|---|---|---|
| 1 | 123456789 | Alice | 26015 | 1 | 84/07/31 | Pharyngitis | Antibiotic |
| 2 | 231546586 | Bob | 26010 | 2 | 82/05/20 | Flu | Aspirin |
| 3 | 378565241 | Carol | 50010 | 3 | 20/01/30 | Gastritis | Antacid |
| 4 | 489754278 | David | 20015 | 4 | 80/07/02 | Broken Leg | Physiotherapy |
| 5 | 589076542 | Emma | 26018 | 5 | 75/02/07 | Gastritis | None |
| 6 | 675445372 | Fred | 26013 | 6 | 75/02/17 | Asthma | Bronchodilator |
| 7 | 719283746 | Gregory | 26020 | 7 | 70/05/04 | Diabetes | Insulin |
| 8 | 812345098 | Henrik | 20010 | 8 | 65/12/08 | Cancer | Chemoterapy |

Figure 6   A correct fragmentation sample, source: [128]

Encrypted attributes and tuple identifier are contained in $F_1$ as in $F_2$. $F_1$ and $F_2$ are then stored respectively in $S_1$ and $S_2$.

Figure 7 schematizes the resulting structure.

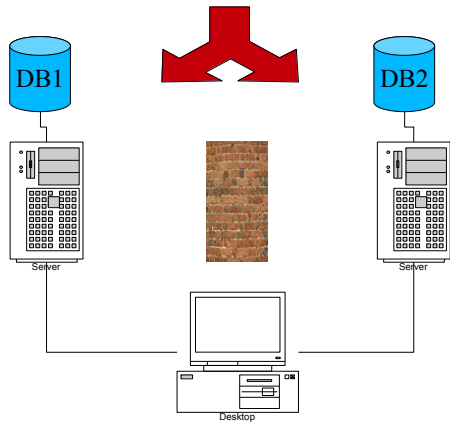

Figure 7   Non-communicating servers

### 1.1.1.1.4  *Query execution*

The client decomposes each query Q in two subqueries: $Q_1$, which is executed in $S_1$ on the fragment $F_1$ giving the result set $RS_1$, and $Q_2$, which is executed in $S_2$ on the fragment $F_2$ giving the result set $RS_2$. If needed, the join between $RS_1$ and $RS_2$ is made choosing one of these strategies:

- Execute $Q_1$ and $Q_2$ in parallel, and then the client joins $RS_1$ and $RS_2$;
- Execute first one sub-query (say it is $Q_1$ in $S_1$), then send $RS_1$ to $S_2$ to execute $Q_2$ and to make the join with $RS1$. $RS_2$ is the final result set to return to client.

While the first strategy is heavier since are larger and the client needs local computing, the second exposes some plaintext data to one of the server, and then is potentially dangerous [110].

## 2.2.2.2  *Unlinkable fragments*

In practice, it is not easy to ensure that split servers do not communicate; therefore the previous technique may be almost inapplicable. A possible remedy is to divide the information in two or more fragments. Each fragment contains all the fields of the original information, but some are in clear-text while the others are encrypted [19].

E.g., given the relation R=(SSN, Name, DoB, Zip, Treatment, Illness), the physical fragments corresponding to its fragmentation F ={ {Name, DoB ,Zip}, {Illness}, {Treatment } } are illustrated in Figure 8:

| | | $F_1$ | | | | | $F_2$ | |
|---|---|---|---|---|---|---|---|---|
| Salt | Enc | Name | DoB | Zip | | Salt | Enc | Illness |
| $s_1$ | $\alpha$ | Alice | 1980/01/01 | 90012 | | $s_5$ | $\epsilon$ | flu |
| $s_2$ | $\beta$ | Bob | 1965/07/23 | 90022 | | $s_6$ | $\varepsilon$ | heart disease |
| $s_3$ | $\gamma$ | Carol | 1971/10/27 | 90010 | | $s_7$ | $\zeta$ | cold |
| $s_4$ | $\delta$ | Dave | 1950/11/22 | 90005 | | $s_8$ | $\eta$ | osteoporosis |

| | | $F_3$ |
|---|---|---|
| Salt | Enc | Treatment |
| $s_9$ | $\theta$ | actifed |
| $s_{10}$ | $\vartheta$ | altace |
| $s_{11}$ | $\iota$ | actifed |
| $s_{12}$ | $\kappa$ | alendronate |

Figure 8    A fragmentation with encryption sample, source: [6]

Here, *enc* contains the encryption of all the attributes that are not in plaintext. To protect *enc* from the so-called frequency attacks, a suitable salt is applied to it. These fragments may be stored in one or more servers. Figure 9 schematizes the resulting structure.

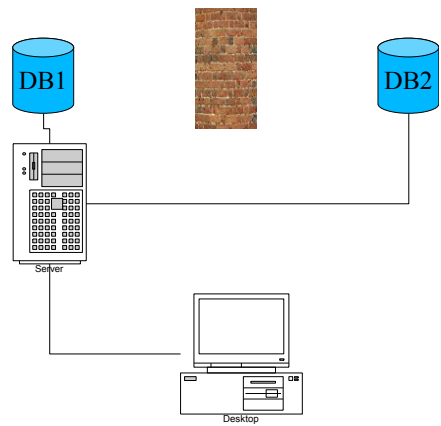

Figure 9    Unlinkable fragments

### *1.1.1.1.5    Choice of fragments*

Generally, given a relational schema $R$ and a set $C$ of confidentiality constraints over it, different fragmentations may exist that are both correct and non-redundant [128]. The designer must choose wisely to ensure that the information cannot be reconstructed. In particular, care should be taken to not leave clear-text primary / external keys that could allow reconstructing the entire information (the chosen fragments need to be *unlinkable*). Moreover, among the correct fragmentations, the designer has to choose a solution that provides minimality. If minimality is characterized by the minimum number of fragments, the problem is NP-hard. An alternative definition of minimality assumes that a solution is minimal if merging any two fragments would break at least a confidentiality constraint; using this definition, a heuristic approach is proposed in [19].

Some tools exist that allow, given a relation table, to produce views (vertical fragments) over it, in such a way to protect the privacy of possible sensitive information while providing maximal visibility over the data. A sample of these tools is Pri-views[9]. It is based on a greedy algorithm designed by University of Bergamo (UNIBG) and University of Milano (UNIMI) to solve the problem of creating unlinkable fragments in the storage of sensitive attributes [20]. The used algorithm departs from the use of encryption, while, usually, in the literature, this kind of problem has been addressed using both fragmentation and encryption.

### *1.1.1.1.6 Query execution*

Having plaintext attributes, the queries that can be resolved using only them are very efficient since they do not need decryption.

At query time, the original query is then decomposed in two subqueries:

- The first, executed on the Server, chooses a fragment (all fragments contain the entire information) and selects tuples from it according to clear-text values. It returns a result set where some fields are encrypted;
- The second, executed on the Client (only if encrypted fields are involved in the query), decrypts the information and removes the spurious tuples.

An example of query in unlinkable fragments scenario is represented in Table 2:

---

[9] http://www.primelife.eu/results/opensource/60-pri-views

Table 2	Query translation using unlinkable fragments [6]

```
SELECT Name, Illness
FROM   Patients
WHERE  DoB<1970/01/01 AND Treatment LIKE 'actifed'
```

(a) Original query $Q$

```
SELECT Salt, Enc, Name
FROM   F_1
WHERE  DoB<1970/01/01
```

(b) Query operating on fragment $F_1$

```
SELECT Name, Illness
FROM   Decrypt(Res_{Q_{F_1}}, k)
WHERE  Treatment LIKE 'actifed'
```

(c) Query operating at the client

### 2.2.3 Data fragmentation with owner involvement

An adaptation of the non-communicating-servers technique consists of storing locally the sensitive data and relations, while outsourcing storage of the generic data [16], as shown in Figure 10.

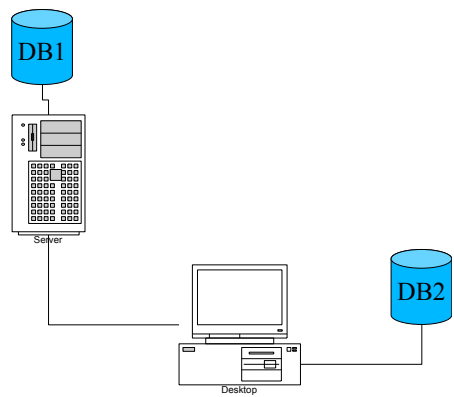

Figure 10	Data fragmentation with owner involvement

The original relational schema $R$ is split in <$F_o$, $F_s$>, where the former is stored at the data owner, and the latter is stored at the external server. Both have a common tuple identifier, to reconstruct the original relation.

Figure 11 shows a fragmentation for the sample in Figure 5.

$F_o$

| ID | SSN | Treatment | Illness |
|---|---|---|---|
| $id_1$ | 123-45-6789 | actified | flu |
| $id_2$ | 652-98-3471 | altace | heart disease |
| $id_3$ | 842-74-9249 | actified | cold |
| $id_4$ | 843-42-8251 | alendronate | osteoporosis |

$F_s$

| ID | Name | DoB | Zip |
|---|---|---|---|
| $id_1$ | Alice | 1980/01/01 | 90012 |
| $id_2$ | Bob | 1965/07/23 | 90022 |
| $id_3$ | Carol | 1971/10/27 | 90010 |
| $id_4$ | Dave | 1950/11/22 | 90005 |

Figure 11    An example of physical fragments with owner involvement, source: [6]

The fragmentation <$F_o$, $F_s$> of the relational schema $R$ is considered correct if it satisfies the following conditions:

- All attributes in $R$ should appear in at least one fragment, to avoid loss of information; and
- $F_s$ should not violate any confidentiality constraint;

It need to be non-redundant ($F_o$ and $F_s$ have no attribute in common) to avoid unnecessary storage at the data owner side and replica management problems [6].

To compute a fragmentation to reduce as possible the data owner's workload, a metric for measuring the cost of a fragmentation is needed. In [16], four metrics are proposed to respond to different minimization goals:

- Min-Attr: minimizes the number of attributes in $F_o$;
- Min-Size: minimizes the physical size of the attributes in $F_o$;
- Min-Query: minimizes the number of queries that involve at least one attribute in $F_o$;
- Min-Attr: minimizes the number of conditions in queries that are evaluated over attributes in $F_o$;

Since the problem is NP-hard, a heuristic algorithm was proposed in [16] to easily adapt to different metrics.

### 1.1.1.1.7    *Query execution*

A user query Q must be translated by the client in the queries $Q_o$ and $Q_s$ on $F_o$ and $F_s$, respectively, plus a $Q_{so}$ that combines the resulting sets of $Q_o$ and $Q_s$. $Q_s$

is a query that operates only on attributes on the server, while $Q_o$ is a query that can be evaluated only by the data owner.

The translation process can use *server-owner* or *owner-server* strategy [128]. In the former, $Q_s$ is evaluated on the server and then the result is sent to the owner, which proceeds with the evaluation of $Q_o$ and $Q_{so}$. In the latter, $Q_o$ is evaluate on the owner, then $Q_s$ is evaluate on the server side and, in the end, $Q_{so}$ is evaluate again in the owner side.

An example of the two different strategies for the query

select Name from MedicalData where Illness='Asthma' and zip='26013'

on fragmentation of Figure 11 is condensed in Table 3 [128]:

Table 3    Query translation in fragmentation with owner involvement

| Server-Owner strategy | Owner-Server strategy |
|---|---|
| $q_s$ := SELECT tid<br>FROM $F_s^e$<br>WHERE Illness='Asthma'<br><br>$q_{so}$ := SELECT Name<br>FROM $F_o^e$ JOIN $R_s$ ON<br>$F_o^e.tid=R_s.tid$<br>WHERE ZIP='26013' | $q_o$ := SELECT tid<br>FROM $F_o^e$<br>WHERE ZIP='26013'<br><br>$q_s$ := SELECT tid<br>FROM $F_s^e$<br>WHERE Illness='Asthma' AND<br>tid IN {6}<br><br>$q_{so}$ := SELECT Name<br>FROM $F_o^e$ JOIN $R_s$ ON<br>$F_o^e.tid=R_s.tid$ |

## 2.3  Selective access

In many scenarios, access to data is selective, with different users having different views over the data. Access control can discriminate between read and write operations on an entire record or only on a part of it.

Key management solutions for outsourced databases can be classified in three categories: owner side policy enforcement solutions, user-side policy enforcement solutions, and solutions where access policy is shared among actors (owner/user/server) [7].

An intuitive way to handle this issue is to encrypt different portions of data with different keys that are then distributed to users (according to their access privileges) [120].

Limiting our scope to read operations, the access rights defined by the data owner can be represented by using an access matrix A, where rows correspond to subjects, columns correspond to objects, and entry A [ s; o ] is set to 1 if *s* has permission to access *o*; 0 (zero) otherwise [29]. An example is represented in Figure 12:

|       | $t_1$ | $t_2$ | $t_3$ | $t_4$ | $t_5$ | $t_6$ | $t_7$ |
|-------|-------|-------|-------|-------|-------|-------|-------|
| Alice | 0 | 0 | 1 | 0 | 0 | 1 | 1 |
| Bob   | 1 | 1 | 0 | 1 | 0 | 1 | 1 |
| Carol | 1 | 0 | 1 | 1 | 1 | 0 | 0 |
| David | 0 | 1 | 1 | 1 | 0 | 1 | 0 |

Figure 12    An example of access matrix

The column *i* of matrix corresponds to the Access Control Lists (ACL$_i$) of the tuple t$_i$, while the row *j* corresponds to the capability list (CAP$_j$) of user u$_j$.

The simplest solution for access control consists in encrypting each tuple in the outsourced database with a different key and assigning to each user the set of keys associated with the tuples she can access; but this solution brings to the proliferation of a lot of keys per user.

Access control policies can be translated into equivalent encryption policies guided by two basic requirements [6][120]:

- No more than one key is released to each user, and
- Each resource is encrypted not more than once.

In [29][120], to achieve these objectives, a hierarchical organization of keys has been envisioned. Basically, users with the same access privileges are grouped and each resource is encrypted with a key that corresponds to the set of users that can access it. In this way, a single key can be used to encrypt more than one resource. The authors consider a user hierarchy whose elements are all the possible sets of users in the system together with the partial order (≤) naturally induced on it by the subset containment relationship. Each user group has associated the tuples whose ACL, defined in the access matrix, corresponds to the group itself. A directed acyclic graph (DAG) with a node for each set of users, and an edge from node Y to node X if X ≤ Y represents the user hierarchy (an example of it is shown in Figure 13).

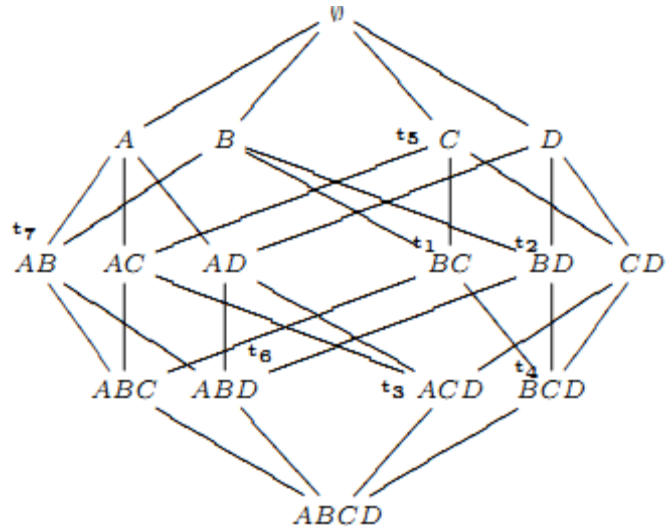

Figure 13 An example of user hierarchy, source: [29]

By assuming that each node of the hierarchy is associated with a key, the access control problem could be solved by encrypting each tuple with the key of the node corresponding to its ACL and by assigning to each user the set of keys of the nodes to which she belongs. On this basis, a heuristic algorithm that minimizes the total number of distributed keys delivered to all the users was proposed in [129].

An alternative solution (that is based on the same kind of above rights matrix, but builds a binary tree whose intrinsic properties contribute to reduce key management complexity) was proposed in [130]. Each level of the binary tree corresponds to the rights of a user profile. Each user receives the key corresponding to her hierarchical level; using it, she can derive all the keys for allowed data using derivation mechanisms. Since few keys are assigned for user at the top of the hierarchy, while those situated in the bottom receive a bigger number of keys, a preliminary sort of user profiling, based on the right importance, is used to minimize the average number of delivered keys.

A solution that organizes data in a binary tree and use also derivation mechanisms was proposed in [131]. Since it does not follow any defined data placement strategy to group data according to user rights, as the previous works did, and allowed data could be disseminated on different parts of the binary tree (to avoid data disclosure), the number of received keys could grow very quickly especially in a scenario of billion of data blocks [7].

A two level encryption scheme, one before outsourcing data, done by the owner and the second, in case of user or role revocation, done by the service provider, was proposed in [132].

## 2.4 Dynamic rights management

A user's rights may change over time (e.g., the user changes role or department). Therefore removing users from group/roles becomes necessary. If the key management is not dynamic, this can be achieved on outsourced databases as follows [134]:

- Encrypt data by a new key;
- Remove the original encrypted data, and
- Send the new key to the rest of the group.

The user, unless the database is re-keyed, continues to have access to the data [7]. Note that the data owner must perform these operations since the untrusted DBMS has no access to the keys. This active role of the data owner, however, goes somewhat against the reasons for choosing to outsource data in the first place.

In [7], the authors, exploiting the never-ending trend to lower price-per-byte storages, propose to replicate $n$ times the source database, where $n$ is the number of different roles having access to the database. Each database replica is a view, entirely encrypted using the key created for the corresponding role. Each time a role is created, the corresponding view is generated and encrypted with a new key, expressly generated for the newly-created role. Users do not own the real keys, but receive a token that allows them to address a request-to-cipher to a set KS of key servers on the Cloud.

Another important issue, common to many access control policies, concerns time-dependent constraints on access permissions [9]. In many real situations, it is likely that a user may be assigned a certain role only for a limited time. In such case, users need a different key for each time period. A time-bound hierarchical key assignment scheme is a method to assign time-dependent encryption keys and private information to each class in the hierarchy. Key derivation will depend also on temporal constraints; once a role's time period expires, users in that role need to be re-authorized.

# 3 CRYPTOGRAPHY IN DATABASES

*In this Chapter I analyse the different levels and granularity of database encryption, and the types of attack to database security.*

Confidentiality, integrity and availability are the main properties of database protection [135]. Confidentiality has been defined by the International Organization for Standardization (ISO) in ISO-17799[10] as "ensuring that information is accessible only to those authorized to have access"; data integrity assures that none can modify the information without a trace; availability provides access to data by authorized users in a reasonable time. Along the years, a lot of ACP (Access Control Policy) has been defined, based on database model (relational rather than object) and policy control (i.e., DAC-Discretionary Access Control, RBACC-Role Based Access Control, MAC-Mandatory Access Control)[139]. Traditionally, ACPs are based on the assumption that the DBA (DataBase Administrator) is trusted, but this assumption no longer holds in outsourced data centers and in the Cloud, where the platform-as-a-service (PaaS) provider is external to data owner. A solution to this problem is that the DBMS treats only raw-data, encrypted in such a way that DBA (or another intruder) cannot read the information. There are three main categories of database encryption [28][136]: storage level encryption, database level encryption, and application level encryption (as represented in Figure 15).

---

[1] ISO/IEC 17799, Jan 4, 2009

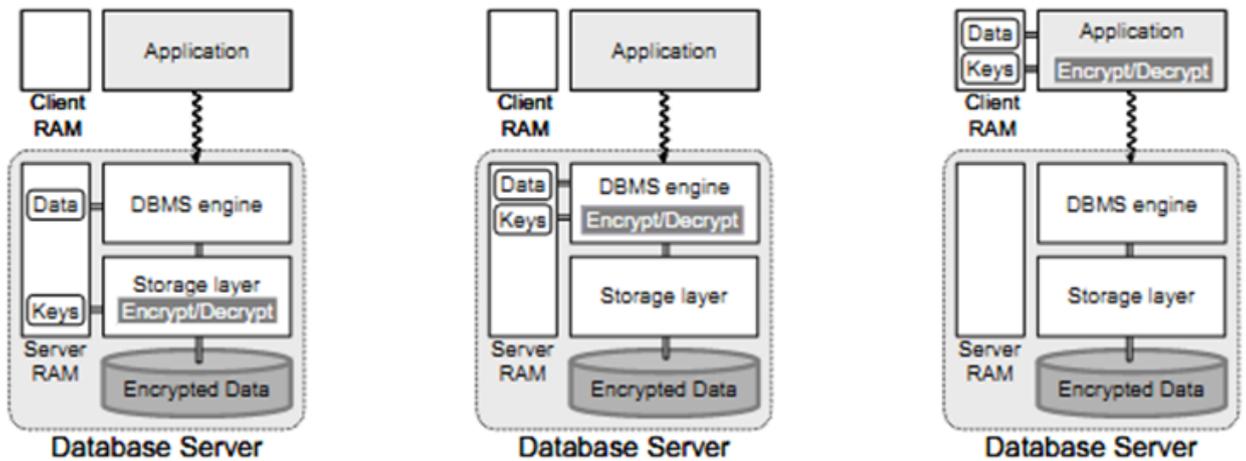

Figure 15    The three options for database encryption level, source: [28]

## 3.1   Storage-level encryption

In Storage-level encryption (SLE), data is encrypted either at the file level (NAS/DAS) or at the block level (SAN) [152]. A short while ago, Toshiba has released a hardware implementation of SLE, a family of hard drives - called Self-Encrypting Disk[11]. The system is based on the Opal specifications of Trusted Computing Group[12], supports native encryption AES 256 and can automatically delete its contents if not used by the rightful owner.

This encryption is not selective; since the storage subsystem has no knowledge of database objects and structure, it encrypts an entire support or portions of support (i.e. directories or files), without respect to user privileges or to data sensitivity. When encrypting only portion of support, there is the additional risk that logs and temporary files remain plaintext. It prevents theft of storage but it is unsuitable for preventing unauthorized access by an honest-but-curious system administrator, who must know the encryption key. On the other hand, it is entirely transparent to the system, so it needs no database modification [28].

---

[11] http://storage.toshiba.com/main.aspx?Path=StorageSolutions/PCNotebook/MKxx61GSYGSeries

[12] http://www.trustedcomputinggroup.org/developers/storage/specifications/

## 3.2 Database-level encryption

Database-level encryption (DLE) secures data as it is written to and read from a database. The encryption is applied to the Db at various granularities, such as database, tables, columns (most frequently), and rows. It can be related with some logical conditions for selecting affected data, too.
Several database encryption schemes have been proposed in the literature, e.g. [136]:

- In [146] a scheme based on the Chinese-Reminder theorem is proposed, where each row is encrypted using different sub-keys for different cells. This scheme enables encryption at the level of rows and decryption at the level of cells;
- An extension of the previous scheme that supports multilayer access control is proposed in [151]. It classifies subjects and objects into distinct security classes that are ordered in a hierarchy, such that an object with a particular security class can be accessed only by subjects in the same or a higher security class;
- The scheme presented in [147] proposes encryption for a database based on Newton's interpolating polynomials;
- The scheme presented in [148] is based on the RSA public-key scheme and suggests two database encryption schemes: one column oriented and the other row oriented;
- The SPDE scheme is presented in [149]. It encrypts each cell in the database individually together with its cell coordinates (table name, column name and row-id) to obtain different cipher-text values for equal plaintext values (against static analysis), and to prevent a tuple is moved to a different location (against splicing attacks) [150].

One disadvantage of all but the last these schemes is that the basic element in the database is a row and not a cell. Thus the structure of the database is modified. In addition, all of those schemes require re-encrypting the entire row when a cell value is modified. Thus, in order to perform an update operation, all the encryption keys should be available [28][136].

DLE is not transparent to application as SLE, so it involves some modifications to the indexed encrypted data and in stored procedures and triggers. The system is slowed down by the encryption overhead. Usually, it is not a defence from the curious DBAs [152].

## 3.3 Application-level encryption

In Application-level encryption (ALE), data is encrypted/decrypted by the application that generates it. Plain-text data is made available only at client side, while data sent over the network is encrypted [153][95][13].
The main advantages of this solution are:
- The encryption keys and the encrypted data stored in the database are separated, since the keys never have to leave the application side; and
- A high flexibility since the encryption granularity and the encryption keys can be chosen depending on application logic.

The price to pay is that this scheme usually involves returning to the client larger result sets, which are then filtered at client side, when decrypted. To accomplish this result, applications need to be modified and the network traffic increases [28].

## 3.4 Granularity in database-level encryption

Database-level encryption is the most common solution for data protection. It can have different types of granularity, namely [28][136]: database, tables, columns, and rows.
In [29] the granularity is divided into relation level, attribute level, tuple level, and element level, while [154] uses the taxonomy: attribute value, record/row, attribute/column, and page/block.

### 3.4.1 Database

In this case, the whole database is encrypted using only one key, as if it was a single file. The cons of this technique are:
- It does not allow to define different privileges on each table;
- The schema definition becomes particularly complex;

- The system performance suffers considerable degradation (an improvement can be achieved with appropriate caching) [152];
- Its security is closely linked to the physical security of the device with which the master key is kept.

For these reasons, the database granularity solution is seldom used.

### 3.4.2 Tables

A specific key encrypts each table separately. Performances are better than the previous solution, but still very far from those of a clear-text database, because encrypting an existing table can be slow. Encryption affects performance only when data is retrieved from or inserted into an encrypted column. No reduction in performance occurs for operations involving unencrypted columns, even if these columns are in a table containing encrypted columns. Accessing data in encrypted columns involves small overheads (e.g., in an Oracle 11 database, the overhead associated with encrypting or decrypting an attribute is estimated to be around 5%) w.r.t. clear text data [13].

The total performance overhead depends on the number of encrypted columns and their frequency of access. The columns most appropriate for encryption are those containing the most sensitive data. The definition (and enforcement) of integrity constraints, foreign keys and indexes is very complex (see Section 2.2.1. on Data encryption).

### 3.4.3 Columns

All the data in a column (or set of columns) of a table is encrypted with the same key. This is the solution adopted by most DBMS suppliers, e.g.

---

[13] http://download.oracle.com/docs/cd/B28359_01/network.111/b28530/asotrans.htm

Transparent Data Encryption in Oracle 11G[14] or Microsoft SQL Server[15], as it allows encrypting only sensitive data. However, it needs to build ad-hoc indexes customized for the expected queries (again, at the expense of performance). With this approach, it is also not possible to define access privileges on "horizontal" portions of a table such as row sets (e.g., allowing access only to rows with id> 100), as it is awkward to encrypt rows with different keys depending on the user. These mechanisms usually rely on third-party applications, or otherwise are implemented using database triggers or stored procedures[152].

Although column-level encryption permits to reduce the encryption to only sensitive columns, it often results in worse performance because it breaks many of the indexing, query, and relational tools used by the DBMS [155].

### 3.4.4 Rows

Each single row in a table is encrypted using a different key. The main advantage of this technique is the capability to define access control to a subset of data (rows) of a table basing on the distribution of decryption keys. Let us assume that we have a table that includes the data of all students in a university and we want to grant access to the secretary of each course only to data of students enrolled in that course. If we were using database or table-level encryption, we would have to create a view for each course and grant the rights to the corresponding secretary, with the problems outlined above (also, data stays readable by the DBAs). Using column-level encryption, the permissions must be specified at the field level and, unless appropriate indexes or cumbersome procedures are implemented (which may also expose the data to inference or statistical attacks), it would be impossible to make the information instantly accessible to authorized users. Using row-level encryption, instead, it is possible to make available to the authorized user the keys (or the key) that can be used to decrypt only the allowed rows. This technique, besides ensuring a better management of access permissions, prevents any kind of statistical analysis on the table, since every relation between similar rows is nullified by the different key encryption. In a standard RDBMS, however, this technique has significant disadvantages in terms of performance and functionality: querying would be possible only through the construction of appropriate indexes for each column of the table (with a considerable waste of resources both in terms of time and space), while foreign keys would be unusable, since the different key encryption breaks every relation between equal values. Another major issue concerns the management of keys: row-level encryption needs generation and distribution of a key for each row of each table encrypted with this method. To solve (or alleviate) this problem, some key management techniques can be used, such as:

- Broadcast (or Group) encryption [13]: rows are divided into equivalence classes, based on recipients. Every class is encrypted using an asymmetric algorithm where the encryption key is made in a way that each recipient can decrypt the information using only its own private key. Both the public and the private keys are generated by a trusted entity.
- Identity Based Encryption [11]: it bounds the encryption key to the identity of recipient. Each recipient generates by itself a key pair used to encrypt/decrypt information.
- Attribute Based Encryption [12]: it bounds the encryption key to an attribute (a group) of recipient. Each recipient receives by a trusted entity the private key used to decrypt, while the encryption key is calculated by the sender.

---

[14] http://www.oracle.com/technetwork/database/options/advanced-security/index-099011.html

[15] http://msdn.microsoft.com/en-us/library/bb934049.aspx

However these techniques are complex and therefore, to the best of my knowledge, no research prototype or commercial database encryption system adopt row encryption.

## 3.5 Attacks to database security

The database security may be compromised by an attacker that can be categorized into three classes [156]:

- Intruder - A person who gains access to a computer system and tries to extract valuable information.
- Insider - A person who belongs to the group of trusted users and tries to get information beyond his own access rights.
- Administrator - A person who has privileges to administer a computer system, but uses his administration rights in order to extract valuable information.

The different types of attack can be classified in passive and active [136]. The first category contains the attacks that only read the data without altering it and includes [136]:

- Static leakage: it is an attack that gains information on the database plaintext values by observing a snapshot of the database at a certain time. E.g., if the database is encrypted in a way that equal plaintext values are encrypted to equal ciphertext values, statistics about the plaintext values, such as their frequencies can easily be learned.
- Linkage leakage: it is an attack that gains information on the database plaintext values by linking a table value to its position in the index. E.g., if the table value and the index value are encrypted the same way (both ciphertext values are equal), an observer can search the table cipher text value in the index, determine its position and estimate its plaintext value.
- Dynamic leakage: it is an attack that gains information about the database plaintext values by observing and analyzing the changes performed in the database over a period of time. E.g., if a user monitors the index for a period of time, and if in this period of time only one value is inserted (no values are updated or deleted), the observer can estimate its plaintext value based on its position in the index.

The second category contains the attacks that modify the database and includes [157]:

- Spoofing: replacing a ciphertext value with a generated value. Assuming that the encryption keys are secure, a possible attacker might try to generate a valid ciphertext value, and substitute the current valid value stored on the disk. Assuming that the encryption keys were not compromised, this attack poses a relatively low risk.

- Splicing: replacing a ciphertext value with a different cipher text value. Under this attack, the encrypted content from a different location is copied to a new location under attack.
- Replay: replacing a cipher text value with an old version previously updated or deleted.

# 4 IN-MEMORY DATABASES

*In this Chapter, I illustrate the characteristics of the In-Memory Databases, a proven mature technology that takes advantage of the current wide availability of main memory to reverse the usual storage strategy, storing the information directly into main memory and using the disk only for data backup.*

Magnetic disks are heavily used as primary storage in computer systems since their introduction in 1955 with the IBM 350 Disk File device, announced by IBM as a component of the IBM 305 RAMAC computer system on September 13, 1956 [66].

Two of the most critical parameters for a magnetic disc are capacity and performance. Disc drive capacity refers to the size of storage memory available on a disc drive. Disc drive performance describes the speed and efficiency with which data can be written to and read from a drive[16].

During these years, the capacity has rapidly grown from the initial 5 million 7-bit (6-bits plus 1 odd parity bit) characters (about 4.4 megabytes) [67] to the actual multiple terabytes size. Instead, the performance has not grown at the same speed (see Figure 16 and Figure 17).

---

[16] "Disc Drive Capacity and Performance", Seagate Technical paper, 2001

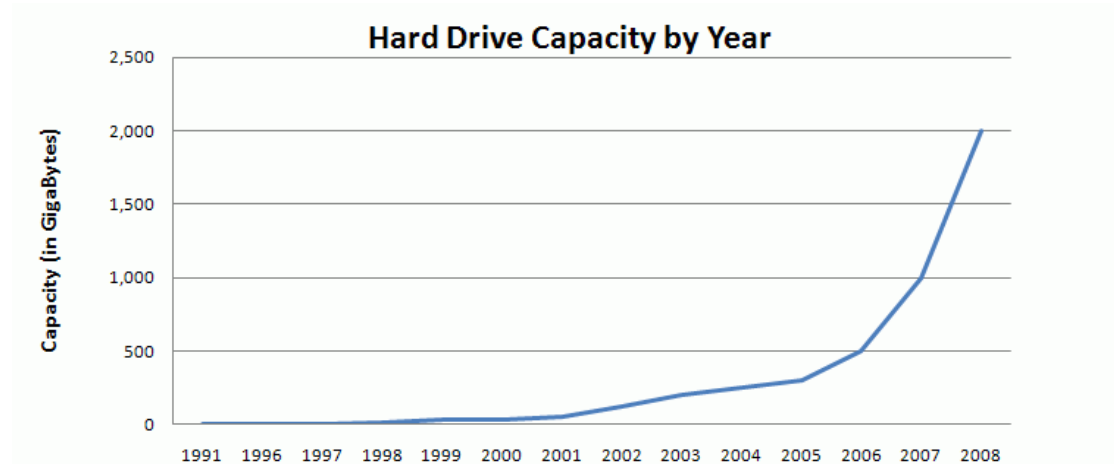

Figure 16 Data sets growth[17]

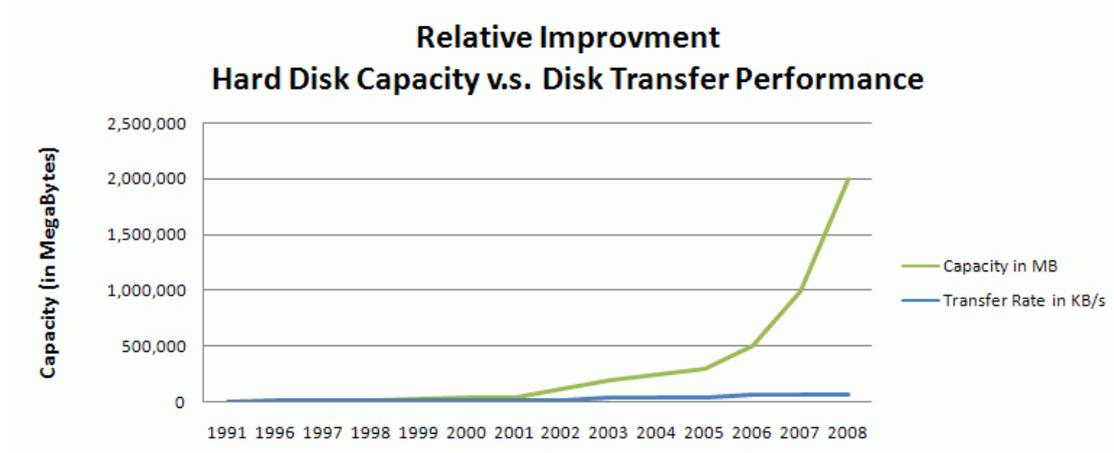

Figure 17 Disk performance improvements vs. capacity growth[18]

With the aim to overcome this limitation, particularly heavy in the case of data centers and transactional systems, many people have proposed new approaches to disk based storage or, as a revolutionary choice, to store data in random access memory, using disk only as backup [63].

Since 1985, some precursor designed a "memory resident DBMS" [68] where data is resident in main memory, but, at that time, RAM capacity was very low to permit the storage of large datasets.

During the last 20 years, the RAM capacity of computers has increased exponentially by a factor of 10 every 4 years, following Moore's Law. The graph below illustrates the typical memory configuration installed on personal computers since 1980.

---

[17] Source: http://wiki.r1soft.com

[18] Source: http://wiki.r1soft.com

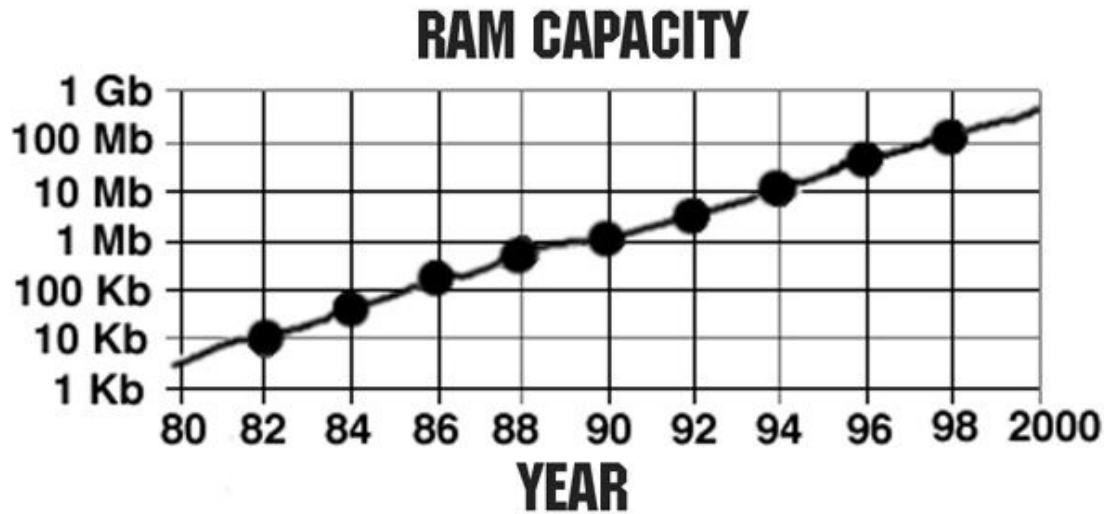

Figure 18   The RAM modules capacity[19]

At the same time, RAM's price per byte is falling down quickly, so today there are the conditions to effectively use the, "main memory database system" (MMDB), also known as in-memory database (IMDB) or as real-time database (RTDB).

"In-memory databases have recently become an intriguing topic for the database industry. With the mainstream availability of 64-bit servers with many gigabytes of memory a completely RAM based database solution is a tempting prospect to a much wider audience."[20]

It is important to remark that, while a conventional database system stores data on disk but caches it into memory for access, in an IMDB the data resides permanently in the main physical memory and there is a backup copy on disk [27].

IMDBs are intended either for personal use (because they are comparatively small w.r.t. traditional databases), or for performance-critical systems (for their very low response time and very high throughput). They use main memory structures, so they need no translation from disk to memory form, and no caching and they perform better than traditional DBMSs with Solid State Disks.

In Table 4, I summarize pros and cons for IMDBs.

---

[19] Source: Intel

[20] http://www.remote-dba.net/t_in_memory_cohesion_ssd.htm

Table 4    IMDBs Pros and Cons

| Pros | Cons |
|---|---|
| Fast transactions | Complexity of durability's implementation |
| No translation | Size limited by main memory |
| High reliability | |
| Multi-User Concurrency (few locks) | |

## 4.1 The impact of in-memory structure

Main memory is a random-access, byte-addressable device while disk is semi-random, block-addressable device. These differences have an impact on different aspects of DBMS.

### 4.1.1 Data representation

To optimize space utilization, necessary to hold the active database entirely in main memory, compact data structures such as T-trees [70] or array structures [71] have been proposed to organize permanent data efficiently in main memory. In particular, the relation between two tuples may be represented using memory pointers instead of external keys [27] (see Figure 19). This permits a smaller storage and a faster access to related records.

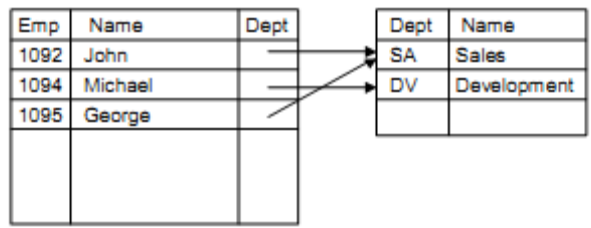

Figure 19    The usage of pointers in relations

In conventional DBMS we need translation from stored representation (e.g. records in files) to applications' form; while in IMDBs, since data is stored in main memory, we have a unique representation of it.

### 4.1.2 Access methods

While disk-oriented index structures, as B-Tree [72] and derived, are intended for block storage and then are designed to minimize the number of disk accesses and the required disk space, main memory index structures are designed to reduce overall computation time while using as little memory as possible. Since data is in main memory, instead to use associative values <key, position> to relate keys to tuples, we can use memory pointers to actual attribute values, which remain in their place, giving some clear advantages [73]:

- A single tuple pointer provides the index with access to both the attribute value of a tuple and the tuple itself, reducing the size of the index;
- This eliminates the complexity of dealing with long fields, variable length fields, and compression techniques in the index,
- Moving pointers will tend to be cheaper than moving the (usually longer) attribute values when updates necessitate index operations, and
- Since a single tuple pointer provides access to any field in the tuple, multi-attribute indices will need less in the way of special mechanisms.

There are two main types of index structures: those that preserve some natural ordering in the data and those that randomize the data. The index structures being studied here are (see Figure 20 and Figure 21): arrays, AVL-Trees, B-Trees, and T-Trees, for the order-preserving class; and Chained Bucket Hashing, Linear Hashing and Extendible Hashing, for the randomizing class.

- Arrays [71] use minimal space, providing that the size is known in advance or that growth is not a problem, but the computational complexity of data movement is $O(N)$[21] for each update, so it useful just in read-only environment;
- AVL-Trees [74] use a binary tree search. Updates always affect a leaf node and may result in an unbalanced tree, so the tree is kept balanced by rotation operations. The disadvantage in AVL-Trees is their poor storage utilization, since each tree node holds only one data item, so there are two pointers and some control information for every data item.

---

[21] N is the number of elements of the array

- B-Trees [75] are a usual structure for disk indexes, since they are broad shallow trees and require few node accesses to retrieve a value. Most database systems use a variant of the B-Tree, the B+-Tree, which keeps all of the actual data in the leaves of the tree, but in main memory this does not speed up the search, while it wastes space. In B-Trees:
    - Storage utilization is good since the pointer to data ratio is small, as leaf nodes hold only data items and they comprise a large percentage of the tree);
    - Searching is reasonably quick since a small number of nodes are searched with a binary search; and
    - Updating is fast since data movement usually involves only one node.

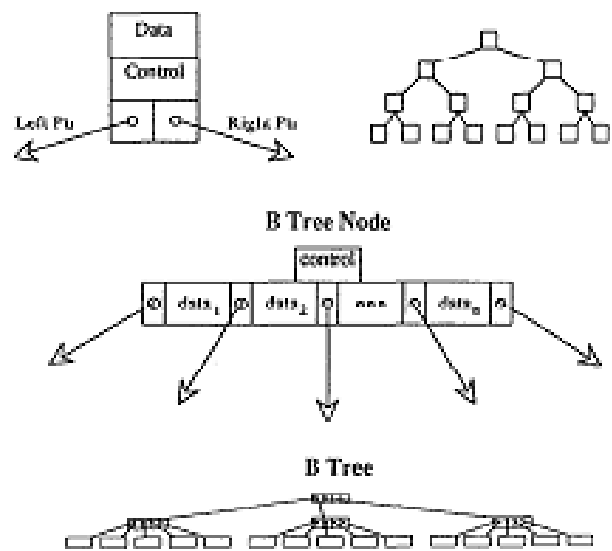

Figure 20    Tree structures, source [70]

- Chained Bucket Hashing [77] is a very fast static structure used both in memory and on disk. Since it is static, it never has to reorganize its data (advantage), but, for the same reason, it may have very poor behaviour in a dynamic environment (disadvantage) because the size of the hash table must be estimated before the table is filled. A wrong estimated size may affect the performance (too small), or the wasted space (too large).
- Extendible Hashing [78] employs a dynamic hash table. With respect to the previous structure, the table size does not need to be known in

advance since a hash node contains several items and splits into two nodes when an overflow occurs. The directory grows in powers of two, doubling whenever a node overloads and has reached the maximum depth for a particular directory size. Since any node can cause the directory to split, the directory can become very large if the hash function is not sufficiently random.

- Linear Hashing [79] uses a dynamic hash table that grows linearly as it splits nodes in predefined linear order. First, the buckets can be ordered sequentially, allowing the bucket address to be calculated from a base address - no directory is needed. Second, the event that triggers a node split can be based on storage utilization, keeping the storage cost constant for a given number of elements.

- Modified Linear Hashing [79] is a variant of the previous structure that uses a directory much like Extendible Hashing, except that it grows linearly, and chained single-item nodes, allocated from a general memory pool. The splitting criteria are based on performance, i.e. the average length of the hash chains, rather than storage utilization. Monitoring average hash chain length provides more direct control over the average search and update times than monitoring storage utilization.

- T-Trees [70] are binary trees, evolved from AVL-Trees and B-Trees, with many elements in a node. They retain the intrinsic binary search nature of the AVL-Tree united to the good update and storage characteristics of the B-Tree. Data movement is required for insertion and deletion, but it is usually needed only within a single node. Rebalancing is done using rotations similar to those of the AVL-Tree, but it is done much less often than in an AVL-Tree due to the possibility of intra-node data movement.

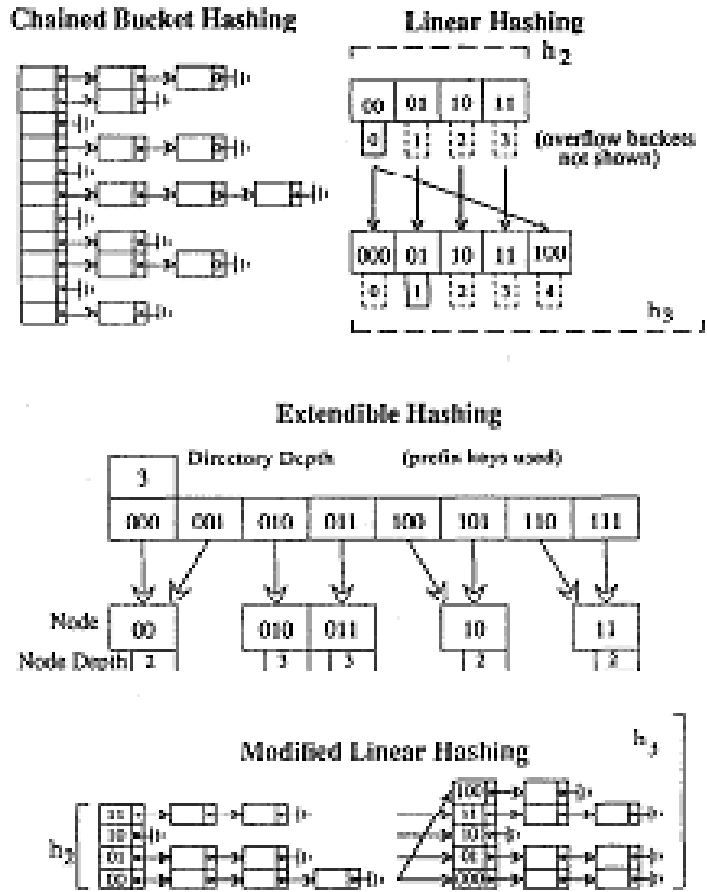

Figure 21    Hash structures, source: [70]

## 4.1.3   Query processing

The use of a random access memory to maintain data has a deep impact on common operations. First of all, we have no advantage in sequential data access, so data contiguity loses value. Operations as sorting become useless because it is more valuable using pointers to create sorted list instead of moving data.

While data may still be represented by the relational model, the use of pointers to express relations naturally brings to a semantic model. The relations are surfing following pointers and joins exploit this possibility, instead of using external keys [27].

The goal is not minimizing the disk accesses, but the processing cost in main memory. This depends on the architecture, and then the optimization techniques need to be evaluated in the running system.

## 4.1.4 Concurrency control

Concurrency is defined as the ability of multiple processes and threads to access and change the data records at the same time. The lower the contention to access and modify data with more users, the better the concurrency, and vice versa.

Database concurrency controls ensure that transactions occur in an ordered fashion to protect transactions issued by different users/applications from the effects of each other. They must preserve the four characteristics of database transactions: atomicity, isolation, consistency and durability, also known as ACID.

Two approaches can be adopted to manage concurrent data access: pessimistic and optimistic [80]. Pessimistic concurrency systems assume that conflict will occur and it avoids conflicts by acquiring locks on data that is being read or modified, so that no other process can modify that data. Optimistic concurrency systems assume that transactions are unlikely to modify data that another transaction is modifying. This is implemented by using versioning technique. This allows readers to see the state of the data before the modification occurs as the system maintains previous version of the data record before it actually attempts to change it. Usually DBMSs use pessimistic approach, while versioned file systems use optimistic approach.

### 4.1.4.1 Granularity of locks in a database with pessimistic approach

The granularity of locks in a RDBMS refers to how much of the data is locked at one time. In theory, a database server can lock as much as the entire database or as little as one column of data. Such extremes affect the concurrency (number of users that can access the data) and locking overhead (amount of work to process lock requests) in the server.

Granularity levels are listed below ordered from large to fine granularity [81]: Database, Table, Disk Block or Memory Page, Record, and Record Field.

Since the best granularity size depends on the given transaction, DBMS should support multiple level so granularity and allows the transaction to pick any level it wants.

By locking at higher levels of granularity, the amount of work required to obtain and manage locks is reduced. If a query needs to read or update many rows in a table:

- It can acquire just one table-level lock
- It can acquire a lock for each page that contained one of the required rows
- It can acquire a lock on each row

Less overall work is required to use a table-level lock, but large-scale locks can degrade performance, by making other users wait until locks are released. Decreasing the lock size makes more of the data accessible to other users. However, finer granularity locks can also degrade performance, since more work is necessary to maintain and coordinate the increased number of locks. To achieve optimum performance, a locking scheme must balance the needs of concurrency and overhead.

In an IMDB, small locking granules are inappropriate, since contention is already low because data is memory resident. Usually IMDBs prefer to use very large lock granules (e.g. table or database) [82].

### 4.1.5 Logging and Recovery

Normally, the use of volatile memory-based IMDBs supports the three ACID properties [83] of atomicity, consistency and isolation, but lacks support for the durability property. To add the latter, when non-volatile random access memory (NVRAM) is not available, IMDBs use a combination of transaction logging and primary database check-pointing to the system's hard disk: they log changes from committed transactions to physical medium and, periodically, update a disk image of the database. Having to write updates to disk, the write operations are heavier than read-only [27] and impact on system performance. Logging policies vary from product to product: some leave the choice of when to write the application on file, others do all the checkpoints at regular intervals of time or after a certain amount of data entered / edited.

Following are some of the desirable properties of logging and recovery algorithms [85]: Reduced log traffic, Speedy logging and recovery, Transaction priority oriented logging and recovery, and Data class oriented logging and recovery.

### *4.1.5.1 Transaction logging and commit processing*

A transaction is a sequence of operations, which may lead to a success or a failure. If successful, the result of the operations must be permanent; while in case of failure, it must return to its previous state when the transaction [84].

In an IMDB, transactions need to be stored in a permanent storage until they are committed and the final result is written to the log. The transaction logging can be divided in *REDO* and *UNDO* tails [84].

To minimize the impact on performance, IMDBs can use some strategy, main of which are:

- Log driven backups [86] : is composed of a stable memory and recovery task, which work in parallel. The stable memory [87] usually consists of a main conventional RAM, a Safe RAM which is called the Safe, and a conventional disk and stores the log tail that includes both *UNDO* and *REDO* logs produced by active transactions. Asynchronically, the recovery task move transaction from stable memory to the checkpoint log.
- Pre-committing [73]: if the system has not a stable memory, it can do a *pre-commit*, releasing the transaction lock without waiting for the information to be written to log. The sequential nature of log ensures that transactions cannot commit before others on which they depend. This way the response time remains the same but the blocking delay of other concurrent transaction is reduced.
- Group commit [73]: a set of updates are grouped together in one log write to amortize the cost of the log write disk I/O over several updates.

### 4.1.6 Performance

Thanks to the lower latency time, IMDBs usually have better performance then conventional DBMSs, especially in read operations.

While in conventional DBMSs the performance is evaluated in terms of disk accesses, in IMDBs the metrics is different, e.g., it is related to measure of checkpointing/logging. The path to performance improvement may be summarized in: Index improvements [70], Parallel and distributed systems [89][64][11][92] , and Checkpointing improvement [90] [91].

The improvement of performance is particularly important when IMDBs are used in Real Time or Multiuser Systems, where they are a good choice for their small latency that allows fast responses to events/requests. But IMDB are often used in embedded application, where compactness is more valuable than performance.

## 4.2 Cache vs. IMDB

Disk cache is a cache or buffer used to hold portions of the disk address space contents to capture a significant fraction of the I/O operations. It can provide access times and transfer rates significantly better than disk, and can improve I/O system performance and thereby postpone or eliminate the predicted I/O system bottleneck [93].

In recent years there has been a surge in the use of DRAM, driven by the performance requirements of large-scale Web applications. For example, both Google and Yahoo! store their search indices entirely in DRAM [63]. Memcached[22] provides a general-purpose key-value store entirely in DRAM, and it is widely used to offload back-end database systems (however, memcached makes no durability guarantees so it must be used as a cache). The Bigtable storage system allows entire column families to be loaded into memory, where they can be read without any disk accesses [11]. Big-table has also explored many of the issues in federating large numbers of storage servers.

The cache is useful only if the page fault rate is low, otherwise the number of disk access is not significantly reduced. The 1000x gap in access time between DRAM and disk means that a cache must have exceptionally high hit rates to avoid significant performance penalties: even a 1% miss ratio for a DRAM cache costs a factor of 10x in performance [87].

---

[22] http://memcached.org/

Modern application, e.g. Facebook, have a very limited locality[23], due to complex linkages between data (e.g., friendships in Facebook). To obtain providing a hit rate of 96.5%, the total amount of memory used by the storage system in Facebook equals approximately 75% of the total size of the data (excluding images) [87]. Using an IMDB, the memory used is incremented of the last 25%, but obtains it guarantees performance independent of access patterns or locality.

---

[23] Locality means that the probability of reference for recently referenced pages is higher than the average reference probability [94].

# 5 BROADCAST ENCRYPTION

*Originally born for signal broadcasting, Broadcast Encryption (BE) schemes are being used and adapted in the emerging field of Online Social Networks (OSN), were communities of users want to securely share data or messages. OSNs present many analogies with distributed databases where data is replicated at more than one node. When a data change happens at a node, the node has to propagate it to all copies of the same data resident at other nodes. In terms of the underlying network protocols, this implies a lot of messages. BE schemes can help to significantly reduce the amount of communications in the network, allowing the transmission of a common encrypted message that is understandable by all nodes involved.*

Broadcast Encryption (BE) is the cryptographic problem of encrypting broadcast content (e.g. TV programs) in such a way that only qualified users (e.g. subscribers who have paid their fees) can decrypt the content.

More formally, "a broadcast encryption system allows a center to communicate securely over a broadcast channel with selected sets of users. Each time the set of privileged users changes, the center enacts a protocol to establish a new broadcast key that only the privileged users can obtain, and subsequent transmissions by the center are encrypted using the new broadcast key" [35]. The following figure shows an example schema:

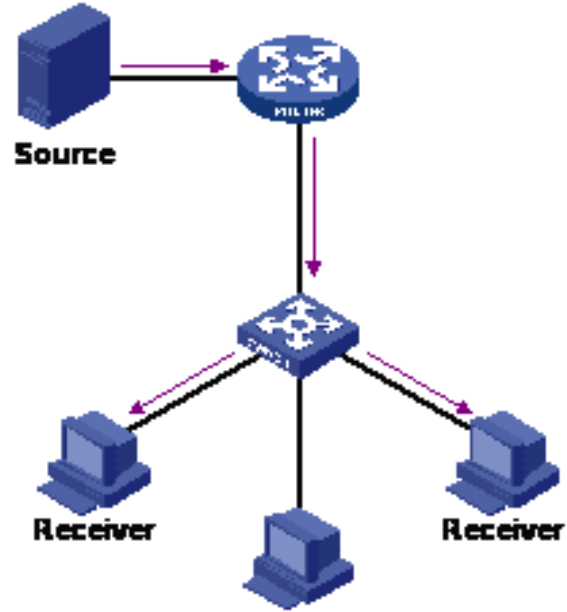

Figure 22　Broadcast encryption

A typical scenario is the transmission of broadcast TV show: the entity responsible for the spread (the International Broadcast Center) wants to send the program to the whole community of viewers. To this end, it is not necessary that the source emits a different signal for each user: a single transmission is sufficient, as each target device will then be able to capture the signal and properly receive the transmitted information.

Generally BE schemes are classified into two types: symmetric key and public key based BE schemes [60]. In the symmetric key setting, the only trusted group center *GC* can generate a broadcast message to users while, in the public key setting, any users are allowed to broadcast a message. I denote by *U* the set of users and by $R \subset U$ the set of revoked users. The following is a formal definition of a symmetric key based BE scheme [60].

A BE scheme *B* is a triple of polynomial-time algorithms (*SetUp*, *BEnc*, *Dec*), i.e., setup, broadcast encryption, and decryption [60]:

*SetUp*: a randomized algorithm which takes as input a security parameter $1^\lambda$ and user set *U*. It generates private information $SKEY_u$ for user $u \in U$. Private information of group center *GC* is defined as the set $SKEY_u$ of private information of all users.

*BEnc*: a randomized algorithm which takes as input a security parameter $1^\lambda$, private information $SKEY_U$ of *GC*, a set *R* of revoked users, and a message *M* to be broadcast. It first generates a session key *GSK* and outputs (*HdrR, CGSK, M* ) where a header *Hdr* is information for a privileged user to compute *GSK* and (*CGSK, M*) is a ciphertext of *M* encrypted under the symmetric key *GSK*. Broadcast message consists of [*R, HdrR , C_{GSK,M}* ]. The pair (*R, HdrR* ) and C_{GSK,M} are often called the full header and the body, respectively.

*Dec*: a deterministic algorithm which takes as input a user index $ind_u$, private information $SKEY_u$ of *u*, the set of revoked users *R*, and a header $Hdr_R$. If $u \in U \backslash R$, it outputs the session key *GSK*.

In public key broadcast encryption, the setup algorithm additionally generates the public keys $PK_U$ of users and $PK_U$ instead of the private information $SKEY_u$ of *GC* is taken as input in the algorithms *BEnc* and *Dec*.

## 5.1 Broadcast Encryption schemas

A first scheme involves an initialization phase, which a distinct secret key is assigned to each user of the system. To transmit a message *M*, the source concatenates as many copies of *M* as the number of receivers, by encoding each copy with a different secret key. The disadvantage of such scheme is that the resulting message has a size that depends on the numbers of intended receivers. Rather than a technique of broadcasting, this pattern can be seen as a way to unify multiple parallel communications between the International Broadcast Center and each of the authorized users [140].

An enhancement can be to encrypt the information using a (newly generated) session key *SK* and then to send the session key to the receivers using the previous scheme and transmitting the information. Each receiver decrypts the message using its own copy of SK. The advantage is that, usually, the length of the key is shorter than that of information [140].

Another scheme divides the users in groups. At each group is assigned a secret key. A user receives the secret keys of all the groups he belongs to. The user needs a size store that depends on the number of her affiliation. This scheme is unsuitable if the groups are dynamic, since it requires changing keys with the composition of the group [38] [140].

## 5.2 Threshold cryptosystem

The first real solution for BE were the threshold cryptosystems [38] [39].The (s, n)-threshold scheme (threshold scheme "s on n") permits to break up a secret information I among N participants, dividing *I* in *N* parts I$_1$,..., In called shares. The shares are constructed so that the knowledge of S (or more) shares allows the reconstruction of I, while the knowledge of S-1 (or less) shares doesn't determine any value or subset of values.

A basic broadcast encryption scheme consists of these phases [37]:
- Registration: to collect the authorized (identifier of) receivers
- Key generation: a new group symmetric key *K* is generated, based on the receivers information
- Encryption: the information *I* is encrypted using *K* obtaining *E*
- Decryption: the authorized receivers decrypt *E*

A broadcast message *M* is composed of two parts: a header, which contains information that can be used to access the content, and the body, which contains the encrypted content.

There are many schemes for BE, among others:
- Stateful schemes
    - Logical Key Hierarchy (LKH) [40]
- Stateless schemes
    - Complete Subtree (CS) [41]
    - Subtree-Difference [42]
    - Layered Subtree-Difference [42]

The original solution of Shamir is based on polynomial interpolation: since, given *S* points $P_1=(x_1,y_1)..P_s=(x_s,y_s)$ with distinct $x_i$, there is one and only one polynomial of S-1 such that $P(x_i)=y_i$ for all i=1..S, Shamir sets P(0)=I and broadcast S-1 point $P_i$. Only the authorized receivers have an additional good point and can recreate the polynomial interpolation and calculate P(0).

## 5.3 Distributed key generation

If the secret key is released by a unique authority, it needs to be trusted. If this is not assured, it is better to use a distributed key generation, where every party contributes to the assembly of the key. The following techniques address this issue.

### 5.3.1 El Gamal

ElGamal is a public key encryption system that is used in many other systems (as IBE and GCC) [141]. It is based on the hardness of computing the discrete logarithm. The three phases of ElGamal are:

1. Key generation: each user generates a public key $K_a^+ = (\rho, \sigma, \sigma^a \mod \rho)$ where:
    a. $\rho$ is a large prime number
    b. $\sigma$ is a primitive radix of $\rho$
    c. a, the true secret key, is randomized in the interval [1.. $\rho$-2]
2. Encryption: to send a message m $\in$ Zp the user randomly chooses a number k in the interval [1.. $\rho$-2] . The ciphertext will be (y,$\Delta$) where:
    a. y= $\sigma^k \mod \rho$
    b. $\Delta$=m* $\sigma^{ak} \mod \rho$
3. Decryption: the receiver obtain the encrypted message (y,$\Delta$) and, using her secret key a, calculate
4. $\Delta y^{-a}$=m* $\sigma^{ak}$* $(\sigma^k)^{-a} \mod \rho$ = m mod $\rho$

ElGamal has the disadvantage that the ciphertext is twice as long as the plaintext. It has the advantage the same plaintext gives a different ciphertext (with near certainty) each time it is encrypted.

### 5.3.2 Identity Based Encryption (IBE)

While IBE is not a BE system, it is fundamental to understand next BE systems as ABE and GCC.

An Identity Base Encryption (IBE) scheme is a public-key cryptosystem where any string is a valid public key. In particular, email addresses and dates can be public keys [55]. IBE allows for a sender to encrypt a message to an identity without access to a public key certificate. The ability to do public key encryption without certificates has many practical applications. For example, a user can send an encrypted mail to a recipient, e.g. bobsmith@gmail.com,

without the requiring either the existence of a Public-Key Infrastructure or that the recipient be on-line at the time of creation.

IBE was proposed by Adi Shamir in 1984 [55] but remained an open problem for many years, until the Boneh/Franklin's pairing-based encryption scheme[24] [30][54] and the Cocks's encryption scheme [56] based on quadratic residues.

Interesting characteristics of IBE are key expiration, management of user credentials (user can choose her PK), and delegation of decryption capabilities.

An identity-based encryption scheme *E* is specified by four randomized algorithms: *Setup, Extract, Encrypt, and Decrypt* [54].

- *Setup*: takes a security parameter *k* and returns system parameters and master-key. The system parameters include a description of a finite message space *M*, and a description of a finite ciphertext space *C*. Intuitively, the system parameters will be publicly known, while the master-key will be known only to the "Private Key Generator" (*PKG*).

- *Extract*: takes as input system parameters, master-key, and an arbitrary $ID \in \{0, 1\}^*$ and returns a private key *d*. Here *ID* is an arbitrary string that will be used as a public key, and *d* is the corresponding private decryption key. The *Extract* algorithm extracts a private key from the given public key.

- Encrypt: takes as input system parameters, *ID*, and $m \in M$. It returns a ciphertext $c \in C$.

- Decrypt: takes as input system parameters, $c \in C$, and a private key *d*. It returns $m \in M$.

One common feature of all Identity-Based Encryption systems is that they view identities as a string of characters.

Figure 23 shows an example of the flow of the IBE operations.

---

[24] The cryptosystem has chosen ciphertext security in the random oracle model assuming an elliptic curve variant of the computational Diffie-Hellman problem

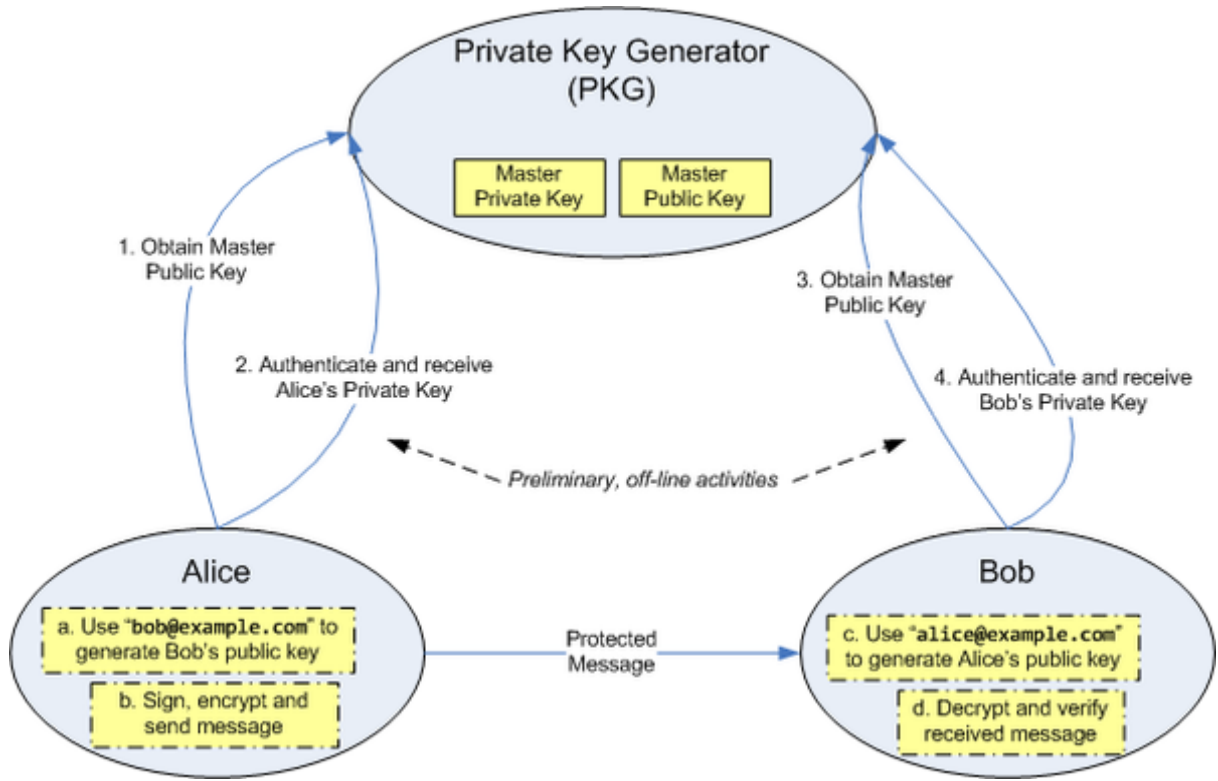

Figure 23    The IBE operations

## 5.4  Attribute Based Encryption (ABE)

As introduced in [44], ABE is a Fuzzy Identity-Based Encryption. In Fuzzy IBE an identity is view as set of descriptive attributes. Each community member receives a private key depending on its attributes (i.e., the subgroup to which it belongs), as showed in Figure 24. Information can be decrypted by people that have a certain attribute. For example, a user can have attributes such as employee and family and therefore is able to decode either the messages sent to collaborators either to relatives.

More formally, "a Fuzzy IBE scheme allows for a private key for an identity, ω, to decrypt a ciphertext encrypted with an identity, ω' , if and only if the identities ω and ω' are close to each other as measured by the set overlap distance metric" .

Known variants of ABE are [45]:

- Secure Attribute Based Systems [52]
- Multi-authority Attribute Based Encryption [47]
- Key-policy Attribute-based Encryption [43]
- Cipher-policy Attribute-based Encryption [46]

- Provable Secure Ciphertext Policy ABE [48]
- Attribute Based Encryption with Non-Monotonic Access Structures [51]
- Predicate Encryption Supporting Disjunctions, Polynomial Equations and Inner Products [49]
- Attribute Based Ring Signatures [50]

Common to these ABE schemes is the existence of a central trusted authority (master) that knows a secret master key and distributes secret attribute keys to eligible users. In [53] was presented Distributed Attribute-Based Encryption (DABE) to allow an arbitrary number of authorities to independently maintain attributes. Even if this scheme is not bound to a central server, it does not use a peer mechanism.

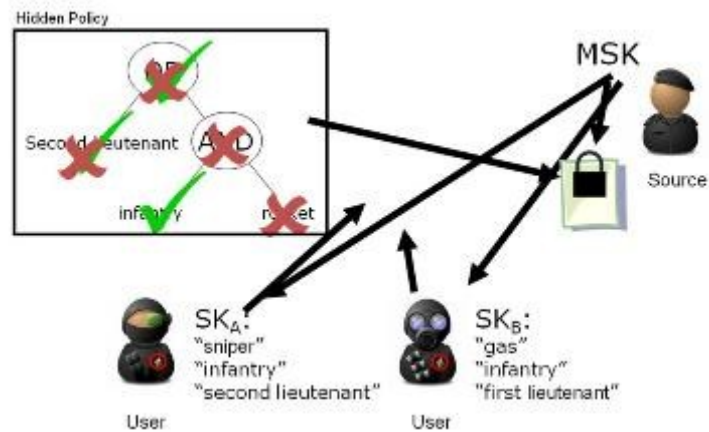

Figure 24    An ABE sample

## 5.5 Encryption in Online Social Network

The widespread success of Online Social Network (OSN) has led to a wide availability of personal data, which are often crawled by other users or OSN itself to target their users. Traditional cryptographic schemes are not adequate to protect data from abuse but, at the same time, permit authorized user the access to it. The use of public key cryptography forces users to store many copies of encrypted data and to know the identities of every community member since it doesn't allow communications based on groups (attributes). To solve this problem, new encryption schemes were studied, specifically designed for use in an OSN [143] [144]. Here I focus on Persona and GCC, which can be evaluated to adopt similar strategies in the development of my system.

### 5.5.1 Persona

Persona [61] is a private OSN that encrypts user data with attribute-based encryption (ABE), allowing users to apply fine-grained policies over users who may view their data. This architecture achieves privacy by encrypting private contents and prevents misuse of a user's applications through authentication based on traditional public key cryptography (PKC).

Persona divides the OSN entities into two categories: users, who generate the content in the OSN, and applications, which provide services to users and manipulate the OSN content.

In Persona, all users store their data encrypted for groups that they define. Any user that can name a piece of data may retrieve it, but they can only read it if they belong to the group for which the data was encrypted.

Persona operations are intended to manage group (defined using ABE) members and access to resources. The areas covered by Persona are:

- Group management with the capability to:
    - Add individuals to a group. The user generates an appropriate attribute secret key, encrypts this key using the target user's public key, and stores the encrypted key on her storage service. The target user can retrieve this encrypted key, decrypt it, and use it as necessary
    - Define groups based on a group defined by another user
    - Provide other users specific rights to named resources. An example of such a right would be the ability to store data on another user's storage service
    - Remove a group member. It requires re-keying: all remaining group members must be given a new key. Data encrypted with the old key remains visible to the revoked member.
- Publishing and Retrieving Data: private user data in Persona is always encrypted with a symmetric key. The symmetric key is encrypted with an ABE key corresponding to the group that is allowed to read this data. This two phase encryption allows data to be encrypted to groups; reuse of the symmetric key allows Persona to minimize expensive ABE operations.

## 5.5.2 Group-oriented Convergence Cryptosystem (GCC)

The previous schemes, based on traditional cryptographic techniques have limitations when dealing with multiple groups in Online Social Networks, since either users must store multiple copies of encrypted data but are unable to give data based on membership in multiple groups, or users must know the identities of everyone to whom they give access.

In [57] was introduced a community key management method based on a group-oriented convergence cryptosystem (GCC). This method leverages the following properties: the community is built on convergence of some users' private keys, the upload and download of resources provide the authentication and integrity checking, as well as there exist efficient mechanisms for access permission delegation and sophisticated revocation.

In this environment, the users in social networks are divided into four categories:

- Kernel members (KM): can create and manage a special community by collaboration and have rights to publish, delete, access or update resources released by other members of the community;
- Full authorized members (FAM): have full rights to publish and access resources in the community, but do not have permissions to delete or update resources;
- Authorized members (AM): can access the resources by using her own access permission, but cannot publish these resources;
- Unauthorized users (UU): may not have permissions to access resources published by community members.

GCC is specified by these algorithms:

- UserRegister: each user can choose a favourite label, generate a private key by herself, and then register her label into the system
- BuildCommunity: when somebody wants to share resources with others, she constructs a community together with a set of trusted

friends. Finally, each member gets a community key, which can be used to access, manage and maintain the resources in this community;

- DelegatePermission: when a user wishes to access a community, her friends hold the community key can delegate an access per-mission key (APK) to her by using this algorithm;
- UploadResource: if one community member wants to post message and resource into the community, she picks the community key, invokes this algorithm to encrypt the resource with her private key, and then transmits the encrypted data to the storage server;
- Download-Resource: anytime one community member can obtain the encrypted data from the server, and invoke this algorithm to retrieve the original post or re-source by her private key and APK.

GCC can be seen as an extension of IBE and ABE:

- As IBE, each user chooses its own keys, but without going through a centralized server
- as ABE, allows the encryption of group but the group key is generated from public keys of the subjects of communication and does not require a trusted server

Moreover, it allows the revocation and delegation.

It is interesting that each user has a single private key; the system has as many public keys as community, but the public keys are used only to create the group key (after this moment, the users decrypt using their secret key).

Table 5    Comparison between Persona and GCC

|  | Persona | GCC |
|---|---|---|
| Cryptosystem | PKC/ABE | ElGamal |
| Autonomy | PKC managed by system manager; ABE managed by group creator | Full autonomy |
| Independence | Yes | Yes, a set of trusted users |
| Collaboration | No | Yes |
| Anonymous authentication | No | Yes |
| Revocation | No | Yes |
| Integrity checking | No | Yes |
| Relationship transitive | From friend to friend | From friend to friend |
| Post message encryption | One-time by client-side | One-time by client-side |

## 5.6  Similarities between Distributed Databases and Online Social Network

Although, at first glance, the concepts of database and OSN seem totally different, I want to investigate their similarities in the field of privacy.
First, it is clear that the OSN has an internal database (see Figure 25) to store the information that dynamically will compose the wall (a space on every user's profile page that allows friends to post messages for the user to see [160]) of the users. The problem of protecting posts hosted in the OSN is then equivalent to protect shared data in an untrusted environment.

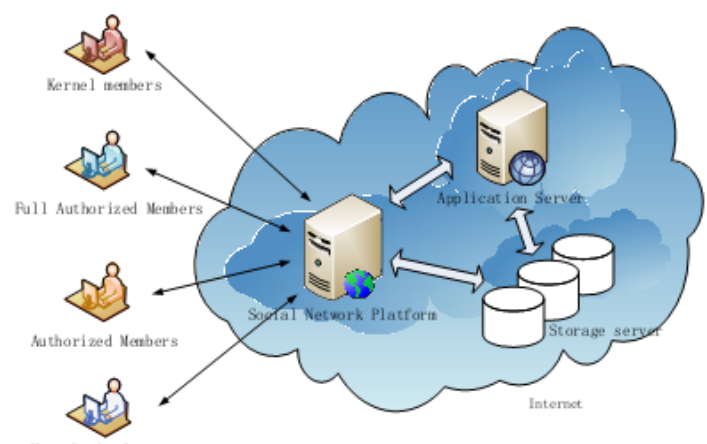

Figure 25    The OSN internal storage

Then, look at the topologies of OSN; traditionally it is represented as a mesh network among users [145], often leaving out the central node on the Cloud (the OSN provider). Considering this central element, instead, the resulting topology is a star, as depicted in Figure 26:

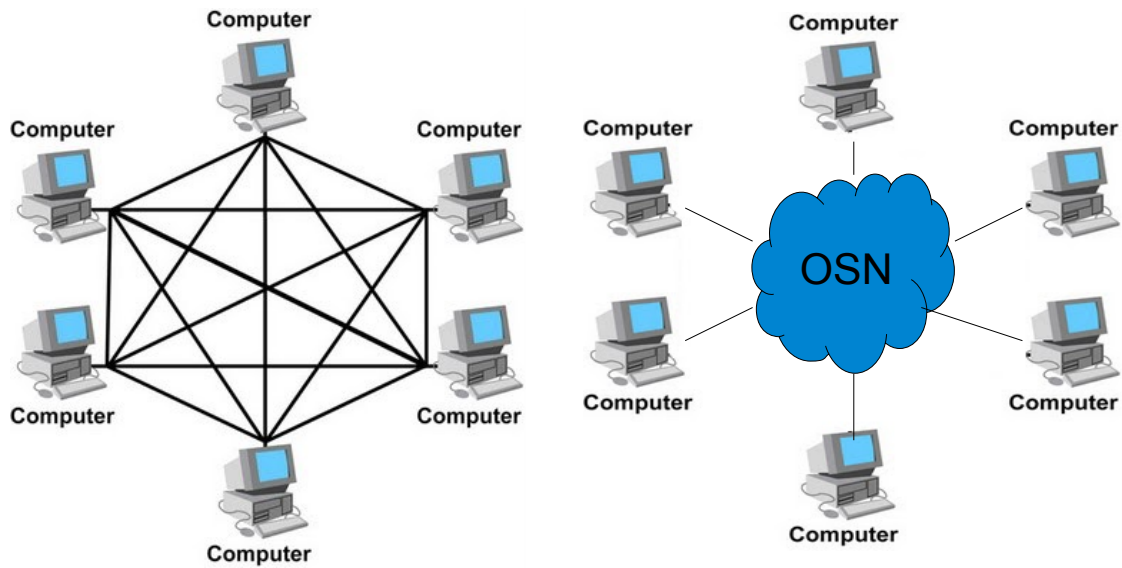

Figure 26    Topology of an OSN, as a mesh and as a star

The same topology is shown in Figure 31, where a Distributed Database Architecture (DDA) is shown. In the latter, at the center of the star is the Synchronizer, which hosts a database. Either the OSN or the Synchronizer is untrusted entity that may not access data content, while the network clients may access information accordingly to the access policy.

As every user leaves a post that may be read by multiple users (based on their permissions), but not from the server itself, so in the DDA, a user can send information to the Central Synchronizer to synchronize the replicated copies ensuring privacy in the central node.

Therefore, it is reasonable to assume that the encryption techniques used in the OSN can be adapted for distributed databases. This topic will be addressed in Section 13.

# 6 PRIVACY WITHIN THE CLOUD

*In this Chapter, I analyse Cloud storage in the wider context of Cloud computing, which is a model for enabling convenient, on-demand network access to a shared pool of configurable computing resources that can be rapidly provisioned and released with minimal management effort or service provider interaction. In this model there are more privacy issues than in Cloud Storage alone, the main of them I identify in server-side presentation layer and semantic model databases.*

Cloud computing includes a plethora of services, usually called XaaS (that stays for "everything-as-a-service"). The most common Cloud computing service models (i.e., from top to bottom, Software as a Service - SaaS, Platform as a Service - PaaS and Infrastructure as a Service - IaaS) are known as SPI. An architectural categorization of Cloud technologies as a stack of service types was proposed in [96] (see Figure 27).

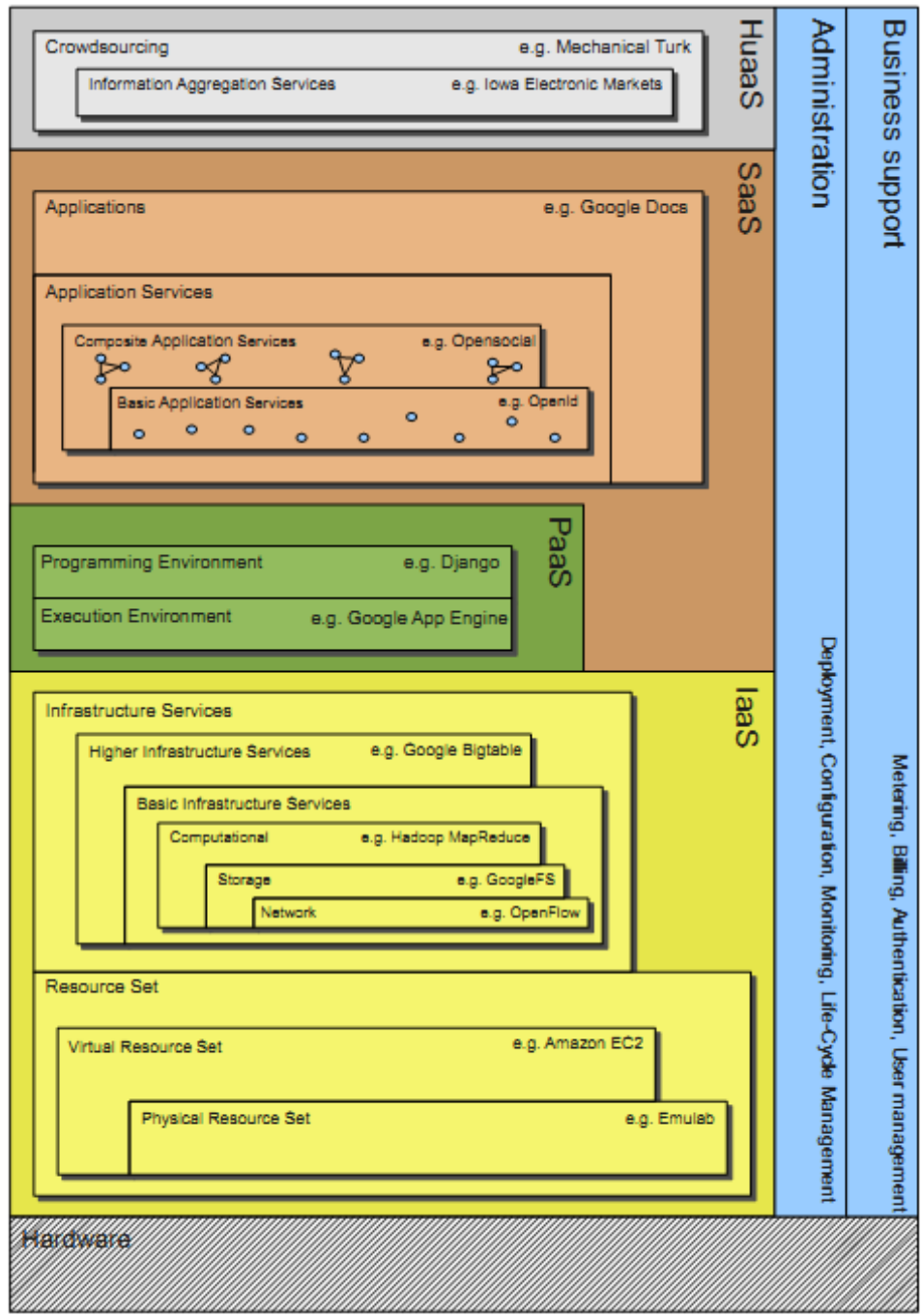

Figure 27    Cloud stack, source: [96]

The lowest level of the stack is IaaS, where a service provider offers computing, storage or networking infrastructure to customers. It contains two sublevels: Resource Set and Infrastructure Services.

- Resource Set may be divided in Physical Resource Set (PRS), which is hardware dependent and therefore tied to a hardware vendor, and Virtual Resource Set (VRS), which can be built on vendor independent hypervisor technology. Examples of PRS services include Emulab [99] and iLO [100]. VRS services include Amazon EC2[25], Eucalyptus [101], Tycoon [98], Nimbus [103], and Open Nebula [104].

- Infrastructure Services include Basic Infrastructure Services (BIS), which combine computational, storage, and network services, and Higher Infrastructure Services (HIS), as Amazon's Dynamo [102], and Google's Bigtable [11].

The second level is PaaS, which contains the provider's resources to run custom applications. It groups Programming Environments, such as Sun (now Oracle) Project Caroline[26] and the Django framework[27], and Execution Environments, such as Google App Engine[28], Joyent Reasonably Smart[29] and Microsoft Azure[30]. Each provider can couple a development and an execution environment to expose its services.

The third level is SaaS, where customers use software that is run on the provider's infrastructure. The application developers can either use the PaaS layer to develop and run their applications or directly use the IaaS infrastructure. Additionally, SaaS can also be hosted in a "conventional" way, i.e. without underlying PaaS or IaaS. SaaS may be articulated in Basic Application Services, such as OpenId[31] [105] and Google Maps[32] services, and Composite Application Services, as My Space[33] or Facebook[34]. A superset of the previous service is the "Application Services" tout-court, such as Google Docs[35], Microsoft's Office Live[36] or Auciti Hangout[37].

---

[25] Akamai Technologies Inc., "Akamai edgecomputing. Enabling applications that grow your business." - http://www.akamai.com/dl/whitepapers/ Akamai_Enabling_Apps_Grow_Business_

[26] http://labs.oracle.com/projects/caroline/

[27] https://www.djangoproject.com/

[28] http://code.google.com/intl/it-IT/appengine/

[29] http://www.joyent.com/

[30] http://www.microsoft.com/windowsazure/

[31] http://openid.net/

[32] http://maps.google.it/

[33] http://it.myspace.com/

[34] http://www.facebook.com/

On top of the traditional SPI scheme, the new level HuaaS has appeared. It indicates a set of Cloud services that are processed by a community that provides, filters, and catalogs information, such as YouTube[38], Amazon Mechanical Turk[39] or Digg[40].

In particular, in the realm of IaaS there are a lot of specialized services, including storage services. Storage services may be classified as Basic Infrastructure Services (BIS), when they only provide basic storage functionality (Amazon S3[41], GoGrid Cloud Storage[42], ExpanDrive[43], Nirvanix Cloud Storage Network[44], Rackspace Cloud Files[45], etc.), or as Higher Infrastructure Services (HIS) when they provide additional functionality, like a query language. In this taxonomy, Database-as-a-Service (DBaaS) offerings like Amazon SimpleDB[46] and technologies, such Google Bigtable [11], 10gen MongoDB[47] and Apache HBase[48] are categorized as HIS.

DBaaS is a managed service, offered on a pay-per-usage basis, which provides on-demand access to a database for the storage of application data [95]. In the Cloud, storage is spanned on multiple servers, usually hosted in large data centres, whose operators manage the infrastructure.

---

[35] http://docs.google.com/

[36] http://www.officelive.com/

[37] http://fun.auciti.com/

[38] http://www.youtube.com/

[39] https://www.mturk.com/

[40] http://digg.com/

[41] http://aws.amazon.com/s3/

[42] http://www.gogrid.com/Cloud-hosting/Cloud-storage.php

[43] http://www.expandrive.com/

[44] http://www.nirvanix.com/products-services/storage-delivery-network

[45] http://www.rackspace.com/Cloud/Cloud_hosting_products/files/

[46] http://aws.amazon.com/simpledb/

[47] http://www.10gen.com/

[48] http://hbase.apache.org/

### 6.1.1 Cryptographic Storage Service

The benefits of using a public Cloud infrastructure are cost reduction, availability (anywhere access), and reliability (backups), but it introduces significant security and privacy risks for the confidentiality and integrity of data. To address these concerns a common solution is the use of cryptographic storage, which protects data using encryption to provide [97]:

- Confidentiality: the Cloud storage provider does not learn any information about customer, and
- Data integrity: any unauthorized modification of customer data by the Cloud storage provider can be detected by the customer;

without giving up to the main benefits of a public storage service:

- Availability: customer data is accessible from any machine and at all times,
- Reliability: customer data is reliably backed up,
- Efficient retrieval: data retrieval times are comparable to a public Cloud storage service, and
- Data sharing: customers can share their data with trusted parties.

The benefits of a CSS can be summarized in:

- Regulatory compliance: often, law makes organizations responsible for the protection of the data that is entrusted to them. Since, in CSS, data is stored encrypted, customers can be assured that the confidentiality of their data is preserved irrespective of the actions of the Cloud storage provider;
- Geographic restrictions: data may be subject to law of Country of physical storing. If data is stored on Cloud, it is not clear its physical location. Since, in CSS, data is stored encrypted, any law that pertains to the stored data has little to no effect on the customer;
- Subpoenas: in case of a legal action, the request of data may be made to the Cloud provider and the latter could even be prevented from notifying the customer. Since, in CSS, data is stored encrypted, the Cloud provider has no access to information and needs to turn request to the customer. Moreover, the request is managed by the customer under investigation and does not affect the others that store data on the Cloud provider;

- Security breaches: a Cloud storage provider may be legally responsible for a security breach. But, in a CSS data is encrypted and data integrity can be verified at any time;
- Electronic discovery: a Cloud provider needs a large amount of additional information to prove data integrity. In a CSS, instead, data integrity can be verified at any time without additional storing; and
- Data retention and destruction: in the Cloud it is difficult to prove the destruction of previously collected data. In a CSS, even in case of data retention, information is not accessible to the provider.

A cryptographic storage service (CSS) can have a Consumer Architecture or an Enterprise Architecture [97].

### *6.1.1.1 A Consumer Architecture*

In a Consumer Architecture, three actors collaborate to share data:
   a. A user Alice that stores her data in the Cloud,
   b. A user Bob with whom Alice wants to share data, and
   c. A Cloud storage provider that stores Alice's data.

The two users run <u>locally</u> a client application that consists of a data processor, a data verifier, and a token generator. At the first step, Alice's application generates a cryptographic key that is stored locally on Alice's system and that it is kept secret from the Cloud storage provider.

Alice can:

- Upload data to the Cloud using the data processor. It attaches some metadata (e.g., current time, size, keywords etc) and encrypts and encodes the data and metadata with a variety of cryptographic primitives. The result is sent to the Cloud storage;
- Verify the integrity of her data using the data verifier. It uses Alice's master key to interact with the Cloud storage provider and ascertain the integrity of the data;
- Retrieve data using keywords. The token generator is invoked to create a token. The token is sent to the Cloud storage provider who uses it to retrieve the appropriate (encrypted) information which it returns to Alice. Alice then uses the decryption key to decrypt the information.

To share data with Bob: (1) Alice's data processor prepares the data before sending it to the Cloud; (2) Bob asks Alice for permission to search for a keyword; (3) Alice invokes the token generator to create an appropriate token, and the credential generator to generate a credential for Bob, and sends them to Bob; (4) Bob sends the token to the Cloud; (5) the Cloud uses the token to find the appropriate encrypted documents and returns them to Bob. (6) At any point in time, Alice's data verifier can verify the integrity of the data.

The process is summarized in the following schema:

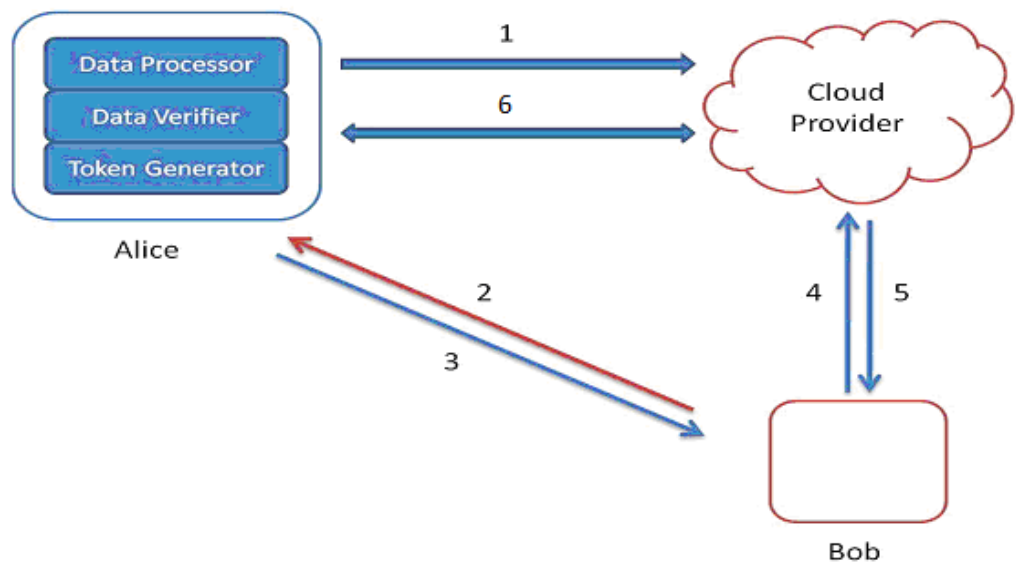

Figure 28    A Consumer Architecture

### 6.1.1.2  *An Enterprise Architecture*

In an Enterprise Architecture, three actors collaborate to share data:

    a. An enterprise MegaCorp that stores its data in the Cloud while using dedicated machines within its network to run a data processor (DP), a data verifier (DV), and a token generator (TG);

    b. A business partner PartnerCorp with whom MegaCorp wants to share data; and

    c. A Cloud storage provider that stores MegaCorp's data.

The process is as follows: (1) each MegaCorp and PartnerCorp employee receives a credential that reflects her organization / team / role; (2) MegaCorp employees send their data, together with an associated decryption policy that specifies the type of credentials necessary to decrypt it, to the internal machine hosting the DP; (3) the latter processes the data using the DP before sending it to the Cloud; (4) the PartnerCorp employee sends a keyword to MegaCorp's internal machine hosting TG; (5) the dedicated machine returns a token; (6) the PartnerCorp employee sends the token to the Cloud; (7) the Cloud uses the token to find the appropriate encrypted documents and returns them to the employee. (8) At any point in time, MegaCorp's DV can verify the integrity of MegaCorp's data.

The process is summarized in the following schema:

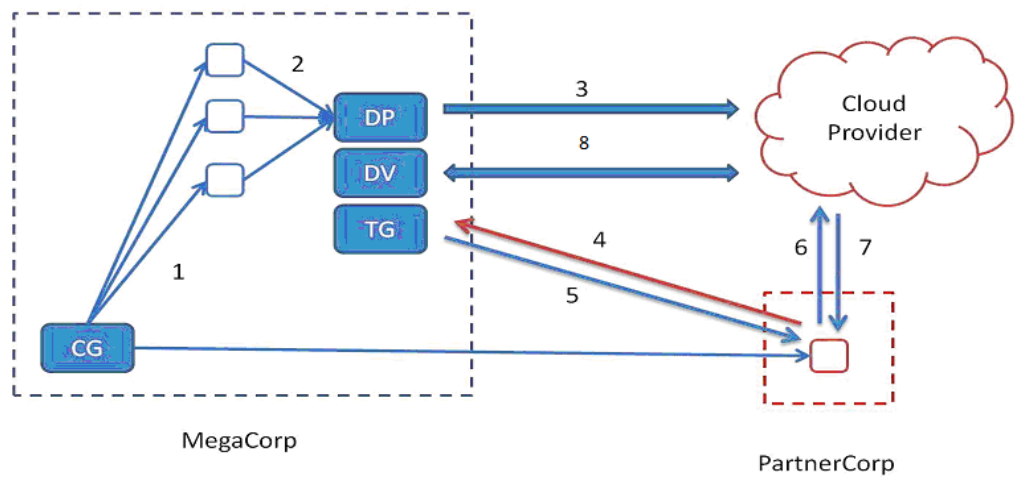

Figure 29    An Enterprise Architecture

## 6.2  Privacy issues in Cloud computing

The CSS assumes the distinction between data storage (on the Cloud) and data manipulation (in the client or corporation side), while Web 2.0 applications are usually hosted by PaaS providers. The most diffused Execution Environments on the Cloud handle both data and application management.

The three main suppliers of Public Cloud Infrastructure (Google App Engine for Business, Amazon Elastic Compute Cloud and Windows Azure Platform) all include a datastore, and an environment for remote execution summarized in Table 6 and Table 7.

Table 6    Datastore solutions used by public Clouds

| Provider | Datastore |
|---|---|
| Google | Bigtable |
| Amazon | IBM DB2<br>IBM Informix Dynamic Server<br>Microsoft SQLServer Standard 2005<br>MySQL Enterprise<br>Oracle Database 11g<br>Others installed by users |
| Microsoft | Microsoft SQL Azure |

Table 7    Execution environments used by public Clouds

| Provider | Execution environment |
|---|---|
| Google | J2EE (Tomcat + GWT)<br>Python |
| Amazon | J2EE (IBM WAS, Oracle WebLogic Server) and others installed by users |
| Microsoft | .Net |

Let us now go back to the scenario from where this dissertation has started: the possibility of an untrustworthy cloud supplier who can intercept communications, modify executable software components (e.g., using aspect programming), monitor the user application memory, etc. Figure 30 describes a typical session, where data travels from data storage (DS) to the final user by passing through database drivers (e.g. JDBC/ODBC drivers), Object-Relational Mapping (ORM) layer, until it arrives to the Presentation Layer (PL). While, as showed before, DS may be managed in a secure manner, the same can not be said about the other layers. The central line in Figure 30 (the *wall*) divides the part in charge of the Cloud and that in charge of User.

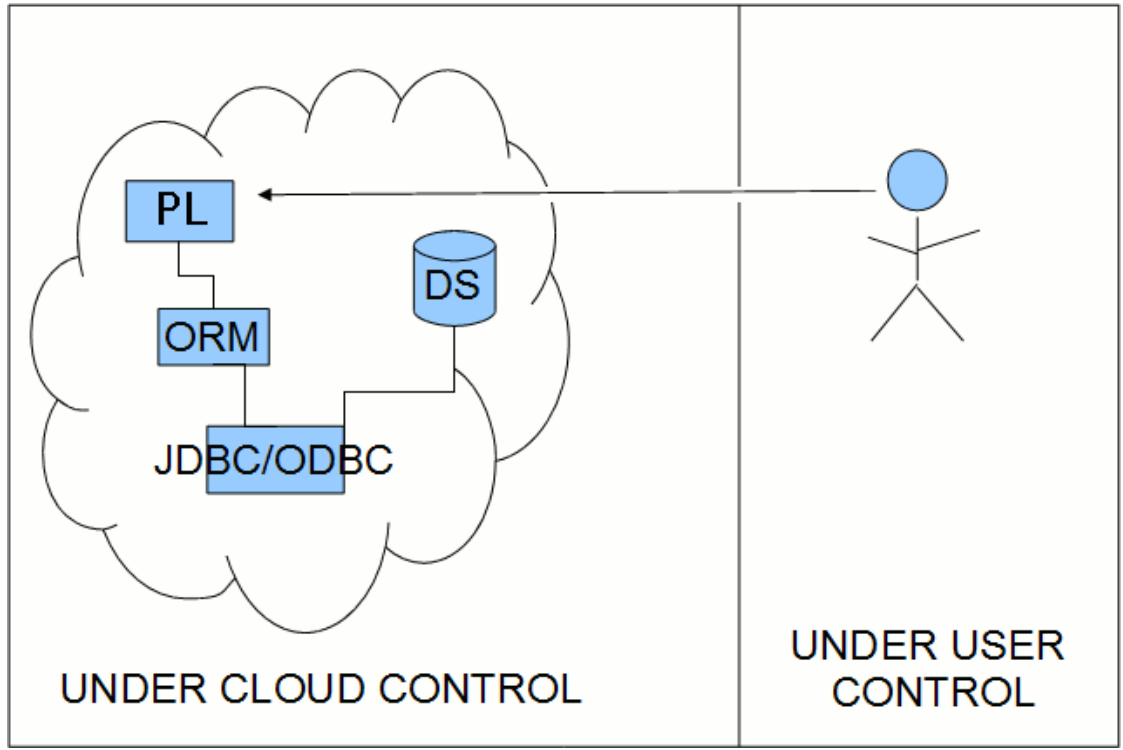

Figure 30    The wall

Hence, available techniques for safely outsourcing data to untrusted DBMS no longer guarantee the confidentiality of data outsourced to the Cloud.

The essential point consists in having the data and the user interface application logic on the same side of the wall. This is a major difference w.r.t. outsourced database scenarios, where presentation was handled by trusted clients. In the end, the data must be presented to the user in an intelligible and clear form; that is the moment when a malicious agent operating in the Cloud has more opportunities to intercept the data. To prevent unwanted access to the data at presentation time, it would be appropriate moving the presentation logics off the Cloud to a trusted environment that may be an intranet or, at the bottom level, a personal computer.

However, separating data (which would stay in the Cloud) from the presentation logics may enable the creation of local copies of data, and lead to an inefficient cooperation between the two parts.

## 6.3 Semantic datastore

In Section 2.2-Data protection techniques, I showed many techniques for *data outsourcing in untrusted servers,* primarily designed for untrusted RDBMS.

But since, today, Cloud computing approaches largely rely on semantic (non-relational) DBMSs, those techniques cannot be applied directly. Semantic DBMSs do not store data in tabular format, but following the natural structure of objects. After more than twenty years of experimentation (see, for instance, [10] for the Galileo system developed at the University of Pisa), today, the lower performance of these systems is no longer a problem. In the field of Cloud computing, there is a particular attention to Google Bigtable.

"Bigtable is a distributed storage system for managing structured data that is designed to scale to a very large size: petabytes of data across thousands of commodity servers. In many ways, Bigtable resembles a database: it shares many implementation strategies with databases." [11]

With a semantic datastore like Bigtable, there is a more strict integration between in-memory data and stored-data; they are almost indistinguishable from the programmer viewpoint. There are not distinct phases when the program loads data from disk into main memory or, in the opposite direction, when program serialize data on disk. Applications do not even know where the data is stored, as it is scattered over the Cloud.

# Part II - Research questions and results

# 7 DESIGN OF A DISTRIBUTED SYSTEM FOR INFORMATION SHARING

*This chapter presents the challenges that arise from the context described by the previous chapters in Part I of the thesis. My research contribution is the design of a new architecture whose goals are data privacy preservation, fine grained access control, and grant-and-revoke capability.*

From the context described in the previous chapters, three main challenges emerge and are addressed by the architecture proposed in this thesis:

- Native Cloud sharing system: in Section 6-Privacy within the Cloud, I exposed the peculiarity of Cloud storage considered not as a singleton but in the more general perspective of Cloud Computing. It brings to state that what presented in Section 2-Background on data protection is not perfectly suitable on the Cloud. My research contribution will consist in the design of a data sharing system thought for the Cloud since birth.
- Safe system with minimal number of trusted components: to prevent privacy leaks in the remote storage, the usual strategies, shown in Section 2-Background on data protection, are: i) introducing some trusted component during data sharing, or ii) dividing information and responsibilities among different untrusted actors. In my research, instead, I want to guarantee privacy in an untrusted environment where only the data owner is trusted.
- Platform independent system, built on available standard technologies: custom tools were developed for privacy in a well-defined platform (e.g.

the schemes for OSN in 5.5-Encryption in Online Social Network). My contribution is a framework that is independent from the Cloud provider, and uses standard industrial technologies and protocols.

## 7.1 Description Language

I will describe my approach to Cloud data privacy by means of the well-known Unified Modeling Language (UML). I chose UML, rather than a more formal description, i.e., Finite-State Automata (FSA), to emphasize the collaboration among the distributed components and their role data exchange. UML[49] is a standardized, general-purpose modeling language for object-oriented software engineering, which implements the concepts of Model Driven Development (MDD) [158], a framework that arose, in the early 1990's, in the telecommunications industry. Although UML is a semiformal language that lacks a well-defined semantics and then hardly allows formal verification, it is widely used in communication protocols design and description (e.g. Broadcast Encryption Protocols, see Section 5).

The strength of MDD is in modelling highly-concurrent processes that are state-based and communicate through well-defined messages [159]. It does so by a broad range of diagrams that covers requirements analysis (called *use case diagrams*), class components (*class diagrams*), algorithmic sequences (*sequence diagrams*), component evolutions (*state diagrams*), etc. The ability to extract sequence diagrams from UML executable models will allow us to verify and validate the system behaviour, as I will detail in Section 10.

## 7.2 The architecture

I build over the notion introduced in [7] of defining a view for every user group/role, but I prevent performance degradation by keeping all data views in the user environment.

---

[49] http://www.omg.org/spec/UML/2.0/

Specifically, I atomize the application/database pair, providing a copy per user. Every instance runs locally, and maintains only authorized data that is replicated and synchronized among all authorized users.

I will consider a system composed of:

1. Local agents distributed at client side;
2. A central synchronization point.

The following figure shows the proposed architecture:

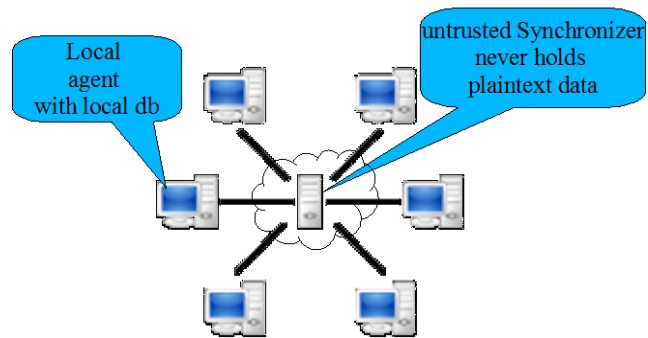

Figure 31    The architecture

## 7.3  The model

Henceforth, I will use the term dossier to indicate a set of related information. My data model may be informally represented by the diagram in the following figure:

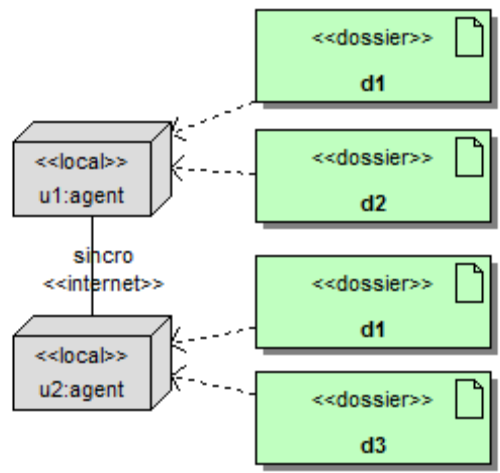

Figure 32    The model

In my model, each node represents a local, single-user application/database dedicated to an individual user ($u_n$). Each node stores only the dossiers that its user may access. Shared dossiers (in this example, $d_1$) are replicated on each node. When a node modifies a shared dossier, it must synchronize, also using heuristics and learning algorithms, with the other nodes that hold a copy of it. In Table 8 and Table 9, I give a simple SWOT analysis [162] of this idea.

Table 8    Strength/Opportunities

| |
|---|
| Information sharing using untrusted Synchronizer; |
| Small amount of local data, less attractive for attackers; |
| Only the final user has clear-text information; |
| Unrestrained individual nodes, that can also work offline (with deferred synchronization); |
| Simplicity of data management (single user); |
| Completeness of local information. |

To clarify the last point, suppose that the user $u_n$ wants to know the number of the dossier she is treating. In a classic intranet approach, where dossiers reside on their owners' servers, in addition to its database, $u_n$ should examine the data stores of all other collaborating users. With my approach, instead, I will simply perform a local query because the dossiers are replicated at each client.

Table 9    Weaknesses/Threats

| |
|---|
| Complexity of deferred synchronization schemes [21]; |
| Necessity to implement a mechanism for grant/revoke and access control permissions |

This last point is particularly important and it deserves further discussion:

- Each user (except the data owner) may have partial access to a dossier. Therefore each node contains only the allowed portion of the information, i.e. the dossier does not contain at all the restricted parts;
- Authorization, i.e., granting to a user $u_j$ access to a dossier $d_k$, can be achieved by the data owner simply by transmitting to each node only the data it is allowed to access, i.e. the restricted attributes are omitted (blanked);
- The inverse operation can be made in the case of a (partial or complete) revocation of access rights. An obvious difficulty lies in ensuring that data becomes no longer available to the revoked node. This is indeed a moot point, as it is impossible – whatever the approach - to prevent trusted users from creating local copies of data while they are authorized and continue using them after revocation. I am evaluating the opportunity to use watermarking for relational databases [26] to provide copyright protection and tamper detection.

# 8 SCENARIOS

*In this Chapter, I define and discuss in depth some business scenarios where the proposed architecture naturally fits practical requirements. I will consider three typical scenarios: a medical, a group of independent professionals, and a nomadic user scenario, as they embody very general and diffuse patterns of usage.*

As we will see, my approach best fit into a typical Business Collaborative Environment scenario, where a group of geographically scattered, associated professionals wants to share some structured information. Each actor can access only partial data of her interest. Data may be located at different sites.

In the following, I will discuss three of these scenarios.

## 8.1 Medical

In this scenario, a family doctor shares some medical records with a specialist as a Radiologist or a Psychologist and the patients' invoices with an Accountant. Clearly, a Radiologist may not access psychotherapy sessions, as well as a Psychologist may not access X-rays. Both of them need access to "Patient Master Data", but only for their patients, a subset of those with the family doctor. After finishing the therapy, access to a patient data may even be revoked. The scenario is depicted in Figure 33.

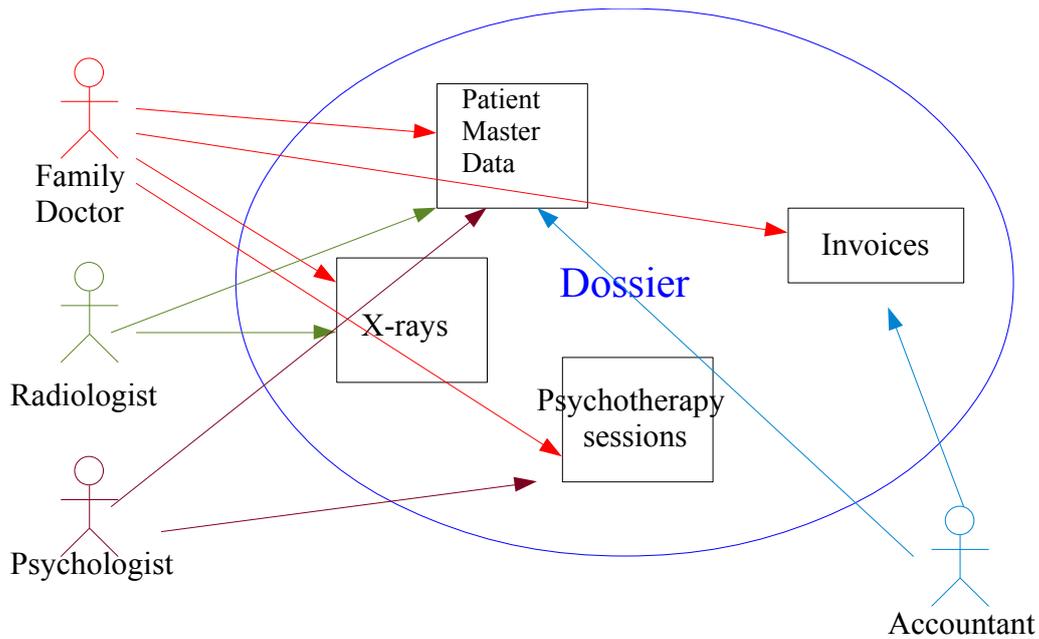

Figure 33    A medical scenario

Let us the following real life example: two Family Doctors $FD_1$ and $FD_2$, two Radiologists $R_1$ and $R_2$, a Psychologist $P_1$, an Accountant $A_1$, and three Patients Alice, Bob, and Charlie. Alice is with $FD_1$, $R_1$, and $A_1$, while Bob is with $FD_1$, $R_2$, and $P_1$, Charlie is with $FD_1$, and $A_1$.

In my distributed system, this is modelled as in Figure 34.

As we can see, each actor stores data locally in her computer; the Patient's data is replicated in each of her doctors; each actor stores locally only the allowed portion of the patient's data.

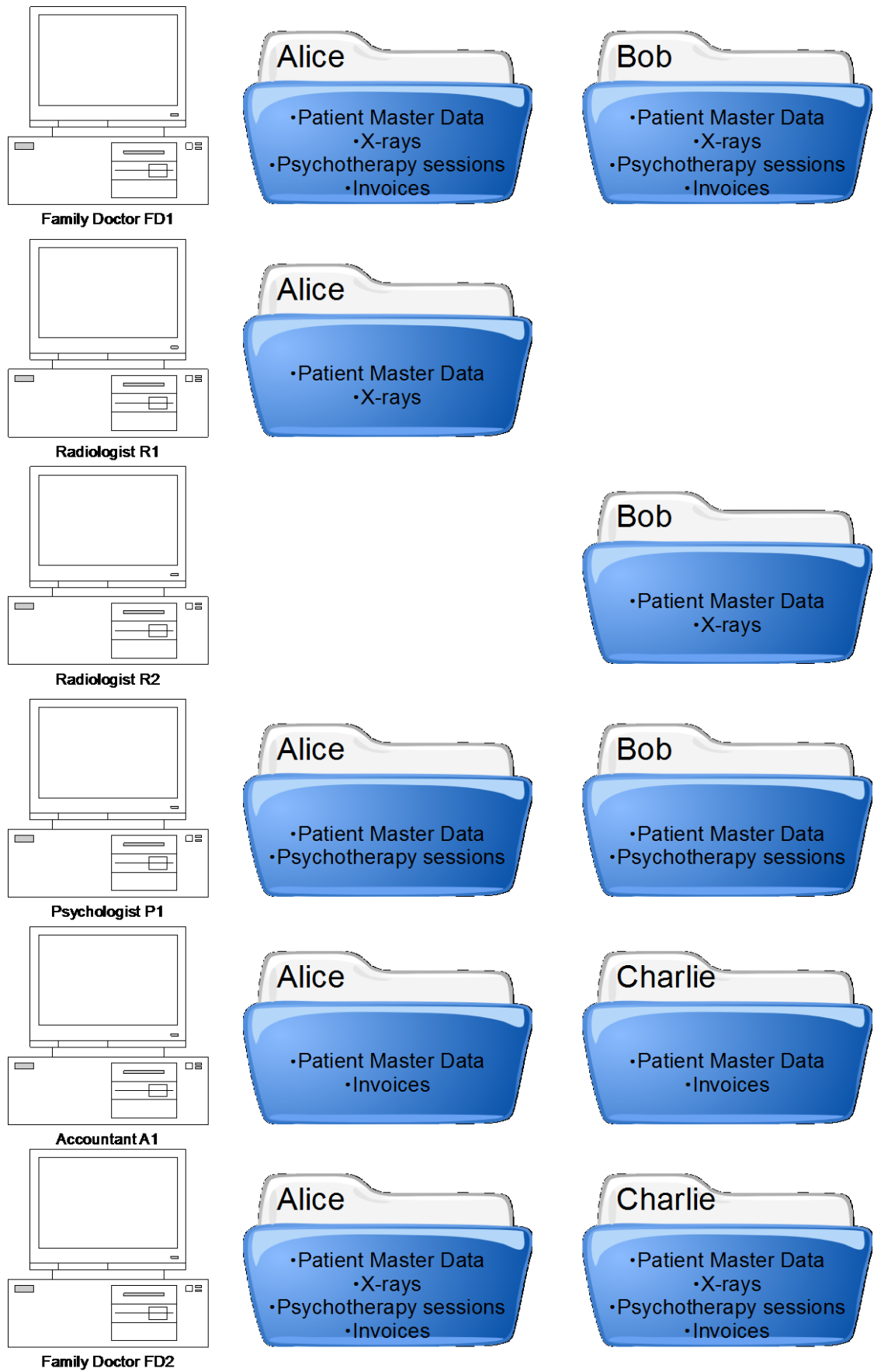

Figure 34    An instance of medical scenario

## 8.2 Collaboration and sharing between independent professionals

The previous scenario can be generalized to capture the case of professionals, such as Lawyers, Engineers, Business Consultants and so on.

This scenario is challenging because each participant may be bound to confidentiality/non-disclosure obligations to her client; these obligations may overlap/conflict with each other. The common characteristics of these figures are strong data locality, uneven connectivity, and, shall we say, patchy cooperation. The framework I develop in this dissertation is designed to adapt nicely to these circumstances, thus allowing the application of Cloud computing.

## 8.3 Nomadic users

Smartphones and tablets are increasingly becoming popular. All these devices now have fairly large storage (for agenda, contacts and other data) that acts as a cache when the device is offline. In the case of data shared among a network of friend, co-workers, etc., synchronization is guaranteed by a central server (e.g. Google Calendar) when connection is up. When shared data is sensitive, my system allows a "protected" synchronization without disclosing information on a Public Cloud (again, Google Calendar is a good example here).

# 9 MY APPROACH: THE iPRIVACY SYSTEM

*In this Chapter, I provide a semi-formal description of my architecture and of the interaction among its component by means of UML diagrams. I show how the architecture satisfies some general requirements, including protecting data privacy, supporting fine-grained access control and providing a grant-and-revoke capability.*

I am now ready to present in detail my approach, that I named iPrivacy. To simplify the discussion, I introduce the following assumptions:

- Each dossier has only one owner;
- Only the dossier's owner can change it.

These assumptions permit the use of an elementary cascade synchronization in which the owner will submit the changes to the receivers. However, they can be relaxed at the cost of a higher complexity in synchronization [34].

My approach consists of two parts: a trusted client and a remote untrusted synchronizer (see Figure 35).

The client maintains local data storage where:

- The dossiers that she owns are (or at least can be) stored as plaintext;
- The others, instead, are encrypted each with a different key.

The Synchronizer stores the keys to decrypt the shared dossiers owned by the local client and the modified dossiers to synchronize.

When another client needs to decrypt a dossier, she connects to the Synchronizer and obtains the corresponding decryption key.

The data and the keys are stored in two separate entities, none of which can access information without the collaboration of the other part.

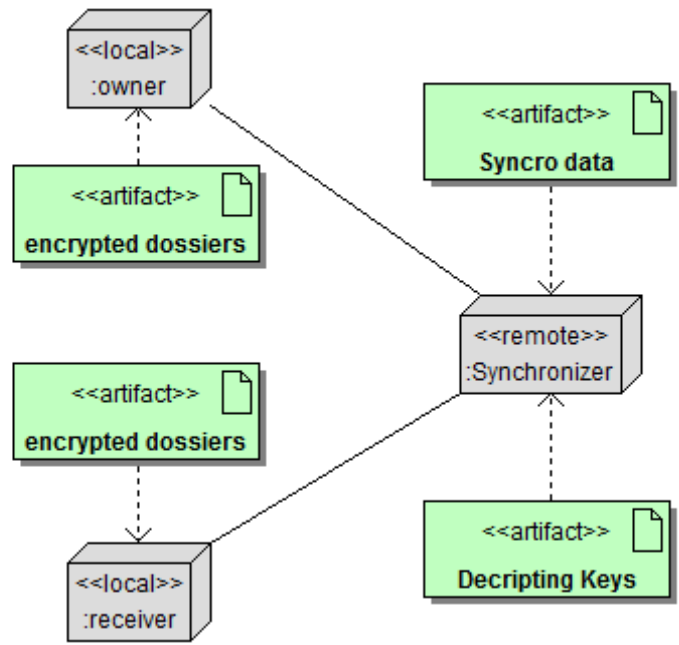

Figure 35    Deployment diagram of the distributed system

## 9.1 Structure

From the architectural point of view, I divide the system into two packages, a local one (client agent), which contains the dossier plus additional information such as access lists, and a remote one (global synchronizer), which contains the list of dossiers to synchronize, their decryption keys and the public keys of clients.
A UML view of involved classes is shown in Figure 36:

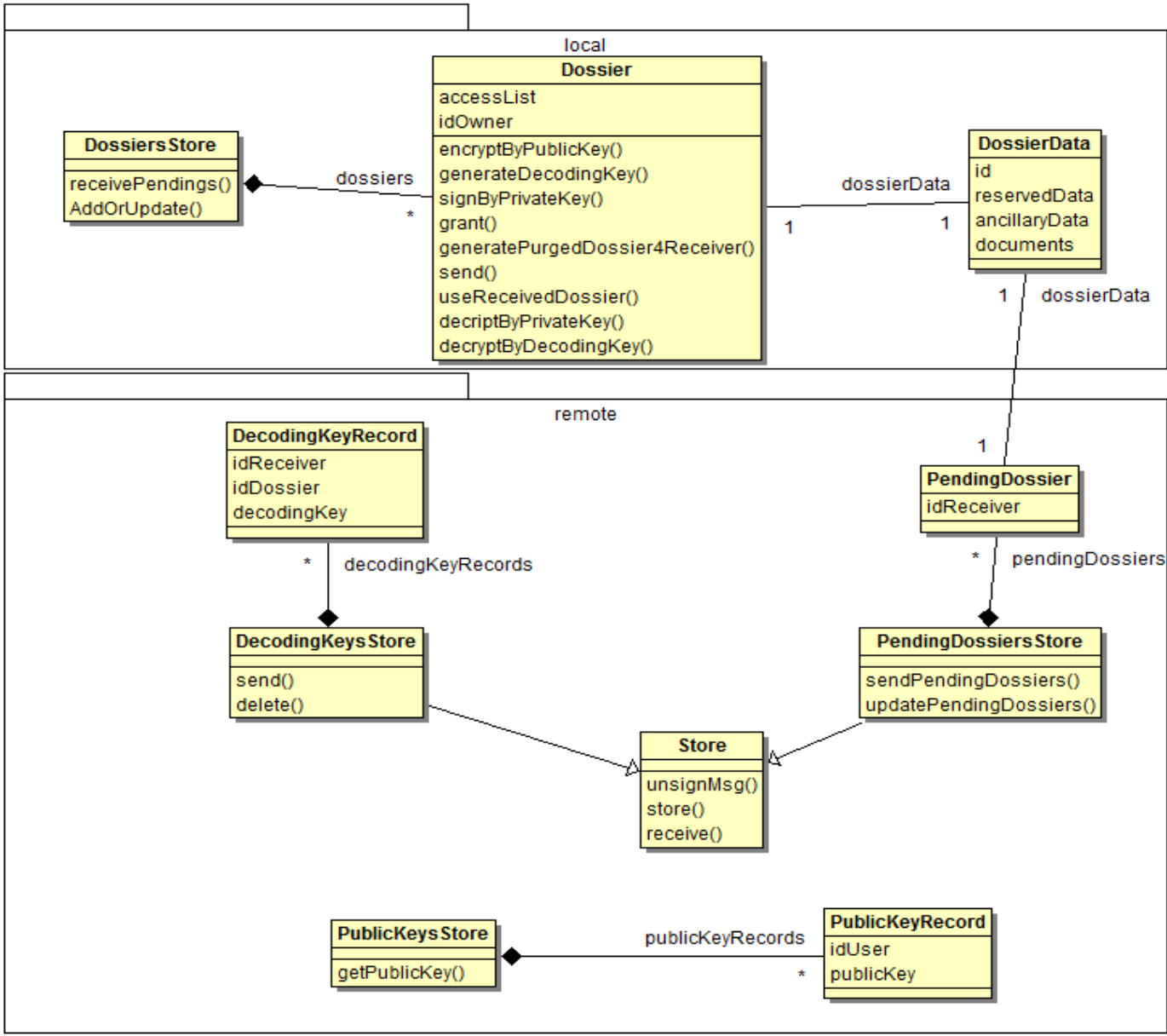

Figure 36    iPrivacy's class view

## 9.2  Grant

An owner willing to grant rights on a dossier must follow the sequence in Figure 37.

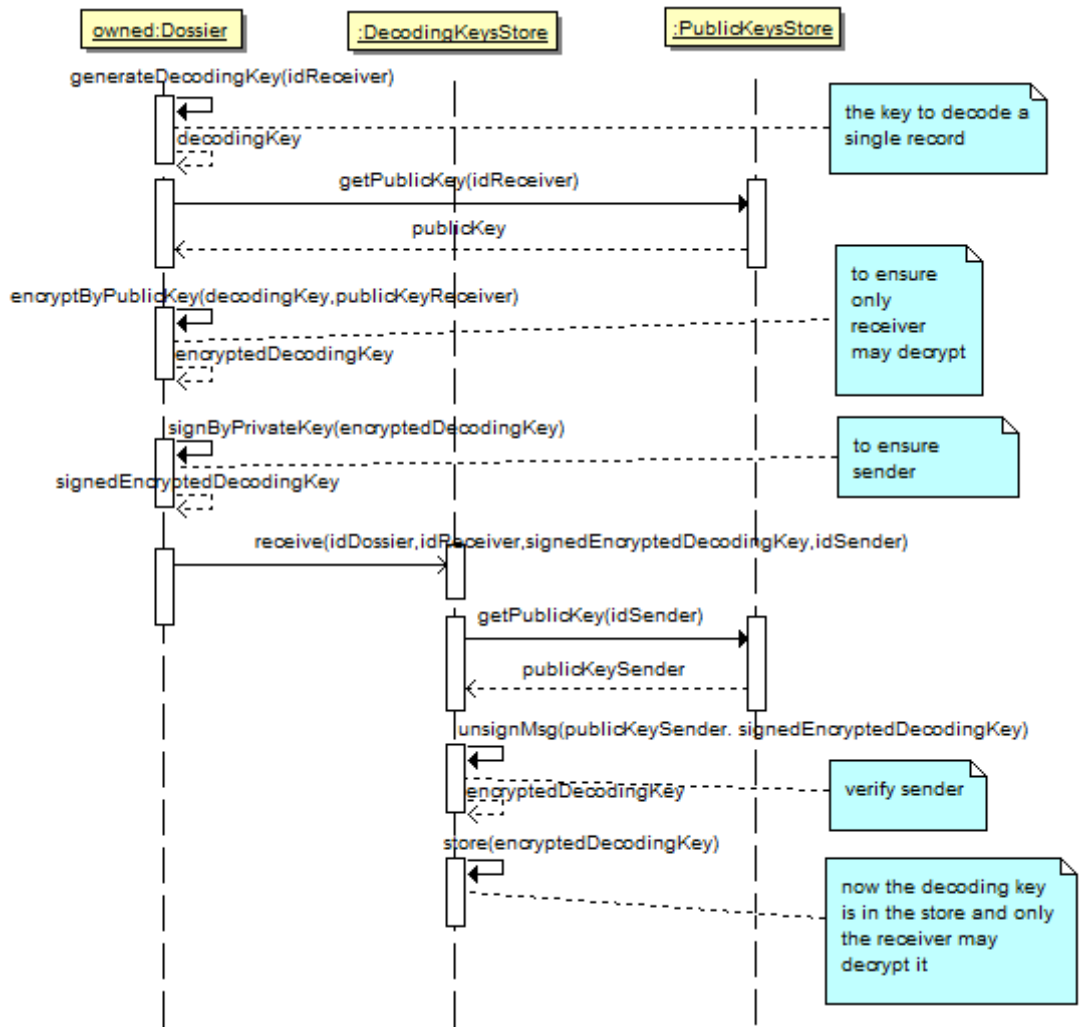

Figure 37  Grant sequence

Namely, for each receiver, the owner:
- Generates the decryption key
- Encrypts it with the public key of the receiver to ensure that others cannot read it
- Signs it with its private key to ensure its origin
- Sends it to the Synchronizer, which verifies the origin and adds it to the storage of the decoding keys. The key is still encrypted with the public key of the receiver, so only the receiver can read it.

## 9.3  Send

When an owner modifies a dossier, she sends it to the Synchronizer using the sequence in Figure 38.

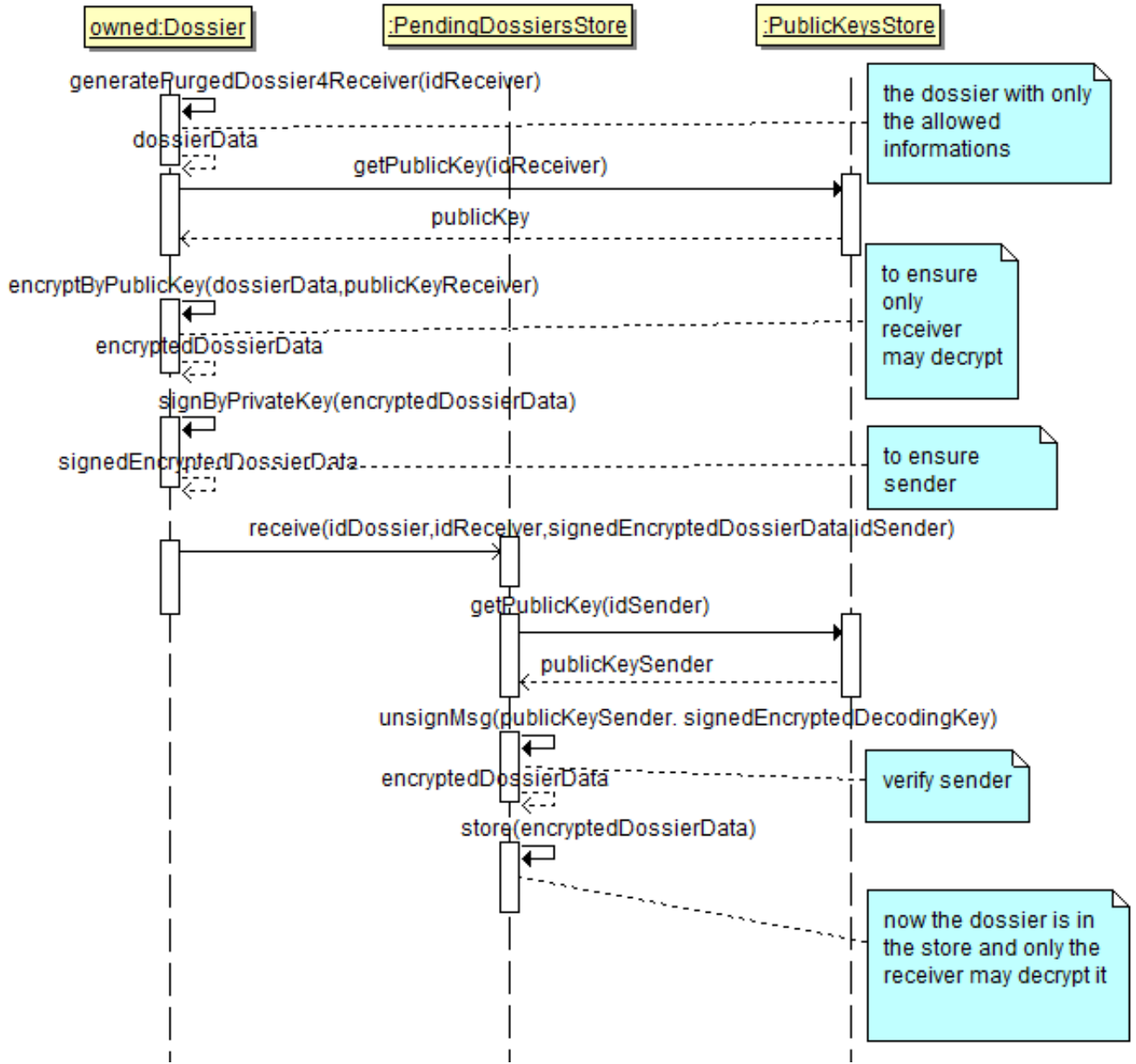

Figure 38    Send sequence

For each receiver, the owner:

- Generates a "pending dossier" by removing information that the receiver should not have access to;
- Encrypts the pending dossier with the previously generated decryption key;
- Signs with his own private key to certificate its origin;
- Sends it to the Synchronizer, which verifies the origin and adds it to the storage of "pending dossiers". Again, the dossier is still encrypted with the public key of the receiver, so only the receiver can read it.

## 9.4 Receive

Periodically, each client updates un-owned dossiers by following the sequence in Figure 39.

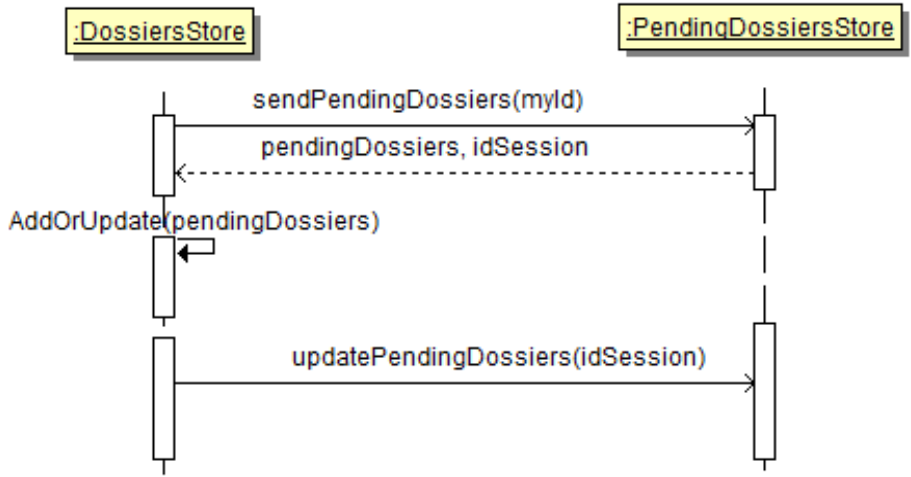

Figure 39    Receive sequence

Each client:
- Requests the "pending dossiers" to the Synchronizer.
- Stores the (still encrypted) dossier in the local storage;
- Removes the received dossiers from the Synchronizer.

## 9.5 Use

When a client needs to access an unowned (encrypted) dossier, the sequence in Figure 40 is used.

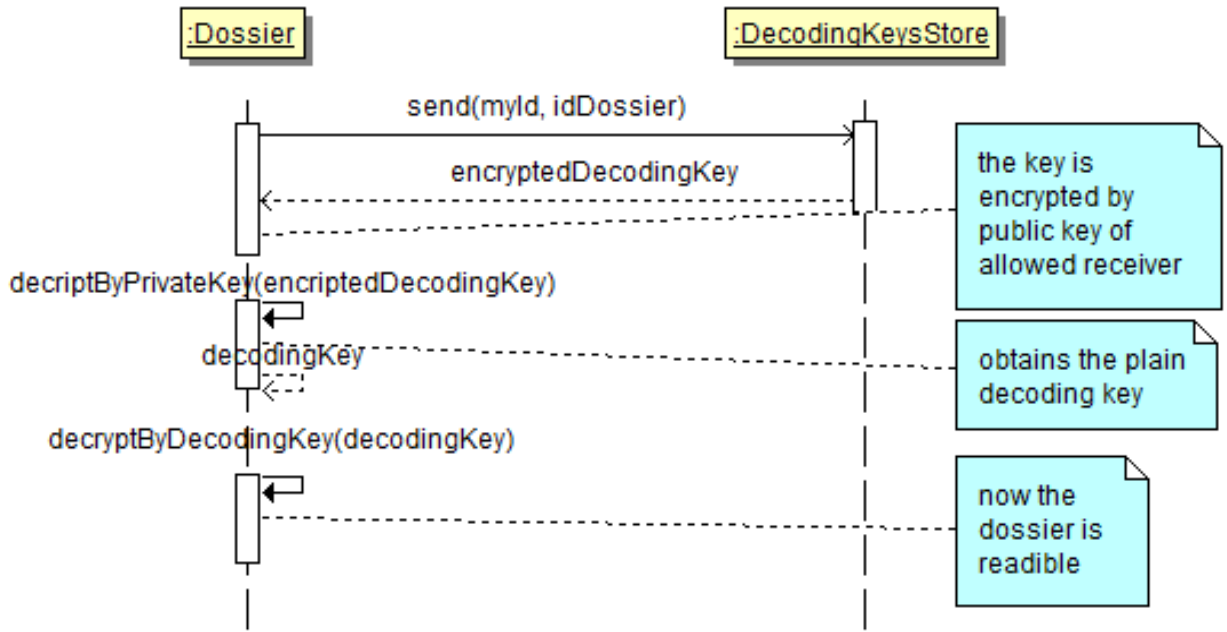

Figure 40    Use sequence

The client:

- Asks the Synchronizer for the decryption key (that is encrypted by her public key);
- Decrypts it with her private key;
- Decrypts the dossier by the resulting decryption key.

If the decryption key does not exist, two options are available:

- The record is deleted from the local datastore because a revoke happened;
- The record remains cached (encrypted) into the local datastore because access rights to it could be restored.

## 9.6  Revoke

To revoke access to a receiver, it is sufficient to delete the corresponding decryption key from the Synchronizer. The sequence is designed in the UML diagram of Figure 41.

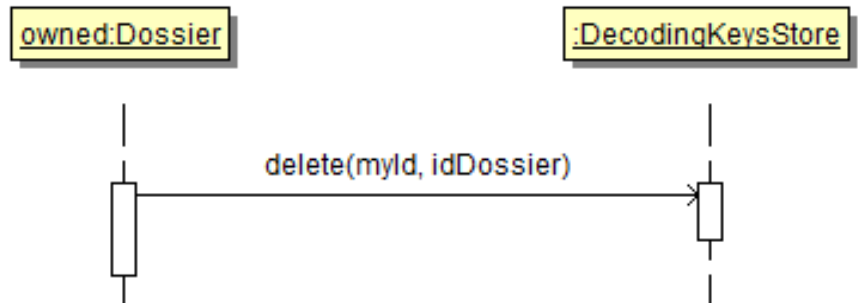
Figure 41    Revoke sequence

# 10 CONGRUENCE BETWEEN PLANNED AND ACHIEVED GOALS

*Using state diagrams of each component of the system, I will asses to which level my architecture satisfies the initial requirements for privacy protection and access control management.*

The initial requirements for my project were: privacy protection towards Cloud platform, fine-grained data access control, and grant-and-revoke permission on shared data.

At the start, I noted that the Presentation Layer may not rest on the Cloud side, exactly to avoid exposing clear-text data to potential inner attacks. For this purpose, I moved the presentation logic to the client side. The next step was moving also the data layer from Cloud to client side, to prevent performance degradation.

In the end I obtained a distributed architecture and a framework for data synchronization, but is the resulting system compliant with the initial requirements?

Let us examine again each component involved in the system:
- Client, distinct by role in:
    - Sender, and
    - Receiver;
- Synchronizer, and
- Network.

## 10.1 The Sender component (client side)

A Client, when acting as *Sender*, manages only its own data. It transmits the modified local data to the Synchronizer. Obviously, the Sender has both the owned dossiers and the related decrypting keys in clear-text form. However, this does not violate the requirements.

The Sender initiated actions are: *Grant*, *Send*, and *Revoke*.

Essentially, when revoking access rights, the sender decides which dossiers have to be shared, who are the receivers, and what part of information each of them receives. This role, therefore, is related only to the last of the initial requirements.

## 10.2 Receiver

A Client, when acting as *Receiver*, manages shared (not owned data), interacting with the Synchronizer to receive public keys, pending dossiers and decrypting keys.

The actions involved are: *Receive*, and *Use*.

Figure 42 shows the state diagram of a Receiver:

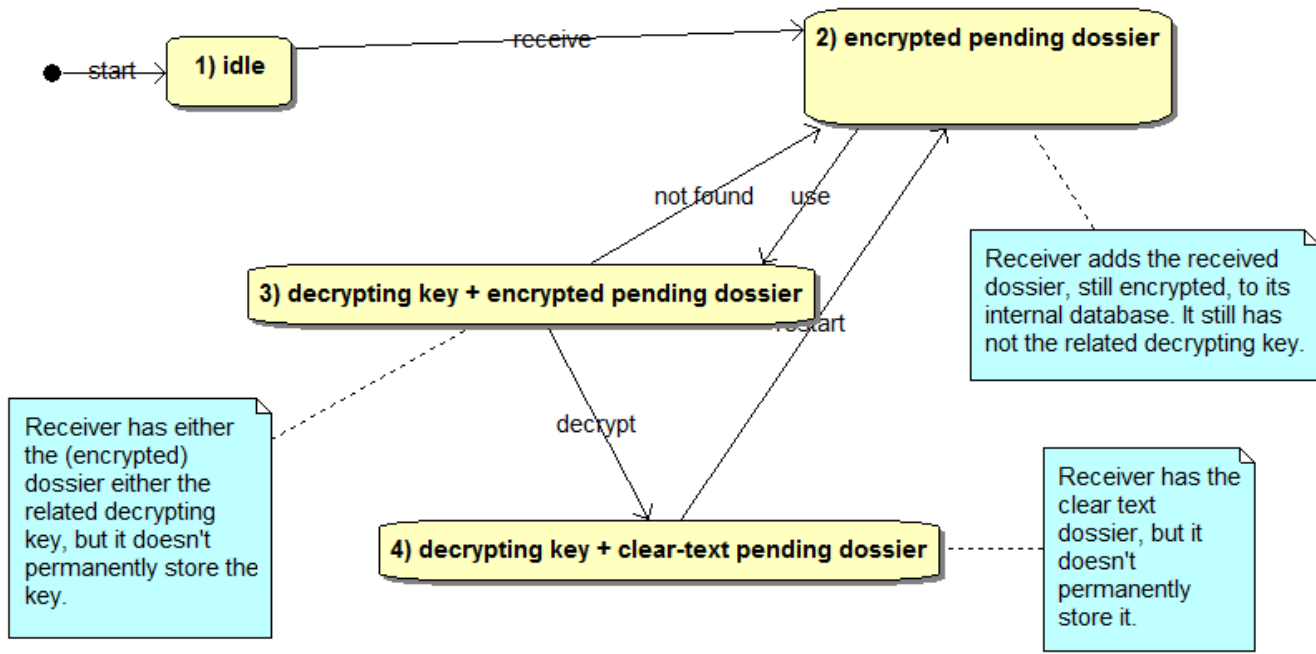

Figure 42    Receiver's state diagram

Let us analyze each state in detail:

1. The receiver starts in idle state (state 1) and there it remains until a synchronization occurs.
2. At that moment, the client goes into state 2 and receives some (encrypted) dossier, which it stores, still encrypted, in the PendingDossiersStore. In this state, the receiver has not yet access to information.
3. When the receiver wants to access a shared dossier, it asks for the decrypting key and, if this is present, receives it from the Synchronizer. Then the Receiver goes into state 3, where it has either the (encrypted) dossier either the related clear-text decrypting key. It will store the latter only in main memory, never in the persistent storage. Hence the dossier is still encrypted; if the decrypting key is not present, then the Receiver goes back into state 2.

4. The next step consists of decrypting the shared dossier by the decrypting key. The Receiver goes into state 4, in which it has the clear text dossier, but it doesn't permanently store this information, limiting itself manipulating clear-text data in main memory.

At DB restart, the Receiver resumes to state 2).

As a result, in no state there is a permanent storage of clear-text data.

The received information is limited to the portion that the owner has sent (this satisfies the "fine-grained access", requirement 0).

If at step 3 the decrypting key is not found, because it was revoked by the owner, then the dossier remains inaccessible. Hence, the "grant-and-revoke permission on shared data", requirement 3, is satisfied).

## 10.3 Synchronizer

The Synchronizer manages public keys, pending dossiers and decrypting keys.

The actions in which it is involved are: *Grant, Send, Receive, Use*, and *Revoke*. Figure 43 describes the state diagram of the Synchronizer:

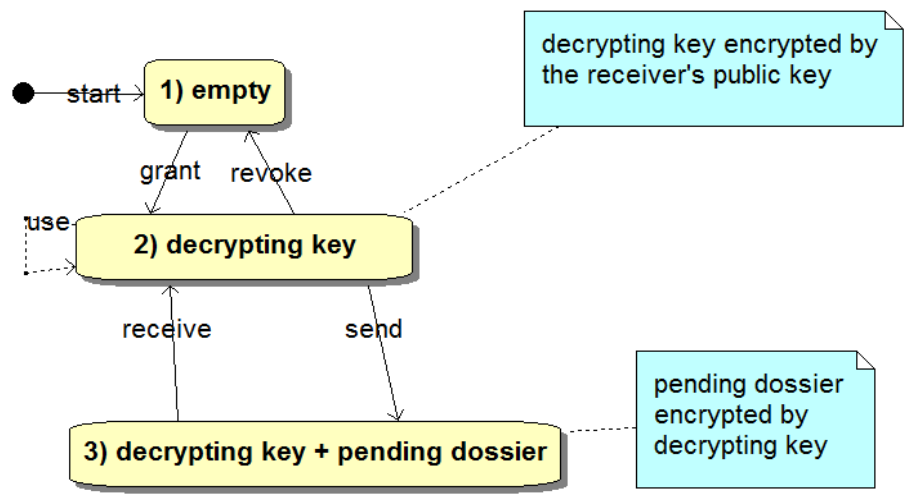

Figure 43    Synchronizer's state diagram

Let us again analyze each state:
1. At the start, the Synchronizer is empty, i.e. it has nothing stored (state 1).

2. When an owner grants the access for a dossier to another client, the Synchronizer goes to state 2, after receiving a decrypting key. That key is encrypted by the receiver's public key, so it is not understandable by the Synchronizer.

3. When the owner sends the pending dossier, this is returned encrypted by *decrypting key*. The Synchronizer passes to state 3, in which it is storing either the (encrypted) dossier either the (encrypted) related decrypting key. The dossier's privacy is guaranteed by the decrypting key, which, in turn, is guarded by the receiver's private key.

4. When the receiver issues a "receive" command, the Synchronizer goes back to state 2 and returns the requested dossier, deleting it from its storage.

5. The Synchronizer remains in state 2 if it processes a "use" command by the receiver, since this more does not alter the remote storage (it is a "read-only" command).

6. At a "revoke" command issued by the owner, the Synchronizer goes back into state 1.

In no state there is a clear-text data or a clear-text decrypting key, so the requirement 1 is satisfied. Requirements 2 and 3 are not applicable to the Synchronizer.

## 10.4 Network

For the network, as for the Synchronizer, only requirement 1 is applicable. I notice that each "message" sent on the network is encrypted:

- Each decrypting key is encrypted by the receiver's public key, and therefore it is readable only by means of the corresponding private key;
- Each pending dossier is encrypted by *decrypting key*, and therefore it is readable only having this. Since I have showed that it is understandable only by the addressed receiver, for the transitive property, also the dossier has the same characteristic.

# Part III - Validation

# 11 EXPERIMENTATION

*A real implementation of my architecture allowed us to test and benchmark the system. The realization is composed of a client application that uses a custom in-memory database with row-level encryption, and a remote synchronizer to manage inter-node communications.*

To experiment with my architecture I implemented the custom client and Synchronizer. The client needs to use row-level encryption. In the usual RDBMSs, however, this technique has significant disadvantages in terms of performance and functionality: querying would be possible only through the construction of appropriate indexes for each column of the table (with a considerable waste of resources both in terms of time and space), while the constraints and foreign keys would be almost unusable.

Another major issue concerns the management of keys: row-level encryption could potentially lead to the generation and maintenance (and / or distribution) of a key for each row of each table encrypted with this method. To solve (or reduce) the concern, I use some advanced techniques of key management, such as:

- Broadcast (or Group) encryption [32]: rows are divided into equivalence classes, based on recipients. Every class is encrypted using an asymmetric algorithm where the encryption key is made in a way that each recipient can decrypt the information using only its own private key. Both the public and the private keys are generated by a trusted entity.
- Identity Based Encryption [30]: it bounds the encryption key to the identity of the recipient. Each recipient generates by itself a key pair used to encrypt/decrypt information.

- Attribute Based Encryption [31]: it bounds the encryption key to an attribute (a group) of recipient. Each recipient receives from a trusted entity the private key used to decrypt, while the sender calculates the encryption key.

The complexity of these techniques is a major reason why conventional RDBMSs do not use encryption at the row-level.

To solve or reduce this limitation, I propose to use IMDBs to store the encrypted rows.

While IMDBs are limited by the amount of main memory in the host computer, they are well suited to be distributed and replicated across multiple nodes to increase capacity and performance.

The proposed approach works around this limitation: not having a single central database containing the whole data, I preferred to give one database for each client application. This database contains only owned data, while external data will be added (or removed) via the synchronizer, based on access permissions.

To minimize cryptography overhead, I encrypt only rows "received" by other nodes, while rows owned by the local node are stored in clear-text form.

Well-known open-source implementations of IMDB are Apache Derby, HyperSQL (HSQLDB) and SQLite. For my implementation, I chose to use HyperSQL rel. 2.0.

## 11.1 HyperSql

HyperSQL[50] is a pure Java RDBMS. Its strength is, besides the lightness (about 1.3Mb for version 2.0), the capability to run either as a Server instance either as a module internal to an application (*in-process*).

A database started *in-process* has the advantage of speed, but it is dedicated only to the containing application (no other application can query the database). For my purposes, I chose server mode. In this way, the database engine runs inside a JVM and will start one or more "in-process" databases, listening requests from processes in the local machine or remote computers.

---

[2] www.hsqldb.org

For interactions between clients and database server, we can use three different protocols:
- HSQL Server: the fastest and most used. It implements a proprietary communication protocol;
- HTTP Server: it is used when access to the server is limited only to HTTP. It consists of a web server that allows JDBC clients to connect over http;
- HTTP Servlet: as the Http Server, but it is used when accessing the database is managed by a servlet container or by an application servlet (e.g. Tomcat). It is limited to using a single database.

Several different types of databases (called catalogs) can be created with HyperSQL. The difference between them is the methodology adopted for data storage:
- res: this type of catalog provides for the storage of data into small JAR or ZIP files;
- mem: data is stored completely in the machine's RAM, so there is no persistence of information outside of the application life cycle in the JVM;
- file: data is stored in files residing into the file system of the machine.

In my work I used the latter type of databases.

A catalog file can use up to six files on the file system for its operations, the most important of which are:
- .log: used to periodically save data from the database, to prevent data loss in case of a crash; and
- .script: containing the table definitions and other components of the DB, plus data of not-cached tables.

Besides these files, HyperSQL can connect to CSV files.

A client application can connect to HyperSQL server using the JDBC driver (.Net and ODBC drivers are "in late stages of development"), specifying the type of database to access (file, mem or res).

HyperSQL implements the SQL standard either for temporary tables either for persistent ones. Temporary tables (TEMP) are not stored on the file system and their life cycle is limited to the duration of the connection (i.e. of the Connection object). The visibility of data in a TEMP table is limited to the context of connection used to populate it. With regard to the persistent tables, instead, HyperSQL provides three different types of tables, according to the method used to store the data:

- MEMORY: it is the default option when a table is created without specifying the type. *Memory table* data is kept entirely in memory, while any change to its structure or contents is recorded in .log and .script files. These two files are read at the opening of database to load data into memory. All changes are saved when closing the database. These processes can take a long time in the case of tables larger than 10 MB.
- CACHED: when this type of table is chosen, only part of the data (and related indexes) is stored in memory, thus allowing the use of large tables at the expense of performance.
- TEXT: the data is stored in formatted files such as .csv.

In my implementation, I use MEMORY tables.

Figure 44 summarizes the structure of HyperSQL.

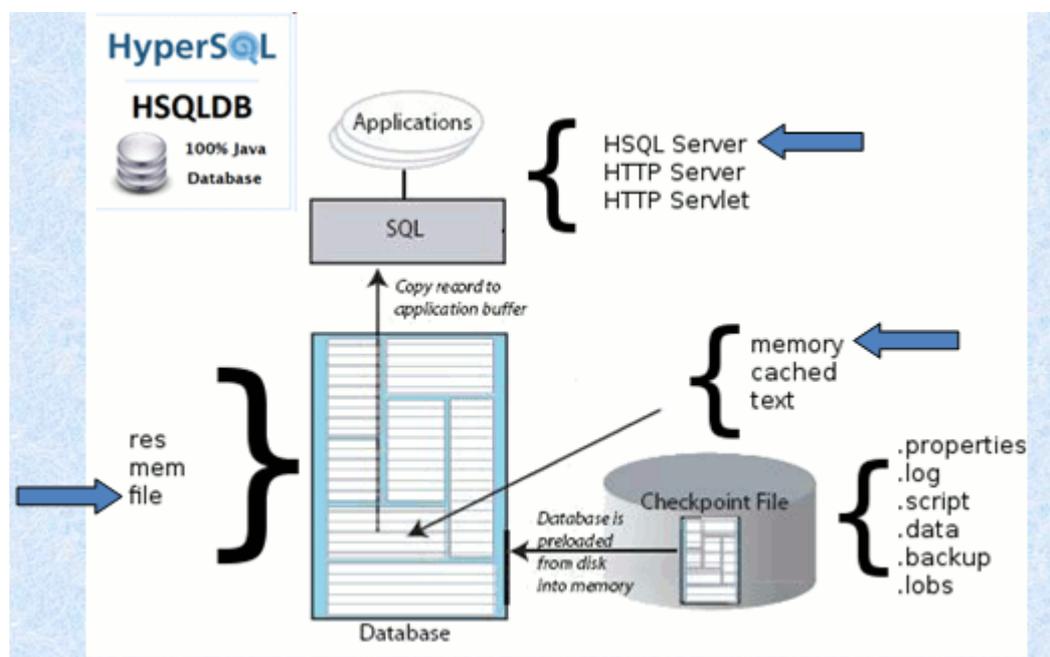

Figure 44    HyperSQL Structure

The Loader and the Serializer are the main parts of HyperSQL that I analyzed and modified. They are the mechanisms that load the data from text files at the opening and save them to the database at closing.

### 11.1.1 Loader

I suppose that the client connects to the DBMS using instructions like:

```
Class.forName("org.hsqldb.jdbcDriver" );
Connection  c = DriverManager.getConnection(
 "jdbc:hsqldb:file:myDB",  "SA", "");
```

Having used a catalog of file type, the static method newSession() of class org.hsql.DatabaseManager is called. Its task is to open the database or to connect to it (if it is already opened). org.hsql.Database is the class that represents the instance of the database in memory, so this is the root of all data structures designed to contain the information of the database. Once the database is loaded into memory, two fundamental classes are used for the parsing of text files: org.hsqldb.ParserCommand (for management of sessions and statements) and org.hsqldb.Scanner (for the recognition of individual SQL tokens). The class responsible for maintaining the database (related to the session) is org.hsqldb.SessionData, whose main attributes are:

```
private final Database database;
private final Session session;
PersistentStoreCollectionSession  persistentStoreCollection;
```

PersistentStore is the data structure that contains all rows in a database table. Specifically, this is an interface implemented by using different classes depending on the type of table represented: in my case I use MEMORY tables, so that the affected class is the org.hsqldb.persist.RowStoreAVLMemory. When the Database object is created, particularly at the invocation of method reopen(), the class org.hsqldb.persist.Logger, which is the class that represents the interface for I/O to and from text files of the database, is instantiated. The starter method of Logger class is openPersistence(), which will open the specified database (if the database is new, the related text files are created). The class org.persist.Log is instantiated after verifying the integrity of the .properties file. My focus is on method open() of this class which checks the status of the Database (if it was closed properly, if it was modified, and so on) and then instantiates the class org.hsqldb.scriptio.ScriptReaderText to read the .script file using the method readAll(Session s). The class org.hsqldb.rowio.RowInputTextLog is used to read a single line of the database and the object that represents a row in the database is the object Row. Two methods of class ScriptReaderText are invoked:

- readDDL(): reads the DDL statements and initialize a class RowInputTextLog for each line read from the *.script* file.
- readExistingData(): it extrapolates the values of each single line, initializes the row and adds it to the *PersistentStore*;

Because of the database file structure, I need to look for Insert statements to find the rows of a table. When one of these statements is encountered, it is managed by the method processStatement(Session s) of ScriptReaderText class. For each field in the row, it checks whether it is primary key and determines the data type, then the value of the field is read by the method readData (DataType t) of RowInputTextLog class.

The sequence of operations is shown in the UML diagram of Figure 45.

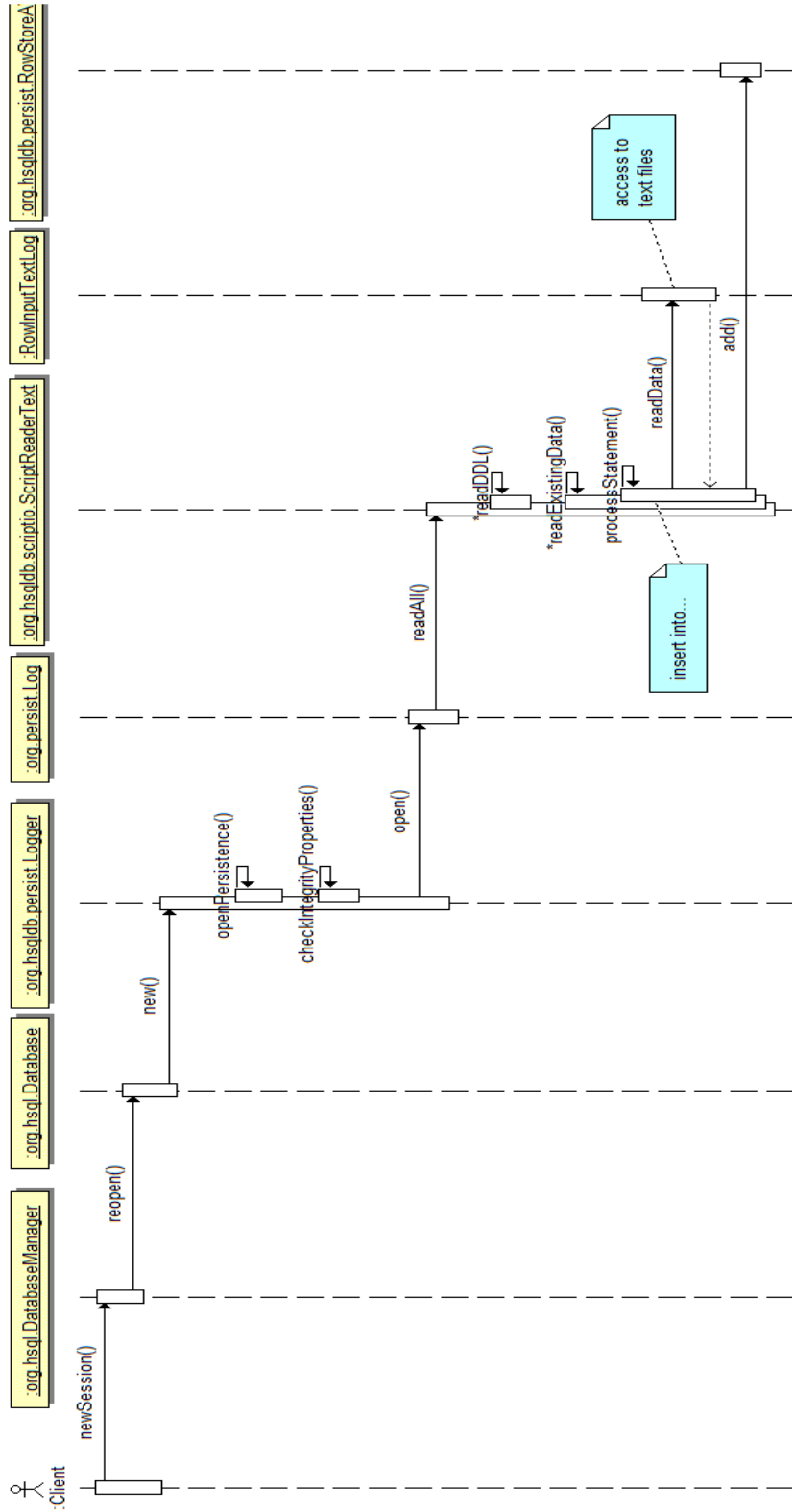

Figure 45    Loader's sequence

### 11.1.2 Insert

When the client issues a SQL Insert command, it is managed by the Table class, which calls the getNewCachedObject() method() of class PersistentStore, which add the node through the RowStoreAVLMemory class. In the end, the new data is in place into the self-balancing binary search tree in main memory. The persistent storage is not involved in this activity. It will be used, instead, in the serializer's actions.

The sequence of operations is shown in the UML diagram of Figure 46.

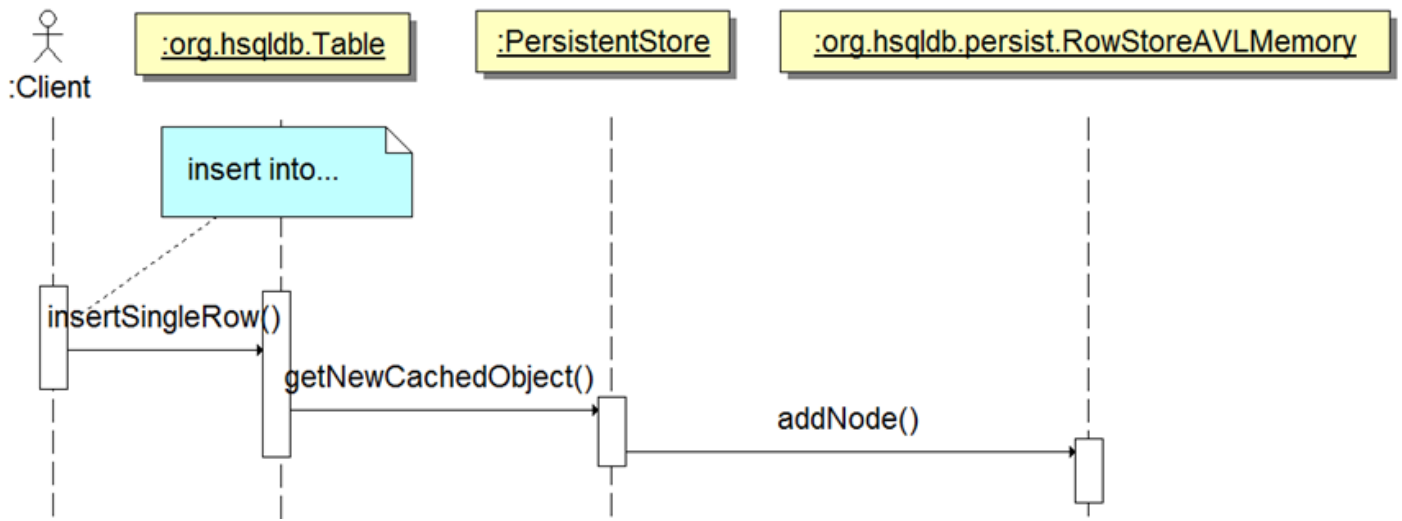

Figure 46    SQL Insert sequence

### 11.1.3 Serializer

The Serializer is the module responsible for saving the modified data into .script and .log files. Changes are initially written in .log file and moved to the .script file, when a shutdown command is issued.

The sequence of operations is shown in the UML diagram of Figure 47.

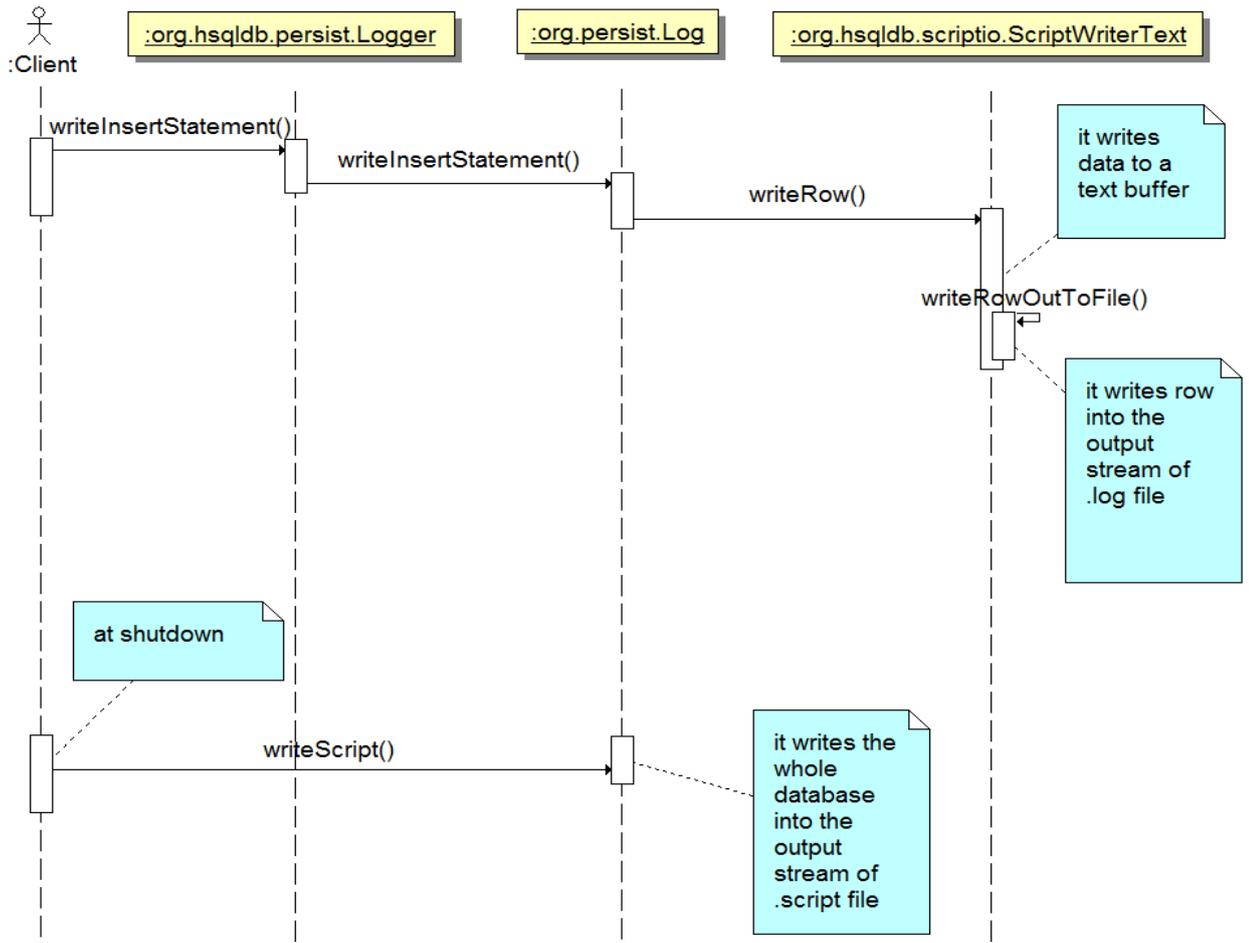

Figure 47    Serializer's sequence

Each database table is represented by an instance of class org.hsqldb.Table, comprising: data structures for the management of content, methods for creating a new table, and operations of insert/select rows. When inserting a new row, the method insertSingleRow() of the Table class is invoked; the first step is to create a new Row object for caching data in memory, which is done by the method getNewCachedObject (Session s, Object [ ] data) of PersistentStore class. Memory-type tables are kept in a balanced tree structure (AVL) implemented in the class org.hsqldb.persist.RowStoreAVLMemory. Once a node (i.e. the row being inserted) is built and added to the AVL (this operation involves several checks on the contents of the fields and of integrity constraints), HyperSQL writes the row into the buffer and then transfers it to the text file (data is written to the .log file until shutdown of the database). To perform this task, the Logger class utilizes the method writeInsertStatement(Session s, Table t, Object [] data), and the method writeInsertStatement() of the Log class. Writing to the file is done using the class org.hsqldb.scriptio.ScriptWriterBase (more precisely, in case of memory-type tables, the ScriptWriterText subclass). The method writeRow(Session s, Table t, Object [] data) of ScriptWriterText class writes data to a text buffer and, at the end of the procedure, transfers it to the file. The buffer (which is only a byte[ ]) is encapsulated in the class RowOutputBase (more precisely, in case of memory-type tables, the RowOutputTextLog subclass), which extends the HsqlByteArrayOutputStream and provides methods to transform any type of data for serializing it into the buffer. Once writing to the buffer is completed, the method writeRowOutToFile() of ScriptWriterText class is used, which calls the method write(byte [ ] b) of the class OutputStream to write into the output stream of .log file. When shutting down the database, method writeScript() of Log class is invoked with the following tasks: creating temporary file for writing .script file, loading each element of the database into memory and writing it to the file by executing the flush() of the OutputStream connected to the file.

In Figure 48 there is an instance of a *.script* file, containing the SQL statements to create and populate the DB.

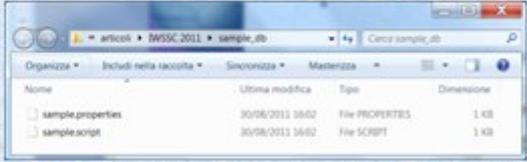

Figure 48    *.script* file's structure

## 11.2 Prototype System

### 11.2.1 Client side

On the client side, using IMDBs, I have only two interactions between each local agent and the Synchronizer (see Figure 49).

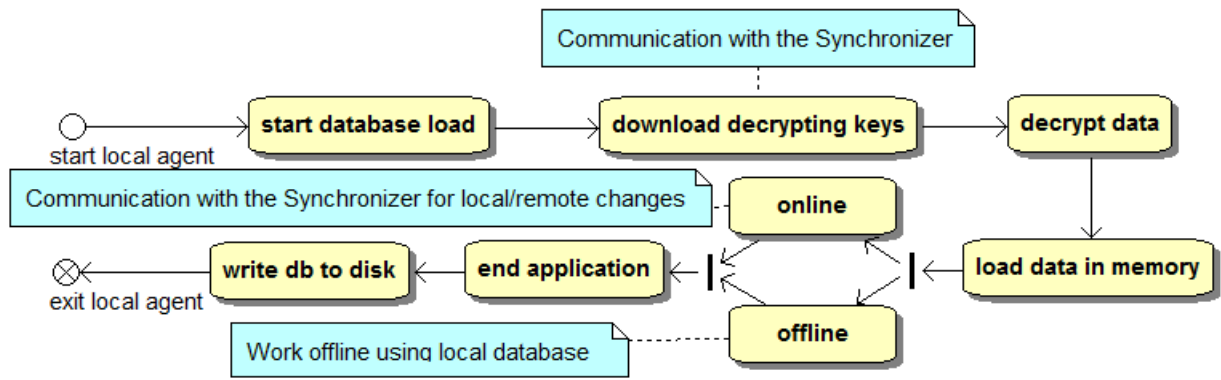

Figure 49    State diagram of client

I have modified the classes included in file hsqldb.jar to handle encryption. The basic idea was to manage encryption in the .log and .script text files. The rows that are owned by the local client are stored in clear-text, while the shared rows "granted" by other owners are stored encrypted.

The values contained in tables are stored in form of SQL insert:

INSERT INTO table_name(field_1, field_2, ..., field_n) VALUES(value_1, value_2, ..., value_n)

Earlier, to obtain control access granularity at the field level, I encrypted field by field. This way, the text contained in the database file is in the form of:

INSERT INTO table_name(field_1, field_2, ..., field_n) VALUES(pk, encrypted_value_2, ..., encrypted_value_n)

The primary key pk needs to be in clear-text, since it is used to retrieve the decrypting keys from the central Synchronizer. I dropped this idea because it requires changing the I/O code for each possible database type and an attacker may obtain some information such as table, primary key and number of rows.

My current solution is to encrypt the whole row by AES symmetric algorithm. The encryption overhead is lower than the previous solution and all information is hidden to curious eyes. To relate the encrypted row (stored locally) to the decrypting key (stored in the remote Synchronizer), I use a new key (id_pending_row). The encrypted row is prefixed by a clear-text header containing the id_pending_row delimited by "$" and "@". The encrypted value is then stored in a hexadecimal representation, so a generic row is of the form:

$27@5DAAAED5DA06A8014BFF305A93C957D

#### 11.2.1.1 Load time

At load time, the .script file will contain clear-text and encrypted rows, as the example shown in Figure 50.

![Modified .script file structure showing INSERT statements and encrypted rows prefixed with $n@ header indicating id_pending_row]

Figure 50   Modified *.script* file's structure

The class whose task is reading the file and loading the appropriate data in memory is ScriptReaderText, whose Class Diagram is the showed in Figure 51.

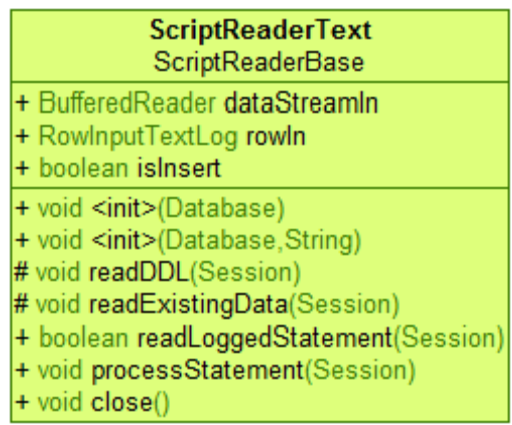

Figure 51   UML of ScriptReaderText class

The readLoggedStatement method parses each line of text in the .log or .script files and forwards the result to the processStatement method, which loads data into memory.

I changed the readLoggedStatement method to make a preprocessing: if it finds a record header (enclosed between $ and @) in the text line, it extracts the id_pending_row_received. Using this id, the client requests to the central Synchronizer the related decoding key, which it uses to decrypt the entire text line and to proceed with normal HyperSQL management. If the decoding key is unavailable, the text line is temporarily discarded (it is not deleted if it was not received for communication problem with the Synchronizer).

The sequence of operations is shown in Figure 52.

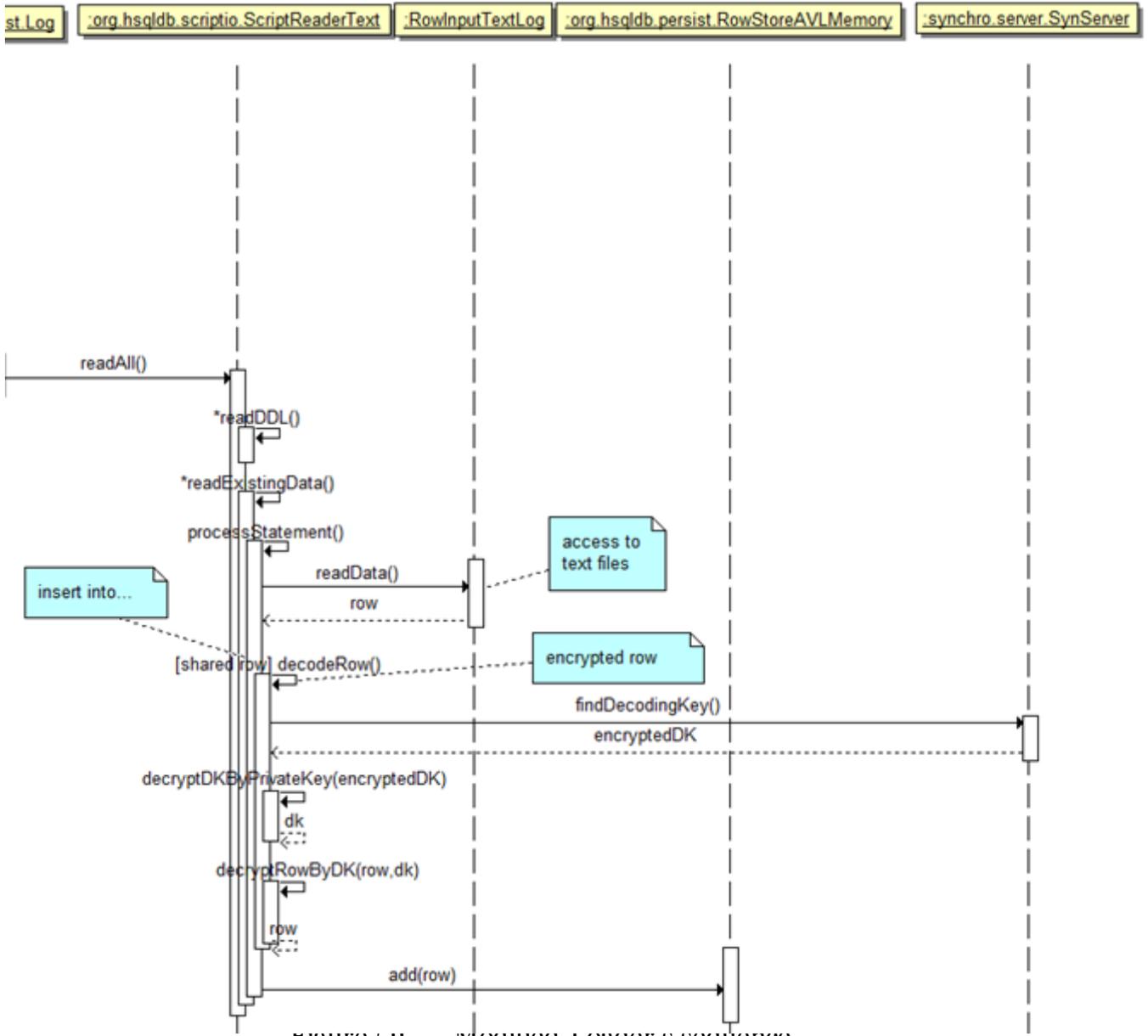

Figure 52 Modified Loader's sequence

### 11.2.1.2 Save time

The class ScriptWriterText, whose Class Diagram is shown in Figure 53, manages the write operations in .log and .script files.

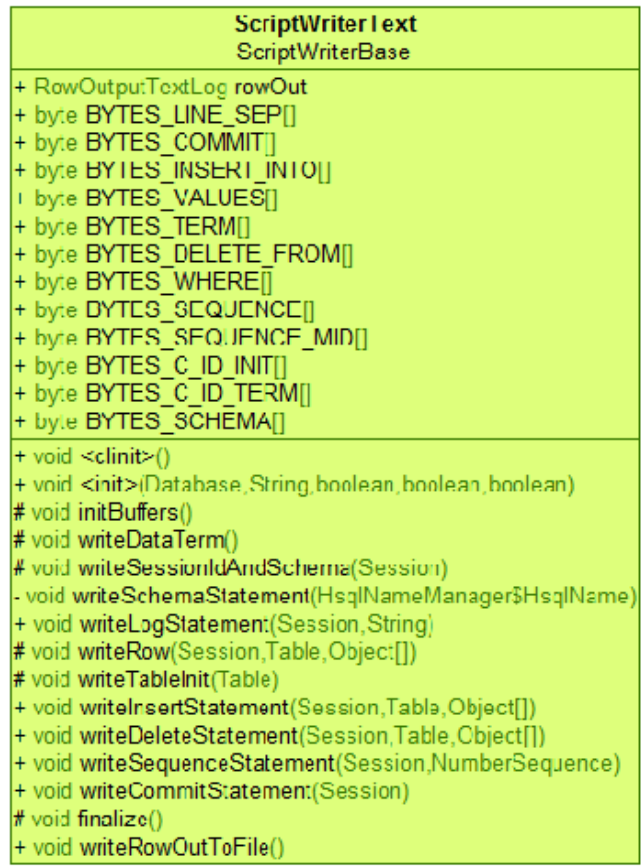

*Figure 53*     UML of ScriptWriterText class

The affected methods are writeRow and writeRowOutToFile.

The former deals with building the string that will be written into the text file (INSERT INTO ….) witch corresponds to the in memory data. A Table instance contains the information about the table structure (table name, field names, types of data, constraints, etc.). The values of fields are in an array of Object. The SQL insert is written in a text buffer that is stored in the .script file by the method writeRowOutToFile. Because each table has an id_pending_row_received column, I modified the writeRow method to check if the row is owned or shared by another user. In the latter case (id_pending_row_received not null), the custom writeRowOutToFileCrypto method is used instead of the writeRowOutToFile method. WriteRowOutToFileCrypto uses the parameter id_pending_row_received to query the related symmetric encryption key from the Synchronizer, needed to encrypt the whole buffer. The result is a hexadecimal sequence which is prefixed by the below header with the id_pending_row_received.

The resulting sequence of operations is shown in Figure 54.

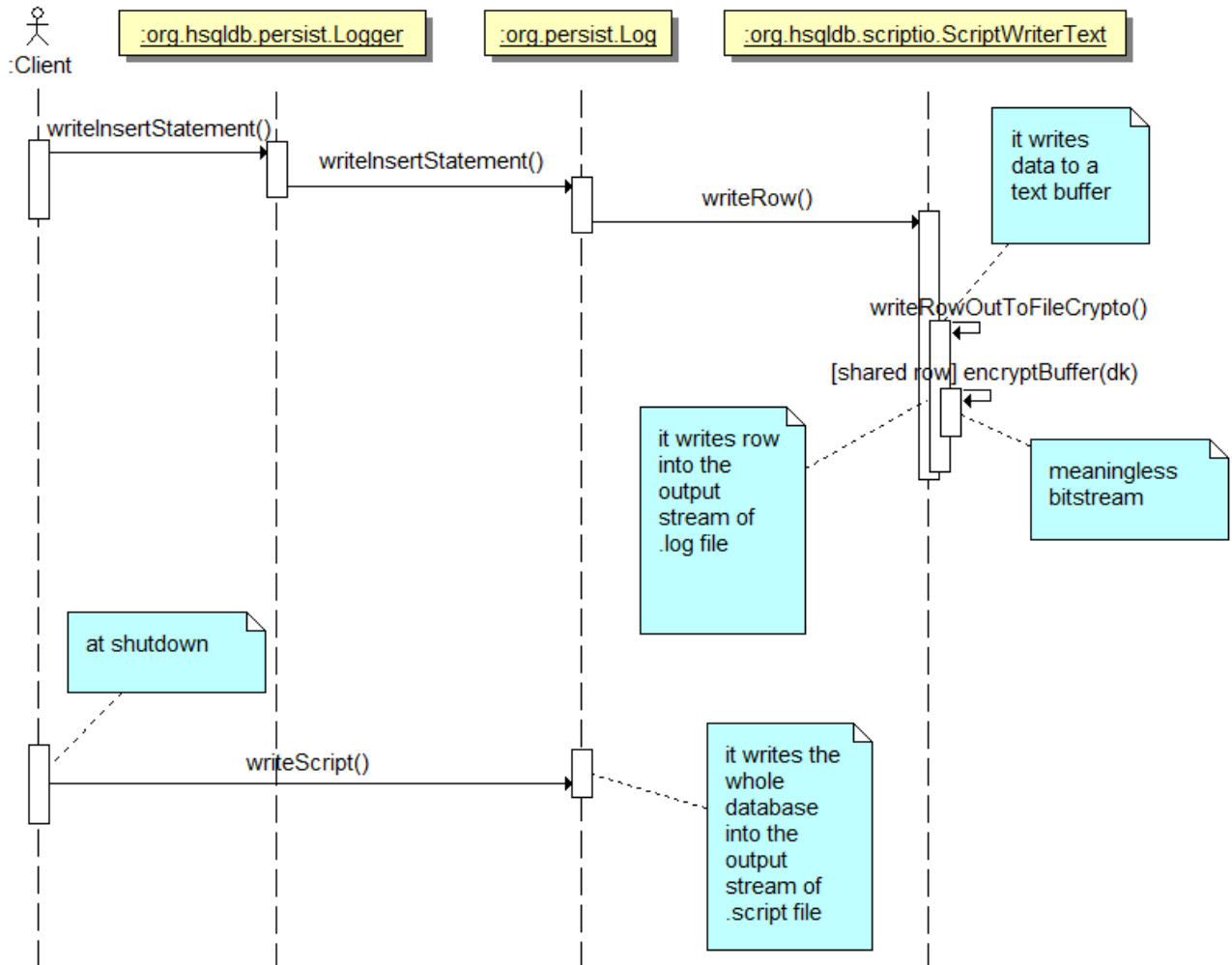

Figure 54    Modified Serializer's sequence

### 11.2.2  Server side

When a data owner adds or updates a row in the local database, it needs to distribute this change to all the related users. To do this, I put in the Cloud a central Synchronizer server that acts as a mailbox.

It uses a simple database with the following tables:

- Users: containing, among others, the id and public key of each user;
- Pending Rows: it contains the rows that were added / modified in the local database of the owner, until they are delivered to destination. A unique row_id is automatically assigned to each pending row. Other information is submission date, sender and receiver. The changed row is stored in encrypted form in field encrypted_row;
- Decrypting keys: contains the keys that are used to decrypt the pending rows. Other information is: sender, receiver, expiry date, id_row.

At modification time, the owner (client side) has to:
- Serialize the row;
- Generate a symmetric key to encrypt it;
- Encrypt the row;
- Encrypt the key by the public keys of receivers;
- Send the encrypted row and the decoding keys to receiver.

Because I store the serialized row, I haven't to worry about columns data types.

The Synchronizer uses RMI to expose its services to clients. The services are grouped into three interfaces:
- KeyInterface with methods related to encryption keys: depositKey, deleteDecriptingKey, getDecriptingKeyByIdPendingRow, getPublicKeyByUser;
- SynInterface with methods for sharing the rows: sendRow, getPendingRowForUser, getAllUsers, resendRow;
- RegistrationInterface to register and manage users: registerUser, SelectUserById, selectUserByIdAndPassword.

Figure 55 shows the Class Diagram of the resulting system.

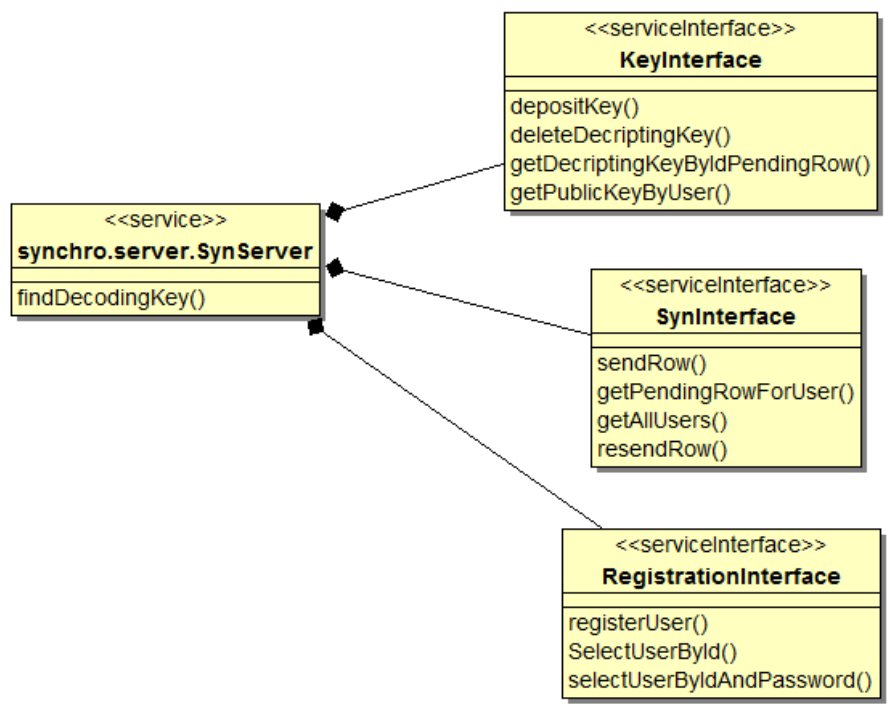

Figure 55    Synchronizer's class diagram

## 11.3  Performances

In contrast to the usual row-level encryption, which needs encryption/decryption at every data access, my solution uses these heavy operations only when communicating with Synchronizer, with a clear advantage, especially in the case of rarely modified databases.

### 11.3.1  Read operations

The system uses decryption only at start time, when records are loaded from the disk into the main memory. Each row is decrypted none (if it is owned by local node) or just once (if it is owned by a remote node), so this is optimal for read operations. Each decryption implies an access to the remote Synchronizer to download the related decrypting key and, eventually, the modified row.

### 11.3.2 Write operations

Write operations occur when a record is inserted/updated into the DB. In principle there is no overhead until the client, once it gets online, explicitly synchronizes data with the central server. At synchronization time, for each modified record, the client needs to:

- Generate a new (symmetric) key;
- Encrypt the record, and
- Dispatch the encrypted data and the decrypting key to the remote synchronizer

In the next Section, I will address the scalability that arises when synchronization/write operations are done massively.

### 11.3.3 Benchmark

To guarantee scalability, it is tantamount to measure the overhead that encryption brings to the custom version of HyperSQL for iPrivacy (henceforth, encHyperSQL). To do so, I wrote a test application that uses the encHyperSQL driver and interacts with the other clients through the Synchronizer. It performs several distinct activities:

- Creation of database and sample tables
- Population of tables with sample values
- Sharing of a portion of data with another user
- Receiving shared dossiers from other users
- Opening the newly created (and populated) database

At the start, the application receives three parameters:

- Number of dossiers
- Number of clients who will be sharing, and
- Percentage of shared dossiers

The above setup is needed to evaluate the inherent overhead that I introduce in my system w.r.t. the original HyperSQL, regardless of communication delays. To minimize the latter, the central Synchronizer and the clients ran on the same computer. During experiments, for testing purposes, it was sufficient to use only two clients (to enable data sharing). For the sake of comparison, I have set up a "competing" application with the following characteristics:

- It uses the <u>original</u> HyperSQL driver;
- It doesn't share data with other clients, since the original HyperSQL has not this capability;
- When populating the database, it creates the same number of dossiers than the previous application; after benchmarking, however, it adds the number of shared dossiers, resulting in the same final number of dossiers.

I benchmarked the system using single-table dossiers of about 200 bytes, in two batteries of tests; the first with 20%, and the second with 40% of shared dossiers, which numbered from 1,000 to 500,000. The results are represented by the graphs from Figure 56 to Figure 58. It is worth noting that the overhead percentage of the modified solution rapidly decreases (with 100,000 dossiers it is around 10%), either in the first battery of tests (Figure 56), and either in the second (Figure 57). In the tests, the total delay (load + create + populate + receive) is linear in the number of dossiers and is limited, even with a huge number of dossiers (Figure 58). Local results can be slightly altered by external events not preventable (e.g., garbage collector).

### 11.3.4 Results

The delay of the system is tightly bound to communications effort with the central Synchronizer. Computing overhead is limited to just one encryption per record at write time and no more than one decryption per record at read time. Since I use symmetric encryption, these operations are very fast. The benchmark demonstrates that the delay is substantially concentrated in database opening, while the subsequent use does not involve additional delays, compared to the unmodified version.

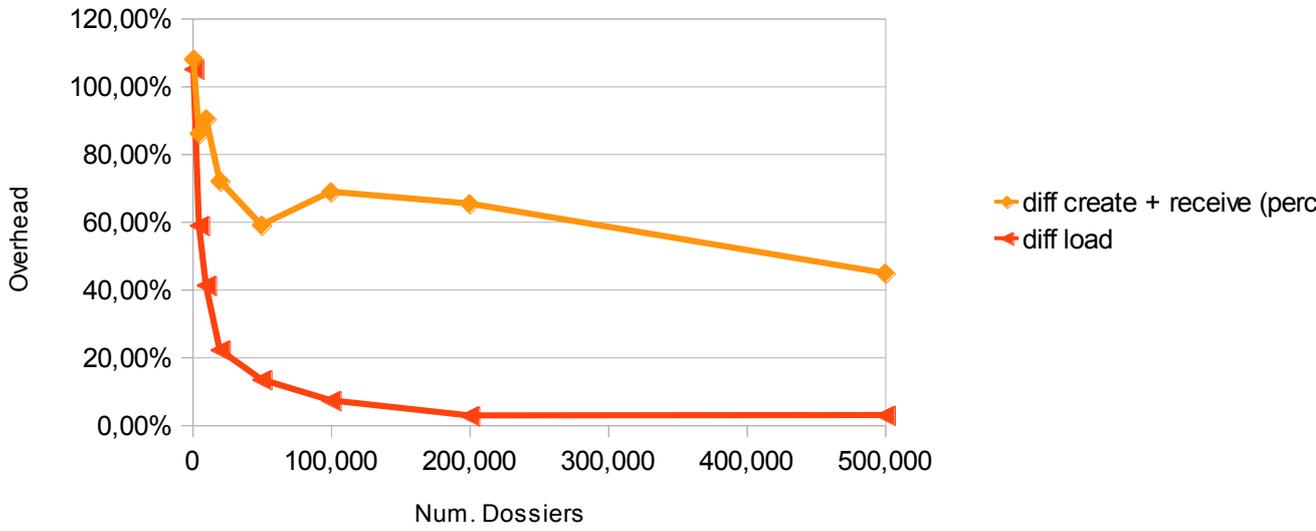

Figure 56     Overhead when 20% of dossiers are shared

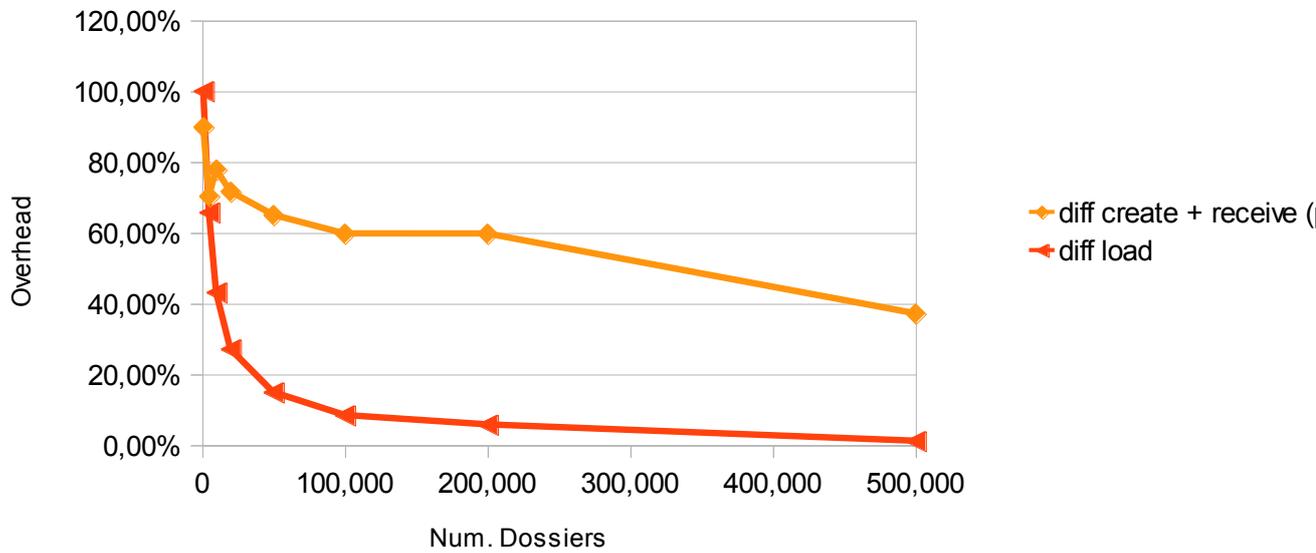

Figure 57     Overhead when 40% of dossiers are shared

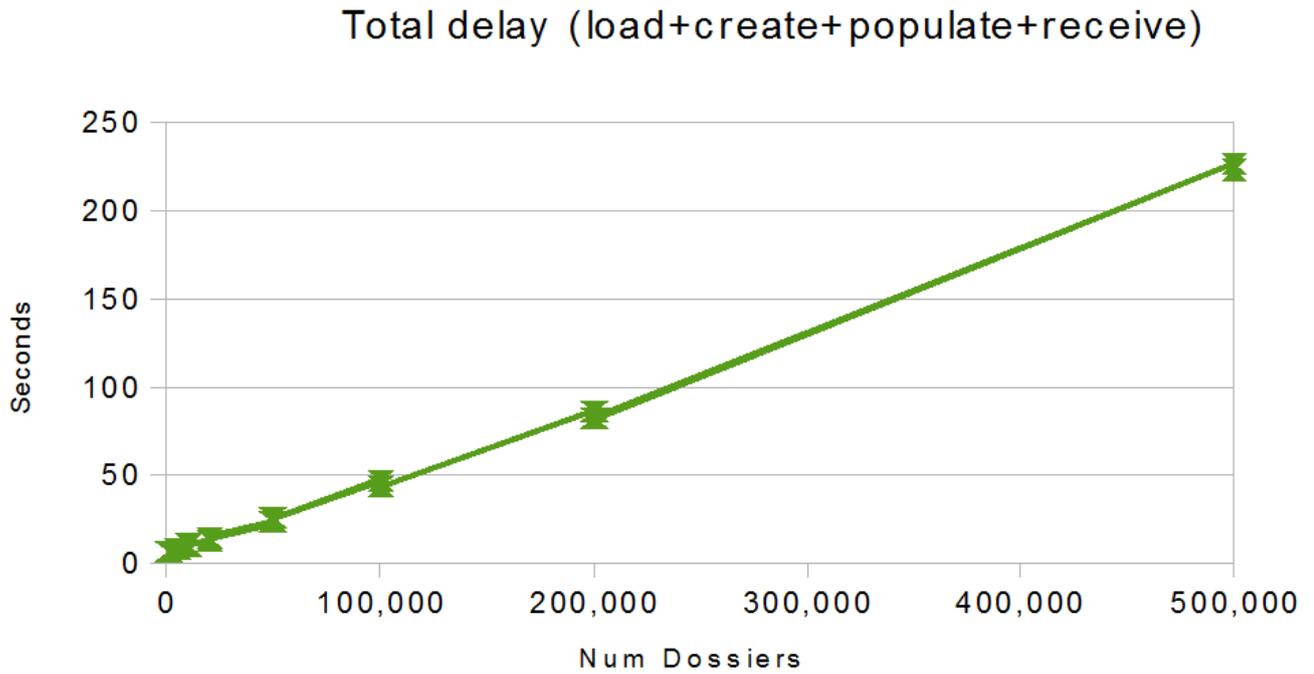

Figure 58　　Total delay

# 12 SCALABILITY

*The previous test implementation used a custom synchronizer that is almost inadequate to stress test. For this purpose, I adapted my protocols to use a general email service like Gmail as synchronizer, and evaluated the system in a large community.*

The scalability tests need a large users' network, so I resolved to use an email service on the Cloud as Synchronizer. For testing purpose I am using Gmail by Google[51]. For enhancing performance, I want to use a synchroClient thread that connects to Gmail to exchange the modified dossiers. Some relevant modifications occur for adequate to the existing email architecture:

1. I have adopted the Internet Message Access Protocol (IMAP) to interact with the server;
2. IMAP lacks the ability for a sender to delete a previously sent message, so the revoke is not on charge of Server, but it is given to client;
3. In the same way, the client is responsible to delete the emails accordingly to my algorithm.

## 12.1 The new architecture

The resulting architecture can be represented by the class diagram in Figure 59, where are represented the following classes:

1. Client, which represents the client application. Important attributes are the structures:

---

[51] www.gmail.com

a. pkHashMap to store the public keys of collaborators
   b. dkHashMap to store the decrypting keys of dossiers
   c. prList to store the pending rows (the received dossiers)
2. Gmail, which represents the email server
3. PK, which represents the public key
4. DK, which represents the decoding key
5. PR, which represents the pending row.

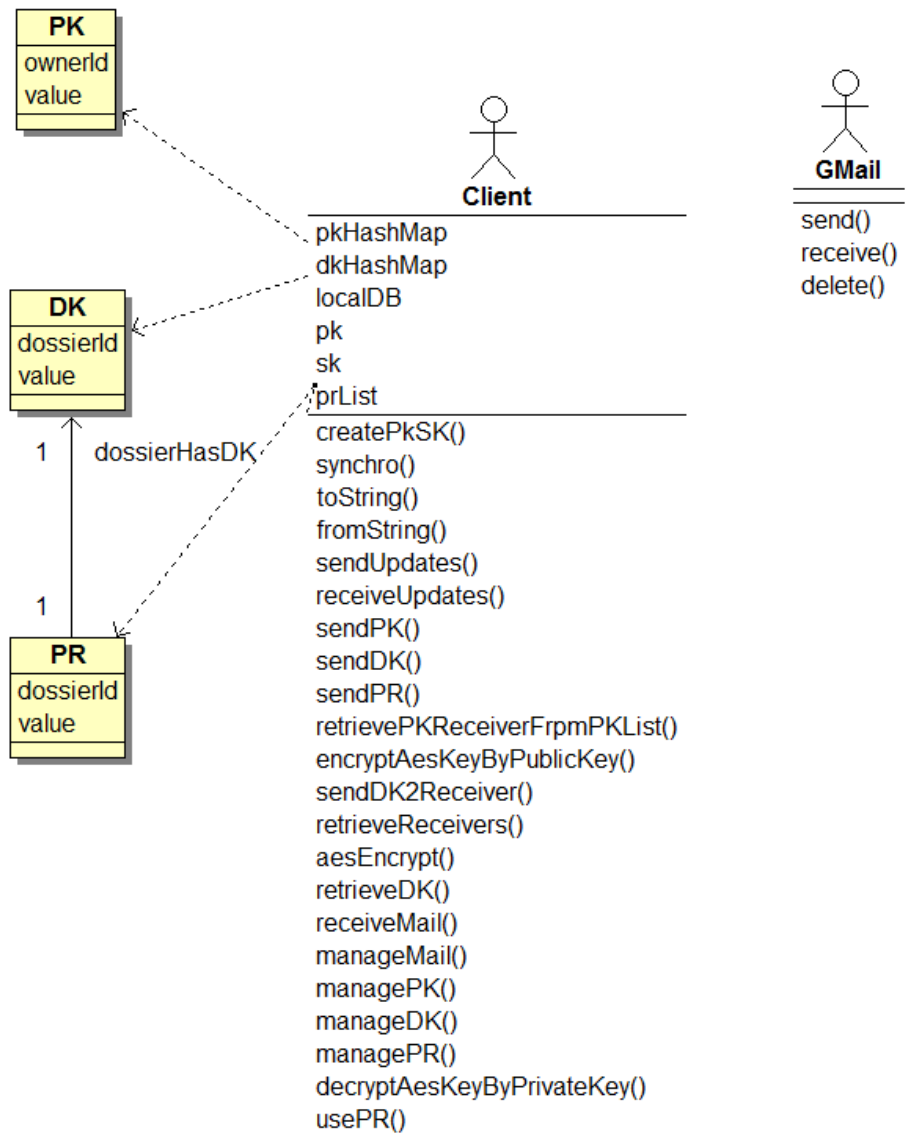

Figure 59    Class diagram

Referencing to Figure 49, the synchroClient runs almost once, at start of client, but can run every time the client is online. While in my original approach the client queries the Synchronizer every time it needs to decrypt a row, now the synchroClient connects to the email server and receives all the messages in a single session.

The client C sends to Gmail three types of messages:

1. C's public key: the message has subject "PK" and the body message contains the public key of C. The message is sent to all the clients collaborating with C.

2. Decrypting keys: for each dossier D, a decrypting keys is created at every change. The key is then sent to all the clients collaborating with C for D, after it has been encrypted using the receiver's public key. The message has subject "DK"+<dossierId> and the body message contains the encrypted key.

3. Modified dossiers: for each dossier D, the message has subject "PR" (Pending Row)+<dossierId> and the body message contains the dossiers, encrypted using a Decoding Key. The message is sent to all the clients collaborating with C for D.

## 12.2 The synchronization phase

To synchronize, a client follows the steps in Figure 60:

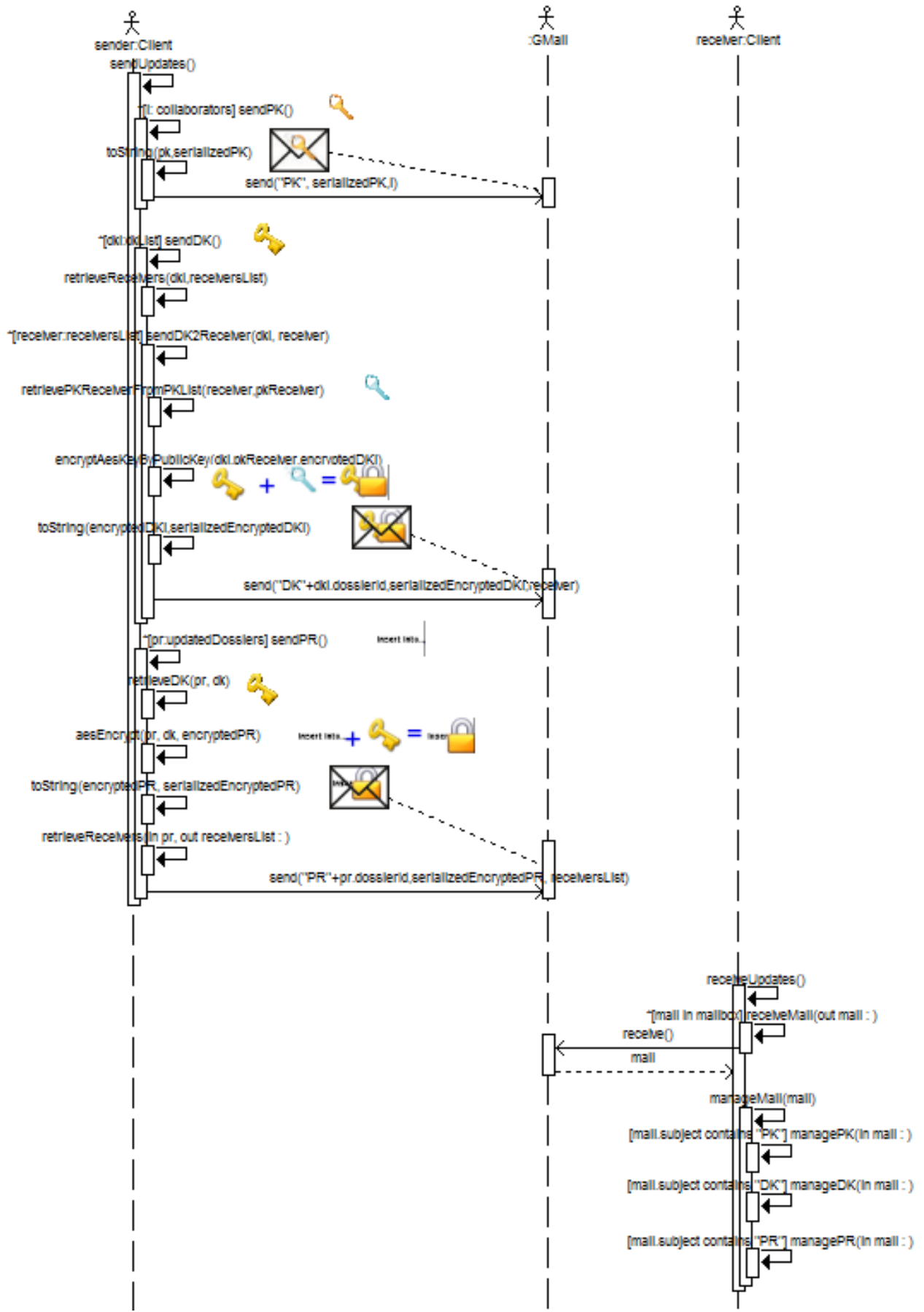

Figure 60  The synchro sequence diagram

More in detail:
1. sendUpdates:
    a. The client sends the PK to all the collaborators (only the very first time the collaboration starts);
    b. For each modified dossier D, the client finds the corresponding decrypting key DKD and the list of collaborators. For each receiver R, the client encrypts DKD using PKR and sends it to R;
    c. For each modified dossier D, the client encrypts it using DKD and sends the result to all the collaborators.
2. receiveUpdate: during this phase, the client receives the emails from Gmail and manage them accordingly to the subject.
    a. Subject "PK": manage the corresponding public key;
    b. Subject starts by "DK": manage the corresponding decrypting key;
    c. Subject starts by "PR": manage the corresponding pending row.

### 12.2.1 managePK

The activity for managing the PK, as shown in Figure 61, consists of:
1. Extracting the public key from the body of the email
2. Adding the couple <sender's email address, pk> to the structure pkHashMap
3. Deleting the message

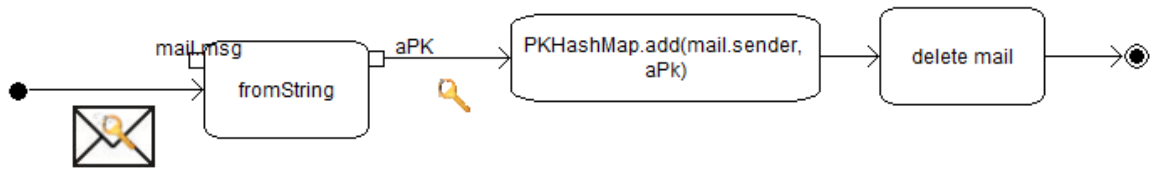

Figure 61    managePK's activity diagram

### 12.2.2 manageDK

The activity for managing the DK, as shown in Figure 62, consists in:
1. Extracting the encrypted DK from the body of the email
2. Decrypting it using the private key of the receiver

3. Extracting the dossierId from the body of the email
4. Adding the couple <dossierId, dk> to the structure dkHashMap

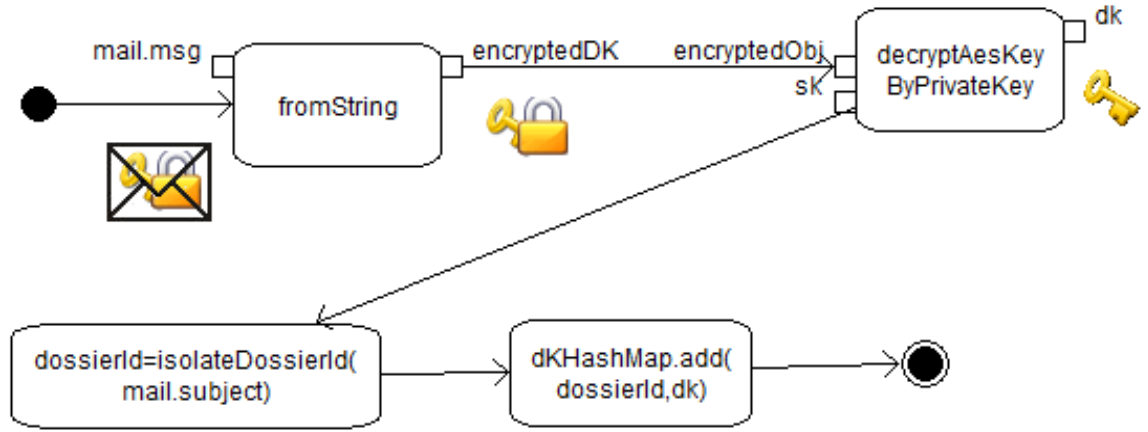

Figure 62    manageDK's activity diagram

### 12.2.3 managePR

The activity for managing the PR, as shown in Figure 63, consists in:
1. Extracting the encrypted PR from the body of the email
2. Extracting the dossierId from the body of the email
3. Adding the string $dossierId@encryptedPR to the structure prList
4. Deleting the message

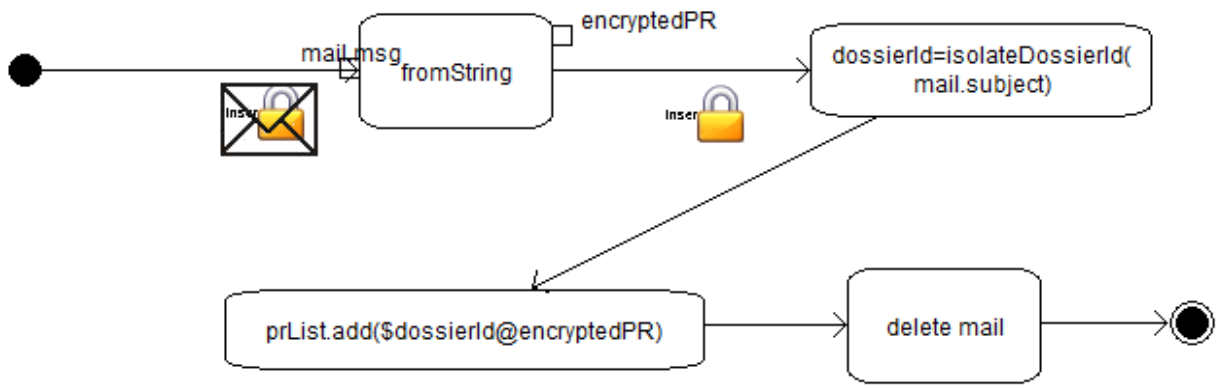

Figure 63    managePR's activity diagram

Please, note that, at the end, the row is not decrypted yet.

The prList is then usually processed, using the dkHashMap to decrypt it and to insert in the memory database the corresponding row.

## 12.3 Considerations

At start time, the client applies the algorithms exposed above to all the messages in the email server. At subsequent synchronization, instead, it can read only the new messages, which contain the changes occurred in the meantime.

Only the DKs are maintained in the email server, the other messages are erased after being read. At a generic moment T the emails' queue in the email server is of the form shown in Table 10.

Table 10    Emails' queue

| | | |
|---|---|---|
| $Dk_1$ $Dk_2$ ... $Dk_n$ | Decrypting keys of previously read pending rows | A large number |
| $Pk_i$ $Pk_j$ | Public key for new collaborations | Some, frequently none |
| $Dk_{d1}$ $Dk_{d2}$ ... $Dk_{dm}$ | Decrypting key of dossier D, encrypted by the public keys of the m collaborators | Some |
| $Pr_d$ | Pending row for document D | Some |

The first band is read only at startup, while the others are read only during the subsequent synchronizations.

Now, let be:

- $N_R$ the number of the (read) messages in the first band
- $N_C$ the number of new collaborators
- $N_{PR}$ the number of just received dossiers
- $N_G$ the average of collaborators for a given dossier
- $S_{pk}$ the size of a public decrypting key (this is a constant quantity, depending on number of bits used for asymmetric encryption)

- $S_{dk}$ the size of a decrypting key (this is a constant quantity, that depends on the number of bits used for symmetric encryption), and
- $S_d$ the size of a dossier

$S_{es}$, the size of email queue (minus the header of the messages) is approximated by the formula:

$$S_{es} = N_R * S_{dk} + N_C * S_{pk} + N_{PR} *( S_{pk} * N_G + S_d)$$

In a static environment, where the workgroup is stable and changes are rare, the order of $S_{es}$ is:

$$S_{es} = N_R * S_{dk}$$

so it is linear in the number of the shared dossiers.

If $N_D$ is the number of the recorded dossiers and $P_s$ is the percentage of shared dossiers, the formula above can be written as:

$$S_{es} = N_D * P_s * S_{dk}$$

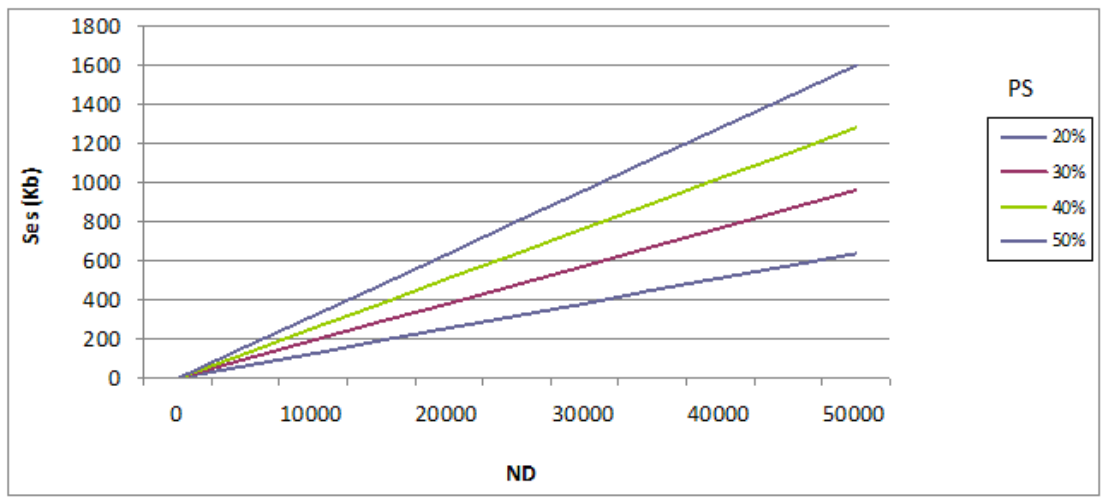

Figure 64    Values of $S_{es}$ for changing values of $P_s$

In Figure 64, $S_{es}$ is plotted for typical percentage values of $P_s$. These results can be easily related with the times one would obtain when downloading emails of the same size through Gmail.

The results of this analysis are particularly interesting in the light of the Business Collaborative Environment scenarios described in Section 8. In such environment, the value of $N_D$ is usually less than 10,000 (e.g., in Italy, the law imposes to the family doctors are bound to have less than 2,000 patients; similarly, the average number of active legal proceedings a law firm handle is 1,500[52]), while the typical value of $P_s$ is less than 20%. As the graph reveals, under these conditions, the value of $S_{es}$ is very small, about some hundred of Kb; hence, the time for downloading the updates from the central server is expected to be only a few seconds.

---

[52] source: P&P Informatics, ICT manager for legal offices in Italy – www.pep.it

# 13 COMPARISON WITH OTHER APPROACHES

*At this time, I don't know similar architectures to compare with, but I want to analyze my data propagation to BE alternatives, to quantify the possible overhead of my approach. I compare not only with pure BE schemes, but also with the OSN adapted schemes.*

My system, after all the evolvements, still uses a Multiple AES (M-AES) propagation of information to all collaborating nodes which sends a different message with the same decrypting key to each receiver. As presented in chapter 4, a better solution consists in using broadcast encryption to minimize communication overhead. In the following I analyze the application of previously seen broadcast methods with my distributed architecture, with a particular attention to the aspects of secret keys generation and revocation capability.

## 13.1 Threshold systems

These systems assume the presence of one trusted Key Generator that assigns (and then knows) the secret key. Moreover, it is intended for spot transmission (as TV broadcasting), where the information is protected only during communication but not afterwards. It lacks the revocation's capability, e.g. a protected TV transmission may be recorded and reused later. These two characteristic make this approach unfeasible for my architecture.

## 13.2 IBE

It is not a BE system, but it allows users to choose their own key, so it doesn't require a trusted service for private key generation. Using IBE we could send encrypted information to each user using her public key. This doesn't diminish the number of messages sent and, since usually information size is greater than key size, the messages are heavier. Instead, it permits time-bounded key, so it would be useful to distribute temporary decrypting keys for offline work.

As a case study, if I use AES 256 bits encryption and a dossier size of 2000Kb, I obtain the following graph:

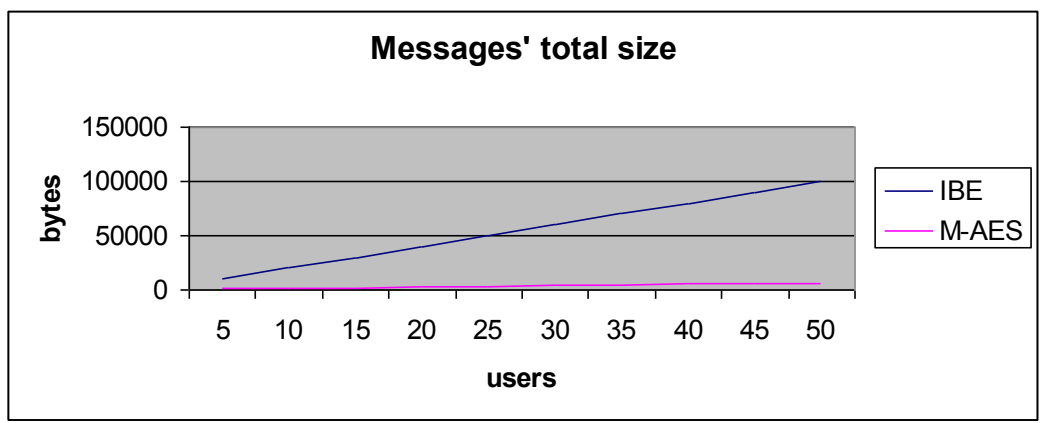

Figure 65     Comparison between using IBE and M-AES

## 13.3 ABE

These systems assume the presence of one (or a group) trusted Key Generator that assigns (and then knows) the secret keys. In my architecture, the Generator cannot be the central Synchronizer, because otherwise the dossiers' protection disappears. The use of DABE can mitigate this problem, since we can use a pool of Key Generators, distributing the responsibility on a larger number of actors to diminish the fault risk, but this is not an absolute assurance. The generator can be the owner, but this leads to the proliferation of keys as every node has to store a different secret key for each possible source of dossier.

Moreover, it lacks the revocation's capability, which was one of the initial requirements.

## 13.4 Persona

Persona was introduced for OSN, but can be a model to evaluate in my system. As this, it encrypts data using symmetric key that is then distributed encrypted by a group key (ABE). This would reduce the number of messages sent to the other users, which would be limited to only one dossier and only one decrypting key. Unfortunately, the use of ABE to manage the group encryption brings back to the previous objections.

## 13.5 GCC

This system, introduced for OSN, can be adapted to my scenario. It seems a perfect solution because:

- As IBE, each user autonomously choose her secret key without requiring a central key generator;
- As ABE, it allows the BE, but the group key is generated from the public keys of the group's component and it does not require a trusted server;
- It allows delegation and revocation of access rights;
- Each user has only a single private key and, although for each user is generated as many public keys as communities, they are used only when creating the group key (that is the key which is really used in encrypted communications) and then they are not locally stored.

Although I was able to verify the formal correctness of the algorithm exposed in [57], during implementation of my test system some problems emerged.

1. I was not able to implement the Converge method since:
   a. it is not clear what is the content of $\Psi = \{x_i, g^{y_i}\}$ (I supposed it is the couples <uid, pk> of interested nodes);
   b. to generate the group key, it needs the computing of
   $$g^{a_i} = \prod_{j=1}^{m} (g^{y_i})^{v_{i+1,j}}$$ where:
      - $v_{i+1,j}$ is obtained from inversion of a Vandermonde matrix

- $g^{y_i}$ is usually implemented using an immutable arbitrary-precision integers (as Java's BigInteger).

  i. in a first time, I inverted the Vandermonde matrix in R, but so the $v_{i+1,j}$ were real and it was impossible to calculate $g^{a_i}$. Then, I realize that I have to invert the Vandermonde matrix in $Z_q$. ;

  ii. after this step, the product to compute $g^{a_i}$ uses BigInteger^BigInteger, that is not implemented in the usual languages (as Java or C++).

2. if I solve the previous difficulty, I obtain a community key $g_k = \{g, (l_1, g^{f(l_1)}), \ldots, (l_{m-1}, g^{f(l_{m-1})})\}$ where each of the (2*m-1) terms is an integer of k bits. If I consider the usual values for AES's key length=256 bits and k=300 bits, I obtain the graph shown in Figure 66.

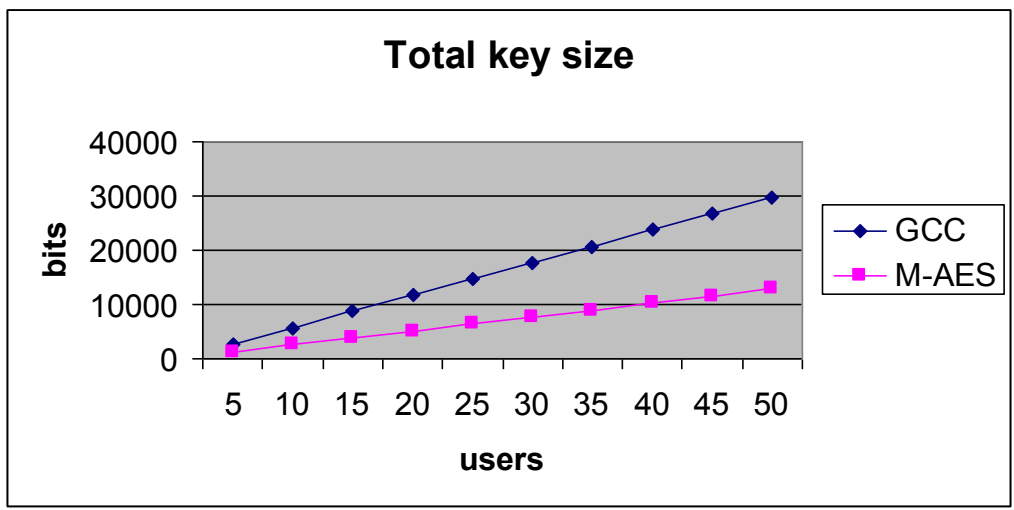

Figure 66    Comparison between AES 256 bits and GCC 300 bits

Even setting k=256 bit in GCC, its performances remains worse than M-AES, as shown in Figure 67.

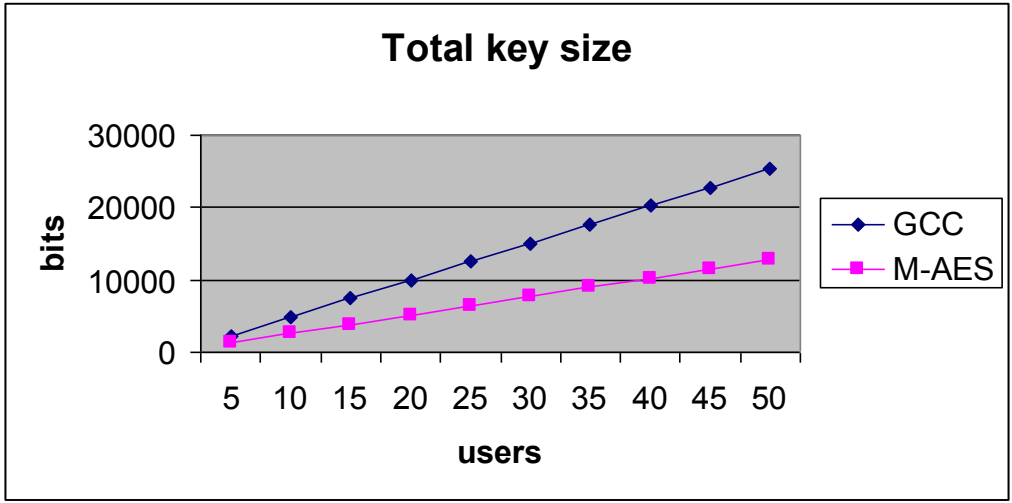

Figure 67    Comparison between AES 256 bits and GCC 256 bits

To make the results comparable, I need to set AES's key length=128 bits and k=64 bits (see Figure 68).

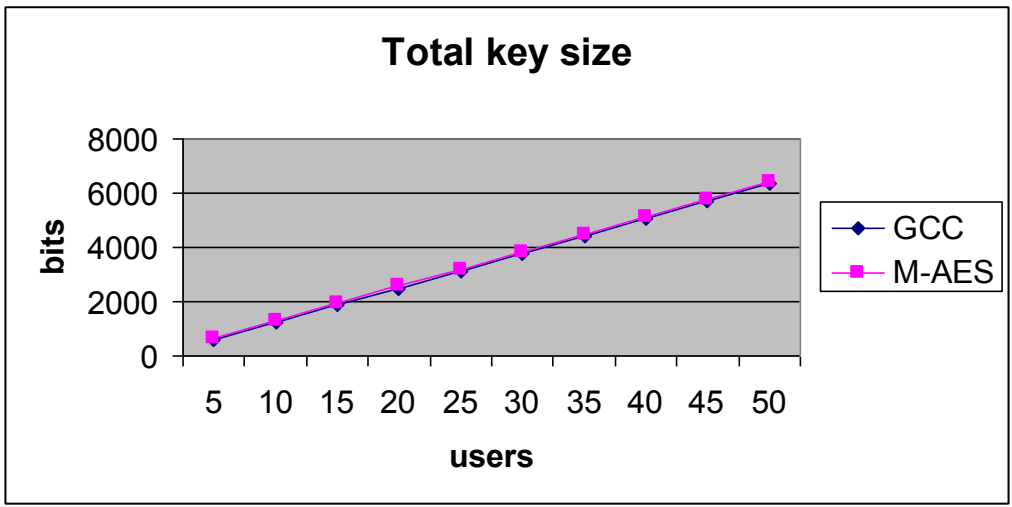

Figure 68    Comparison between AES 128 bits and GCC 64 bits

## 13.6 Conclusions

By tuning BE, one could in theory decrease the amount of data exchanged between the local Clients and the central Synchronizer. In practice, however, the examined solutions are either inadequate to the initial requirements (i.e. Threshold systems, ABE and Persona), or they introduce new complexities that cancel out the benefits of broadcasting (IBE and GCC). This result does not imply that BE is wholly inappropriate, but it stresses that further investigation would be needed before BE is deployed in this scenario.

# 14 VULNERABILITY ASSESSMENT

*My architecture was designed to assure data privacy, but are there possible vulnerabilities? In this Chapter, using UML state diagrams, I present an analytic study of possible system failures.*

Having shown in the previous chapters that my architecture meets the initial requirements (data protection and revocation of permits), in this Chapter, I want to analyze possible vulnerability of my system. I will examine every component involved in the algorithms:
- Client, distinct by role in:
    o Sender;
    o Receiver;
- Synchronizer, and
- Network.

## 14.1 Client

The use of IMDBs can lead to data loss if the client is interrupted (e.g. a hardware or application failure) before it writes data into the permanent storage. If data is owned by the client, there is nothing to do. If data was shared with another user, the client can ask to all collaborators to send again the shared dossiers.

Another problem may be the identity theft: if someone succeeds to discovery the credential of a client, she can communicate with the synchronizer. If she sends her PK to all collaborators, she would have full access to all dossiers, since she can:

- Send fake dossiers
- Can receive the dossiers which other clients sent to her. They were encrypted by her PK, and then she can decrypt them using her own SK.

The problems above analyzed are related to client in general. In the following, instead, I will discuss the peculiar vulnerability of a client acting as sender or receiver.

### 14.1.1 Sender

The sender is autonomous in its work. It manages local data through Data Manipulation Language, even in offline state; sometimes, on demand, it sends the changes to the receivers through the Synchronizer. It has the control of local dossiers and their decrypting key. It receives and stores the public keys (PKs) of the collaborators.

The possible point of failure can be a data crash, which can be resolved using the usual methods (backup and restore), or the change of data that are not under its control: PKs. It cannot send change if it has not the correct PK of each receiver. Let be $S$ the sender and $R$ the receiver. If $R$ changes its PK before $S$ sends the updates, but communicates this change to Synchronizer only after $S$ has already used the old key to encrypt the data, the latter remains not accessible (see the sequence diagram in Figure 69).

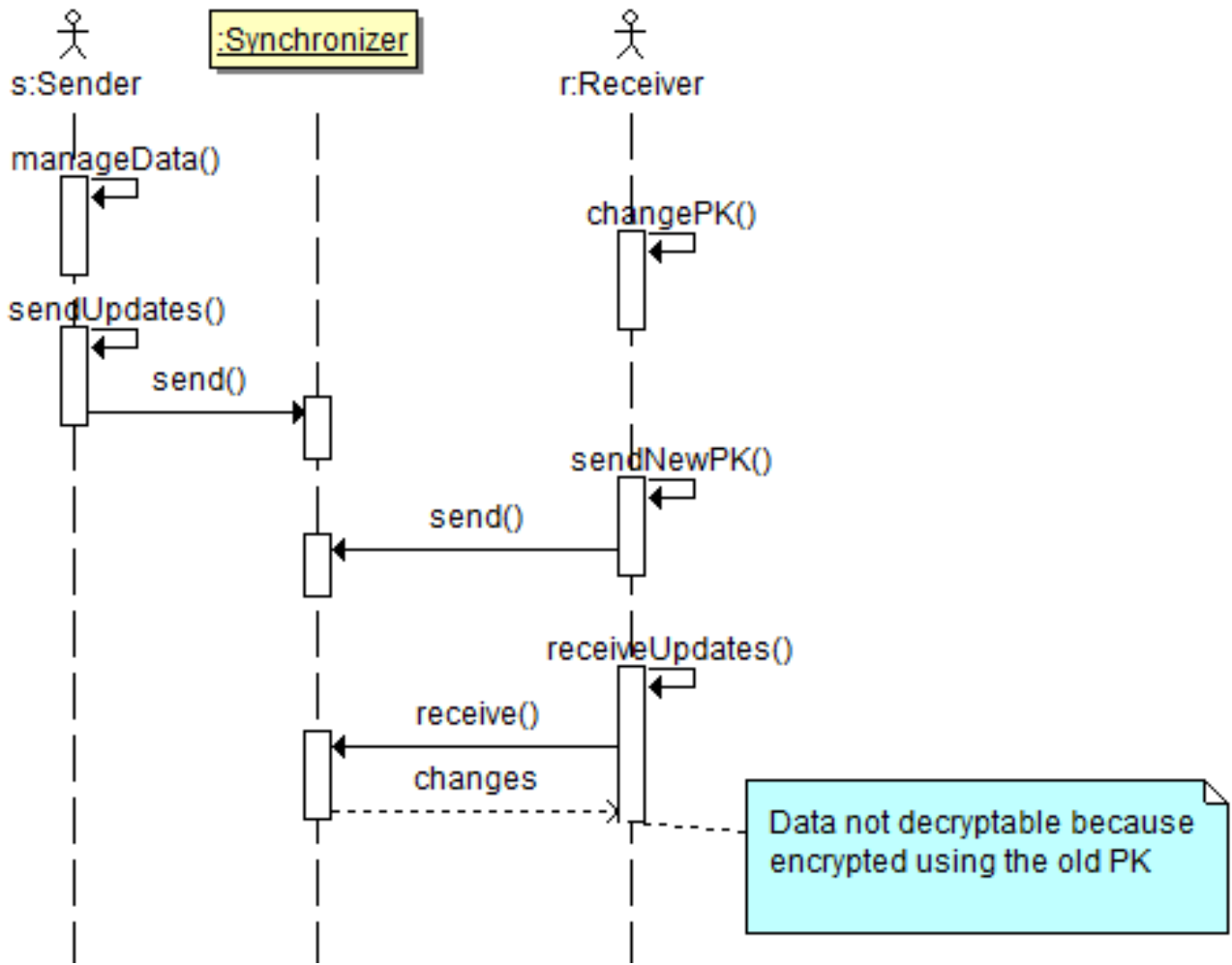

Figure 69    Fault sequence for sender

There are at least two ways to prevent this situation,:

- The receiver doesn't delete the old SK until it receives the explicit confirmation from all the collaborator
- The receiver can request to an owner to send again a dossier (the dossierID is clear text, it is ever accessible).

### 14.1.2  Receiver

Receiver is the most delicate component of the system, because it handles data that does not belong to it. My system relies on the assumption that shared data is never stored in clear form except that in main memory (volatile) and then can be accessed only during a session when the receiver obtains the related decrypting key. There is no way to prevent malicious external software to scan memory to steal the data or to substitute itself to the genuine client program. The other actor in communication, the Synchronizer, has no direct access to client, so it receives only information (as authentication) the client itself evaluates.

The same problem exists if information is remotely stored or if it is on paper: when it arrives at the final user, it has uncensored access to it (e.g., she can photocopy the document or can do a hard copy of the screen).

To discourage thefts and detect tampering, I plan to use database watermarking [26].

Another discouragement is the change of decrypting key at every update of a dossier. If a user is deleted from the receivers' list but she did a fraudulent copy of it, she cannot access the new version which will use a new decrypting key.

## 14.2 Synchronizer

A faulty synchronizer can hide a dossier and the related decrypting key or can generate fakes. The Synchronizer must always simultaneously manipulate a dossier and its decrypting key to not getting caught. To solve this problem, a solution can be the use of two non-communicating synchronizers, one for the dossiers and the other for the decrypting keys. Neither can add/delete a dossier or a decrypting key without the connivance of the other. A synchronizer can refute to send a message, but it is immediately unmasked.

As any other service in the internet, also the Synchronizer can be attacked (e.g., with a Denial Of Service attack), but the prevention of these attacks is not inherent to this thesis.

## 14.3 Network

### 14.3.1 Fault

The network can temporarily go down at distinct moments (see Figure 70):
   a. Before client's initial synchronization: the client cannot receive the decrypting keys and it cannot access the shared dossiers until network returns up. It can still access the owned dossiers, which are stored in clear form.

b. After the client's is synchronized, it has the decrypting keys and the complete database in memory. Although it may not send/receive updates until the network is back up again, it is otherwise fully operational.
c. During synchronization: the client does not receive all the updates, therefore it can have an incomplete DB until network returns up. However, the client can still use the shared dossiers for which it has already received the key. If the network fault happens after the client receives a PK or a PR, but before the Synchronizer deletes it, when network returns up, the PK will be treated as a new one, without consequences nor disruptions.

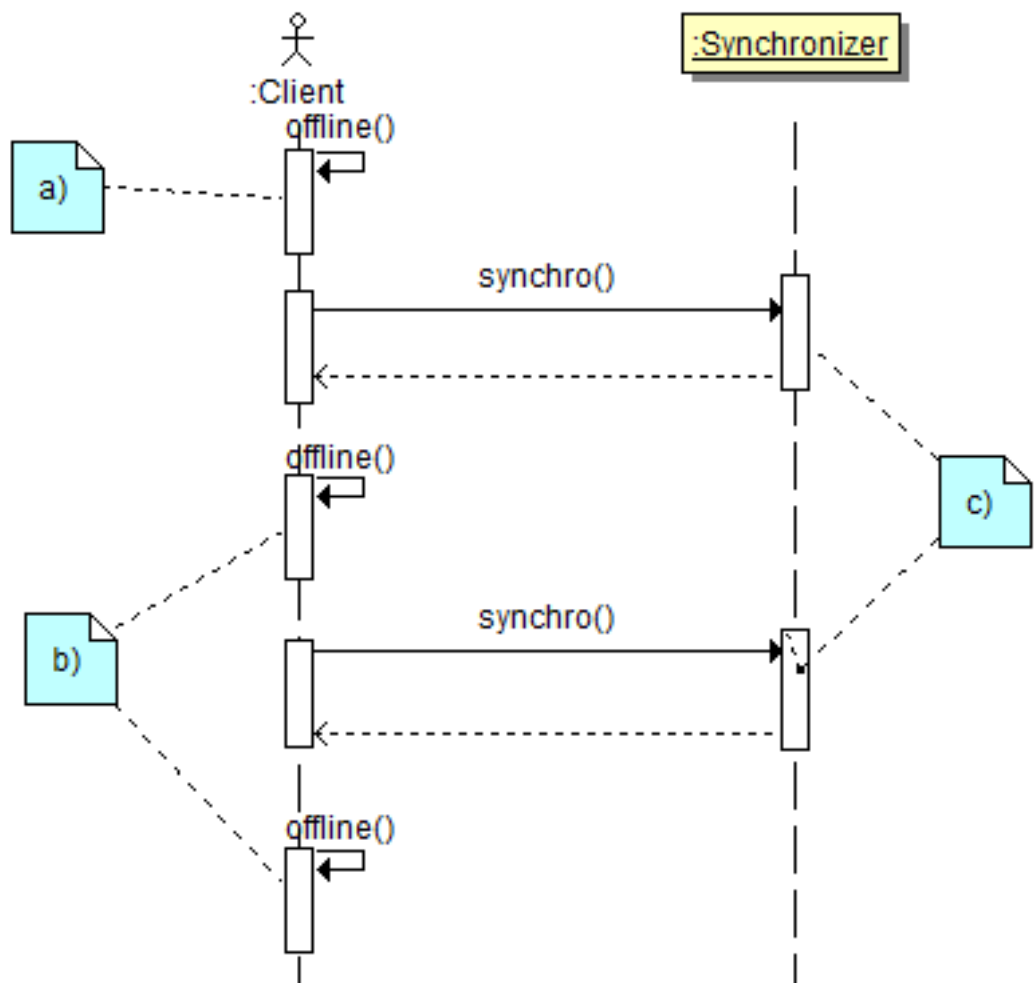

Figure 70    Possible network faults

### 14.3.2 Redirection

Let us now consider the well-known "redirection attack": it happens when the network redirects a client to a fake synchronizer. In this case, the client sends and receives messages to and from an unknown (malicious?) server $S_u$. This type of attack is easily dealt by my architecture, since one of the key assumptions is that server $S_u$ is never trusted. Thus the fake $S_u$ cannot access data at any time albeit it may damage the system acting as a faulty server as described in 14.2.

## 14.4 Conclusions

The detailed analysis in this Section has shown that the proposed architecture is effective in blocking all attacks but one, i.e., the Receiver attack (Section 14.1.2). The final data destination, at some point, has the data in the clear so at that moment there are no protections. Yet, not all is lost. As I discussed before, the use of database watermarking is a good deterrent. Even more importantly, the best defence against those attacks is the reduced interest of the attacker has in a very small portion of the whole database. Since each node contains only the dossiers of its interest, they are a small percentage of the total dossiers in the distributed system. The effort needed to crack open a local node, probably would not be worth the bother.

# 15 EXTENDED SCENARIOS

*My architecture is characterized by data privacy protection, distributed database, revocation capability, offline work. These peculiarities make it suitable for other scenarios; In this Chapter, I show two of them.*

As exposed in Chapter 8-Scenarios, my architecture perfectly fits in a Business Collaborative Environment scenario, but it can be easily extended to very different situations that have some common characteristics:
- The collaborating users are geographically scattered;
- In the network can be present an untrusted element;
- The percentage of data shared among users in the group is limited;
- A simultaneous access to the same record is uncommon;
- The actors are not always online.

In the following, I will analyze some of these scenarios.

## 15.1 Forced cooperation

I want to enforce the collaboration between a central authority and the branch offices. E.g., I want to manage a tender of an untrusted local public administration (LPA), where the bids received by the LPA can be read only after an expiration date (see Figure 71).

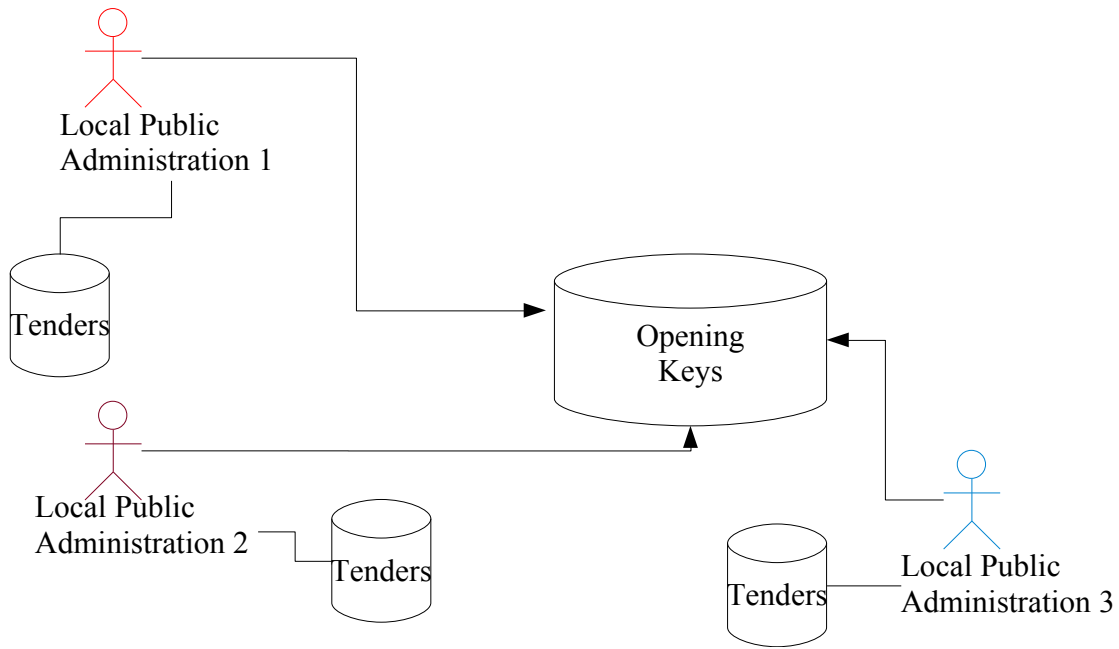

Figure 71   A forced cooperation scenario

The bids can be encrypted and then digitally signed and sent to LPA, while the deciphering keys have to be stored in a different place, e.g., a repository owned by the Central Public Administration (CPA). Neither LPA nor CPA can access the bids without exchanging data, so they control each other. LPA cannot alter a bid because it is digitally signed, nor can delete a bid because CPA knows its existence (it has the related key) and, for the same reason, it cannot add a fake bid.

## 15.2 Vehicle to Vehicle Communication

Even in automotive environment there are private data. E.g., in a car crash, information on location, number or identity of transported people can be used for ulterior motives and then it needs protection (see Figure 72).

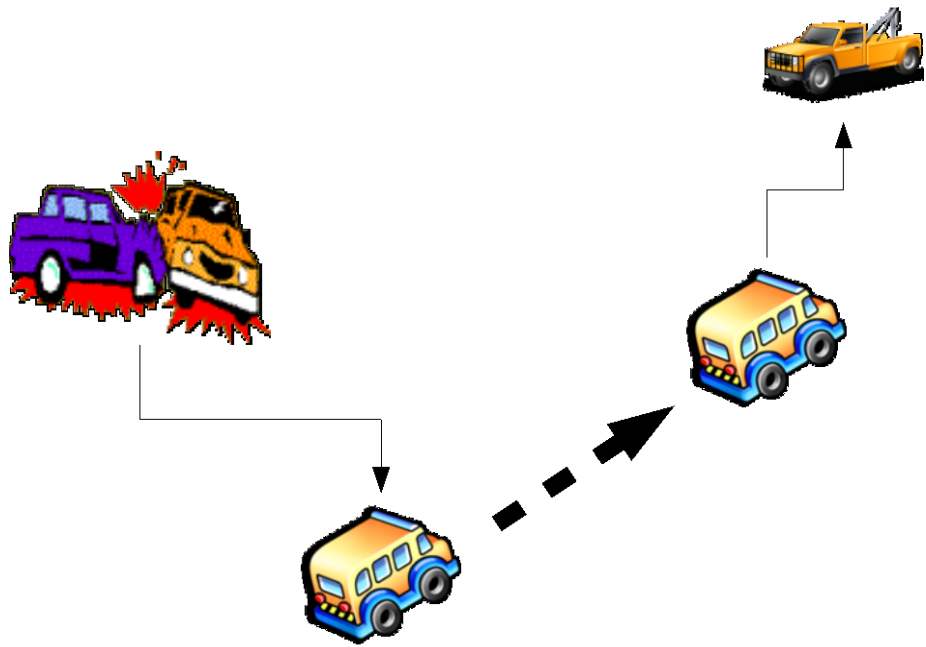

Figure 72    An automotive scenario

When the accident happens, the cars can transmit information on location, damage, garage agreement, etc. to a collector, for example a bus. When the bus meets the right breakdown truck, it can communicate these data. This situation is characterized by limited time rendezvous between actors. My architecture, that allows sporadic access to the network, can be adapted to this scenario.

# 16 CONCLUSIONS AND OUTLOOK

In this paper, I discussed the applicability of outsourced DBMS approaches to the Cloud and provided the outline of a simple yet complete approach for managing confidential data in public Clouds.

I am fully aware that a number of problems remain to be solved. A major weakness of any data outsourcing scheme is the creation of local copies of data after it has been decrypted. If a malicious client decrypts data and then it stores the resulting plain-text data in a private location, the protection is broken, as the client will be available to access its local copy after being revoked. In [22], obfuscated web presentation logic is introduced to prevent client from harvesting data. This technique, however, exposes plaintext data to Cloud provider. The plain-text data manager is always the weak link in the chain and any solution must choose whether to trust the client-side or the server-side. A better solution [26] is to watermark the local database to provide tamper detection.

Another issue concerns the degree of trustworthiness of the participants. Indeed, untrusted Synchronizer never holds plain-text data; therefore it does not introduce an additional Trusted Third Party (TTP) with respect to the approaches described at the beginning of the paper. However, I need to trust the Synchronizer to execute correctly the protocols explained in this paper. This is a determining factor that my technique shares with competing approaches and, although an interesting topic, it lies beyond the scope of this paper.

In experiment phase, I introduced a simple solution to row-level encryption of databases using IMDBs. It can be used in the Cloud to manage very granular access rights in a highly distributed database. This allows for stronger confidence in the privacy of shared sensitive data.

An interesting field of application is the use in (business) cooperative environments, e.g. professional networks. In these environments, privacy is a priority, but low computing resources don't allow the use of slow and complex algorithms. IMDBs and my smart encryption, instead, achieve the goal in a more effective way.

# APPENDIX A PUBLICATIONS

1. **E. Damiani and F. Pagano, "Handling confidential data on the untrusted Cloud: an agent-based approach," Cloud Computing 2010, pp. 61-67**

   *Abstract*

   Cloud computing allows shared computer and storage facilities to be used by a multitude of clients. While Cloud management is centralized, the information resides in the Cloud and information sharing can be implemented via off-the-shelf techniques for multiuser databases. Users, however, are very diffident for not having full control over their sensitive data. Untrusted database-as-a-server techniques are neither readily extendable to the Cloud environment nor easily understandable by non-technical users. To solve this problem, I present an approach where agents share reserved data in a secure manner by the use of simple grant-and-revoke permissions on shared data.

2. **D. Pagano and F. Pagano, "Using in-memory encrypted databases on the Cloud," IWSSC 2011, pp. 30-37**

   *Abstract*

   Storing data in the Cloud poses a number of privacy issues. A way to handle them is supporting data replication and distribution on the Cloud via a local, centrally synchronized storage. In this paper I propose to use an in-memory RDBMS with row-level data encryption for granting and revoking access rights to distributed data. This type of solution is rarely adopted in conventional RDBMSs because it requires several complex steps. In this paper I focus on implementation and benchmarking of a test system, which shows that my simple yet effective solution overcomes most of the problems.

3. **E. Damiani, F. Pagano and D. Pagano, "iPrivacy: a Distributed Approach to Privacy on the Cloud", in press, International Journal On Advances in Security**

*Abstract*

The increasing adoption of Cloud storage poses a number of privacy issues. Users wish to preserve full control over their sensitive data and cannot accept that it to be accessible by the remote storage provider. Previous research was made on techniques to protect data stored on untrusted servers; however I argue that the Cloud architecture presents a number of open issues. To handle them, I present an approach where confidential data is stored in a highly distributed database, partly located on the Cloud and partly on the clients. Data is shared in a secure manner using a simple grant-and-revoke permission of shared data and I have developed a system test implementation, using an in memory RDBMS with row-level data encryption for fine-grained data access control.